\documentclass{aa}

\usepackage{graphicx}
\usepackage{CJK}
\usepackage{enumitem}
\usepackage[varg]{txfonts}
\usepackage{amsmath}
\usepackage{amssymb}
\usepackage{footmisc}
\usepackage{booktabs}
\usepackage[colorlinks = true,
            linkcolor = blue,
            urlcolor  = blue,
            citecolor = blue,
            anchorcolor = blue,
            unicode]{hyperref}

\begin{document}

\begin{CJK*}{UTF8}{gbsn}

\title{Evolved Massive Stars at Low-metallicity \uppercase\expandafter{\romannumeral5}.\\ Mass-Loss Rate of Red Supergiant Stars\\ in the Small Magellanic Cloud}
\titlerunning{Evolved massive stars at low-Z \uppercase\expandafter{\romannumeral5}. MLR of RSGs in the SMC}

\author{
Ming Yang (杨明) \inst{1,2} \and Alceste Z. Bonanos \inst{2} \and Biwei Jiang (姜碧沩) \inst{3} \and Emmanouil Zapartas \inst{2} \and Jian Gao (高健) \inst{3} \and Yi Ren (任逸) \inst{4} \and Man I Lam (林敏仪) \inst{1} \and Tianding Wang (王天丁) \inst{7} \and Grigoris Maravelias \inst{2,8} \and Panagiotis Gavras \inst{5} \and Shu Wang (王舒) \inst{6} \and Xiaodian Chen (陈孝钿) \inst{6} \and Frank Tramper \inst{9,2} \and Stephan de Wit \inst{2, 10} \and Bingqiu Chen (陈丙秋) \inst{11} \and Jing Wen (文静) \inst{11} \and Jiaming Liu (刘佳明) \inst{12} \and Hao Tian (田浩) \inst{1} \and Konstantinos Antoniadis \inst{2, 10} \and Changqing Luo (罗常青) \inst{1} 
}
\authorrunning{Yang, Bonanos \& Jiang et al.}

\institute{
Key Laboratory of Space Astronomy and Technology, National Astronomical Observatories, Chinese Academy of Sciences, Beijing 100101, People's Republic of China\\
                 \email{myang@nao.cas.cn} \and
IAASARS, National Observatory of Athens, Vas. Pavlou and I. Metaxa, Penteli 15236, Greece \and
Department of Astronomy, Beijing Normal University, Beijing 100875, People's Republic of China \and
College of Physics and Electronic Engineering, Qilu Normal University, Jinan 250200, People's Republic of China \and
Rhea Group for ESA/ESAC, Camino bajo del Castillo, s/n, Urbanizacion Villafranca del Castillo, Villanueva de la Cañada, 28692 Madrid, Spain \and
Key Laboratory of Optical Astronomy, National Astronomical Observatories, Chinese Academy of Sciences, Beijing 100101, People's Republic of China \and
Dipartimento di Fisica e Astronomia Galileo Galilei, Universita di Padova, Vicolo dell'Osservatorio 3, I-35122 Padova, Italy \and
Institute of Astrophysics, FORTH, GR-71110, Heraklion, Greece \and
Institute of Astronomy, KU Leuven, Celestijnenlaan 200D, 3001, Leuven, Belgium \and
Department of Astrophysics, Astronomy \& Mechanics, Faculty of Physics, University of Athens, Zografos, 15783 Athens, Greece \and
South-Western Institute for Astronomy Research, Yunnan University, Kunming 650500, People's Republic of China \and
College of Physics, Hebei Normal University, Shijiazhuang 050024, People's Republic of China
}

\abstract{
The mass-loss rate (MLR) is one of the most important parameters in astrophysics, since it impacts many areas of astronomy, such as, the ionizing radiation, wind feedback, star-formation rates, initial mass functions, stellar remnants, supernovae, and so on. However, the most important modes of mass-loss are also the most uncertain, as we are still far from clear about the dominant physical mechanisms of the mass-loss. Here we assemble the most complete and clean red supergiant (RSG) sample (2,121 targets) so far in the Small Magellanic Cloud (SMC) with 53 different bands of data to study the MLR of RSGs. In order to match the observed spectral energy distributions (SEDs), a theoretical grid of 17,820 Oxygen-rich models (``normal'' and ``dusty'' grids are half-and-half) is created by the radiatively-driven wind model of the DUSTY code, covering a wide range of dust parameters. We select the best model for each target by calculating the minimal modified chi-square and visual inspection. The resulting MLRs from DUSTY are converted to real MLRs based on the scaling relation, for which a total MLR of $6.16\times10^{-3}$ $M_\sun$ yr$^{-1}$ is measured (corresponding to a dust-production rate of $\sim6\times10^{-6}$ $M_\sun$ yr$^{-1}$), with a typical MLR of $\sim10^{-6}$ $M_\sun$ yr$^{-1}$ for the general population of the RSGs. The complexity of mass-loss estimation based on the SED is fully discussed for the first time, indicating large uncertainties based on the photometric data (potentially up to one order of magnitude or more). The Hertzsprung-Russell and luminosity versus median absolute deviation diagrams of the sample indicate the positive relation between luminosity and MLR. Meanwhile, the luminosity versus MLR diagrams show a ``knee-like'' shape with enhanced mass-loss occurring above $\log_{10}(L/L_\sun)\approx4.6$, which may be due to the degeneracy of luminosity, pulsation, low surface gravity, convection, and other factors. We derive our MLR relation by using a third-order polynomial to fit the sample and compare our result with previous empirical MLR prescriptions. Given that our MLR prescription is based on a much larger sample than previous determinations, it provides a more accurate relation at the cool and luminous region of the H-R diagram at low-metallicity compared to previous studies. Finally, 9 targets in our sample were detected in the UV, which could be an indicator of OB-type companions of binary RSGs.
}


\maketitle

\section{Introduction}

Red supergiant stars (RSGs) are an "extreme" stellar population that occupies the coolest and most luminous region of the Hertzsprung-Russell (H-R) diagram. They have evolved after their main-sequence phase where they spend most of their lifetime. As Population \uppercase\expandafter{\romannumeral1} stars with an age about $\rm 8-20~Myr$ and moderately high initial masses ($\sim8-40~M_\sun$), they have low effective temperatures of $T_{\rm eff} \sim3500-4500~\rm K$, high luminosities of $\sim4000-400000~L_\sun$, and large radii of $\sim100-1500~R_\sun$ \citep{Massey1998, Massey2003, Levesque2010, Levesque2017, Neugent2020}. Given all the distinct physical properties, RSGs play a critical role in massive star formation and evolution \citep{Humphreys1979, Levesque2005, Ekstrom2013, Massey2013, Davies2017}. 

There are several possibilities for the end fate of the RSGs. However, these pathways are highly dependent on the basic physical parameters, like initial mass, metallicity, chemical composition, and more importantly (with more uncertainties), the mass-loss rate (MLR). The most common fate (to say) of RSGs is to explode as hydrogen-rich Type \uppercase\expandafter{\romannumeral2}-P core collapse supernovae (CCSN). More than half of the discovered CCSNs are Type \uppercase\expandafter{\romannumeral2}-P, and RSGs have been directly identified as progenitors in pre-explosion images of nearby events \citep{Smartt2009, Smartt2015}. Another end point for RSGs is a SN explosion at the blue end of the H-R diagram, as it may evolve backwards and become a Wolf-Rayet star (WR) for a short time \citep{Humphreys2010, Ekstrom2012, Meynet2015, Davies2018}. Recent studies suggest that some RSGs may directly collapse into a black hole without the fabulous supernova explosion, i.e. a so-called ``failed supernova'' \citep{Kochanek2008, Adams2017}. Moreover, in rare cases, episodic or eruptive mass loss events in RSGs a few months or years before the SN explosion, which likely create a thick circumstellar envelope, may also lead to Type \uppercase\expandafter{\romannumeral2}-n SNe \citep{Yoon2010, Zhang2012, Smith2015}. In any case, the final fate and the properties of the possible explosion of the RSGs must be largely impacted by their MLR. Apart from the final fate, the imprint of MLR on the shape of the wind bubbles around the stars and the chemical feedback at the surrounding environment will also be different. In that sense, stellar winds of RSGs drive the evolution of the host region so they are important in a more general way than only the final fate \citep{vanLoon2006b, Javadi2013}.

However, ``the most important modes of mass loss are also the most uncertain'' \citep{Smith2014}, as we are still far from clear about the dominant physical mechanisms of the mass-loss of RSGs (e.g., episodic mass loss, stellar winds, luminosity, metallicity, binarity, and the role of pulsation, convection, rotation on them, etc.; \citealt{Macgregor1992, Harper2001, Yoon2010, Mauron2011, Beasor2016}). The commonly used mass-loss model for the RSGs is similar to that used for the asymptotic giant branch stars (AGBs; e.g., \citealt{vanLoon2006a, Hofner2008}), since they have comparable physical properties (relatively high luminosity, low effective temperature, large radius, low surface gravity, etc.). The general physical picture is that, as the stellar atmosphere is convective and extended (gravitational binding is relatively weak at the outer boundary), pulsations (and/or convection) may lift gas to a distance (e.g., a few stellar radii) where a substantial amount of dust is able to condense with relatively low equilibrium temperature (e.g., $\sim$1,000-1,500 K). The dust envelope expansion is then driven by the radiation pressure on the dust grains, which collides with the surrounding gas and drags it along \citep{Ivezic1995, Willson2000, Elitzur2001, vanLoon2005, Verhoelst2009, Smith2014, Goldman2017, Hofner2018}.

In the context of the European Research Council project of ``Episodic Mass Loss in Evolved Massive Stars: Key to Understanding the Explosive early Universe'' (ASSESS)\footnote{http://assess.astro.noa.gr/}, which aims to determine the role of episodic mass loss in the most massive stars, we have set out to determine the MLRs of RSGs and their impact. In this paper, we present the analysis of MLR for the most comprehensive RSG sample in the SMC to date. The sample selection and data analysis are described in \textsection2 and \textsection3, respectively. The results and discussion are presented in \textsection4 and \textsection5. The summary is given in \textsection6.

\section{The red supergiant sample in the SMC}

The initial RSGs sample used in this work was taken from \citet{Yang2020} and \citet{Ren2021}, which included 1,239 and 2,138 RSG candidates in the SMC derived by using stellar evolutionary tracks and H-band related color-color diagram methods (both were constrained by Gaia astrometry; see more details in \citealt{Yang2019, Yang2020, Ren2021}), respectively. The two samples were crossmatched to remove duplication, which resulted in 2,332 unique targets. Following \citet{Yang2019, Yang2021a}, we retrieved data in 53 different bands including 2 ultraviolet (UV), 26 optical, and 26 infrared (IR) bands as listed in Table~\ref{dataset}, ranging from the UV to the mid-IR. See \citet{Yang2019} for the content of the photometric catalogs we used.

\begin{table*}
\caption{Datasets for the spectral energy distribution fitting} 
\scriptsize
\label{dataset}
\centering
\begin{tabular}{cc}
\toprule\toprule
Datasets & Filters\\
\midrule 
Galaxy Evolution Explorer (GALEX) & FUV, NUV\\
SkyMapper & u, v, g, r, i, z\\
M2002 & U, B, V, R\\ 
Magellanic Clouds Photometric Survey (MCPS) & U, B, V\\ 
NOAO Source Catalog (NSC) & u, g, r, i, z, Y\\ 
\textit{Gaia} & BP, G, RP\\ 
Optical Gravitational Lensing Experiment (OGLE) & V, I\\ 
Deep Near-Infrared Survey of the Southern Sky (DENIS) & I, J, $\rm K_S$\\ 
Vista Magellanic Cloud Survey (VMC) & Y, J, $\rm K_S$\\ 
Two Micron All-Sky Survey (2MASS) & J, H, $\rm K_S$\\ 
InfraRed Survey Facility (IRSF) & J, H, $\rm K_S$\\ 
\textit{AKARI} & N3, N4, S7, S11, L15, L24\\ 
Wide-field Infrared Survey Explorer (\textit{WISE}) & [3.4], [4.6], [12], [22]\\
\textit{Spitzer} & [3.6], [4.5], [5.8], [8.0], [24]\\
\midrule 
\end{tabular}
\end{table*}

During the initial analysis, we discovered that previously used \textit{Spitzer} Enhanced Imaging Products (SEIP) source list (including 2MASS, \textit{WISE}, and \textit{Spitzer} photometric data) might not be the best choice, while SAGE-SMC provided better photometry and quality assessment \citep{Gordon2011}. Thus, we replaced the SEIP with SAGE-SMC data and removed blended targets according to the SAGE-SMC close source flag, which resulted in 2,304 targets. Additionally, one more target was also removed due to lack of 2MASS data (visual inspection indicated that the target was contaminated by the nearby environment), since we normalized the spectral energy distribution (SED) at this wavelength. In total, the intermediate RSG sample contains 2,303 sources. Meanwhile, the \textit{WISE} data were also updated accordingly from the ALLWISE catalog, with the same photometric quality criteria ($nb=1$, $ext\_flg=0$, $w?cc\_map=0$ or $w?flg=0$) but different signal-to-noise ratios (SNRs; $\rm SNR_{WISE1}>5$, $\rm SNR_{WISE2}>5$, $\rm SNR_{WISE3}>7$, and $\rm SNR_{WISE4}>10$) for each band. Moreover, in order to take advantage of the newly released data, we updated several datasets to their latest versions. For example, Gaia data was updated to EDR3 \citep{Gaia2021}, SkyMapper data was updated to DR2 \citep{Onken2019}, NSC data was updated to DR2 \citep{Nidever2021}, VMC data was updated to DR5. New data were also added to the data pool, e.g., MCPS and DENIS. 

As we compiled the catalogs and checked the data, we also discovered several issues with respect to some datasets. For instance, the saturation flag in the VMC pipeline was not correctly applied, which resulted in a bunch of saturated stars marked as stellar instead of saturated objects (these data were discarded; Nicholas Cross, private communication). The MCPS I-band data suffered from poorly modeled point spread function (PSF; \citealt{Zaritsky2002}) and hence were discarded. Some targets fainter than about $V=15.5$ mag in M2002 were discarded since they had large uncertainties. The high SNR criteria for \textit{WISE} [12] and [22] band data were not sufficient in some cases ($\sim$1\%) identified by visual inspection of images, due to the low sensitivities and angular resolutions at long wavelengths. 

\begin{figure}
\center
\includegraphics[scale=0.47]{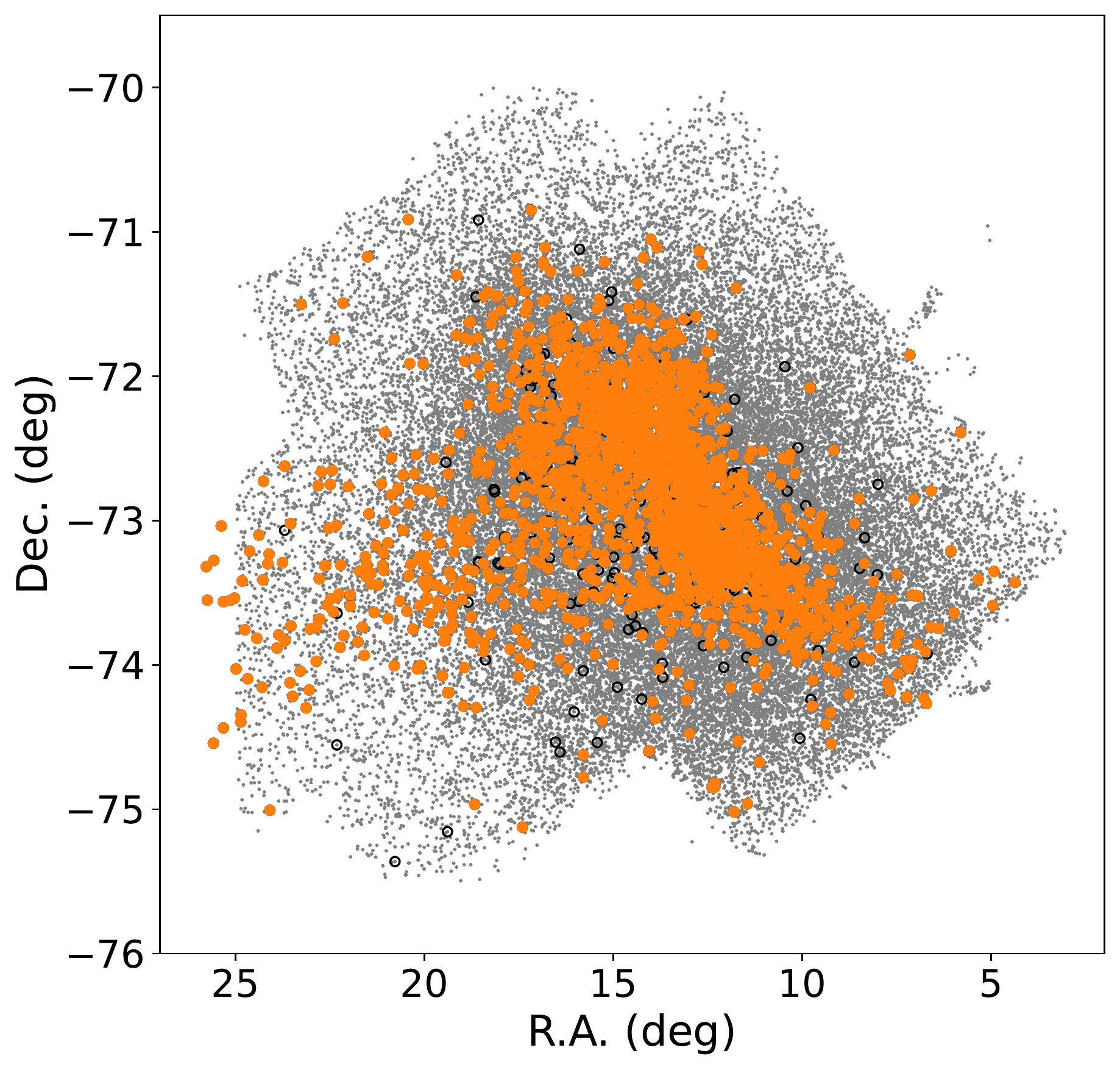}
\caption{The spatial distribution of the final RSG sample (orange color) in the SMC. Targets removed from the initial sample are marked as open black circles (see text and Figure~\ref{sample_cmd} for details). Background targets (gray dots) are from \citet{Yang2019}.
\label{sample_spatial}}
\end{figure}

\begin{figure*}
\center
\includegraphics[scale=0.43]{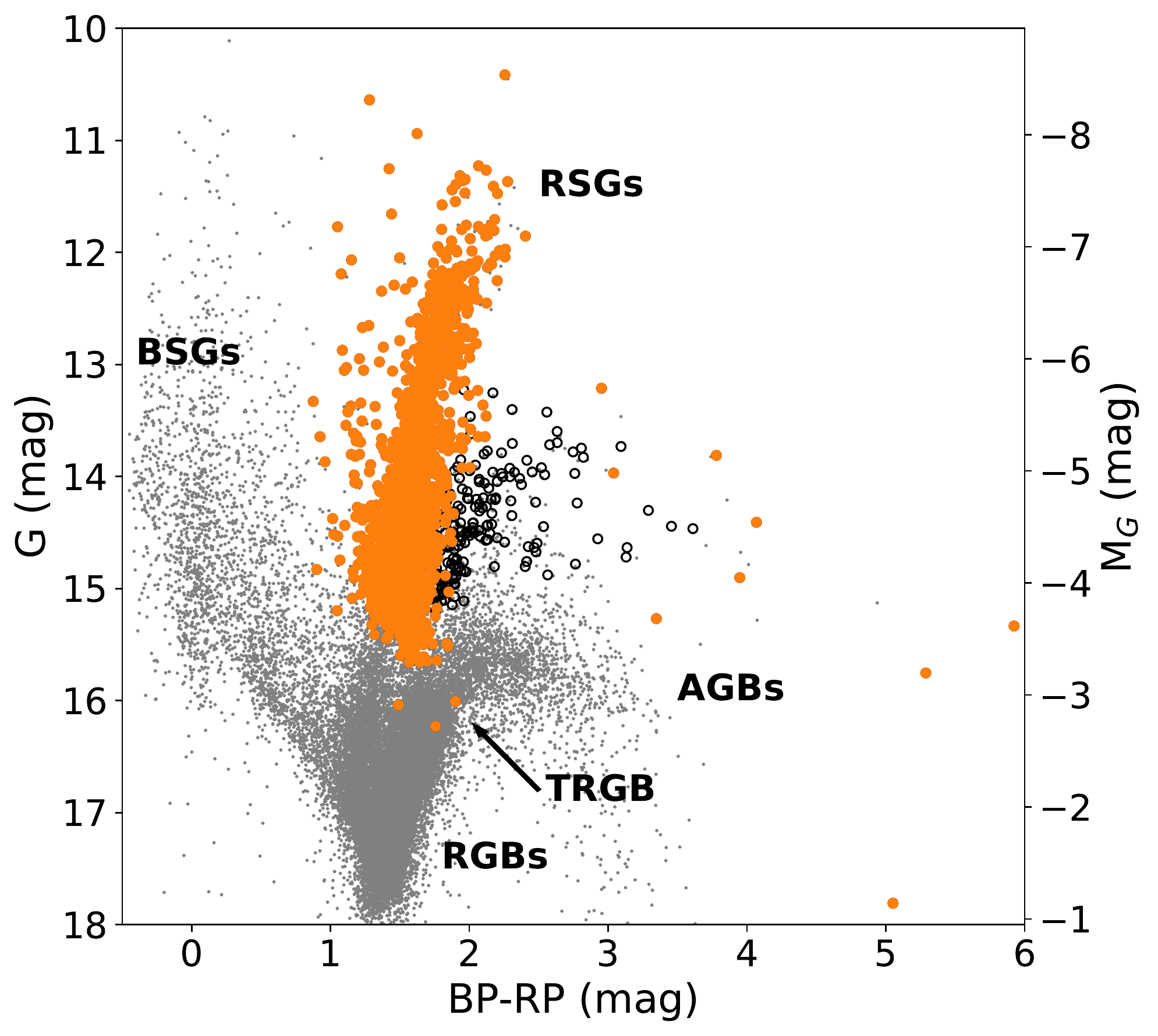}
\includegraphics[scale=0.43]{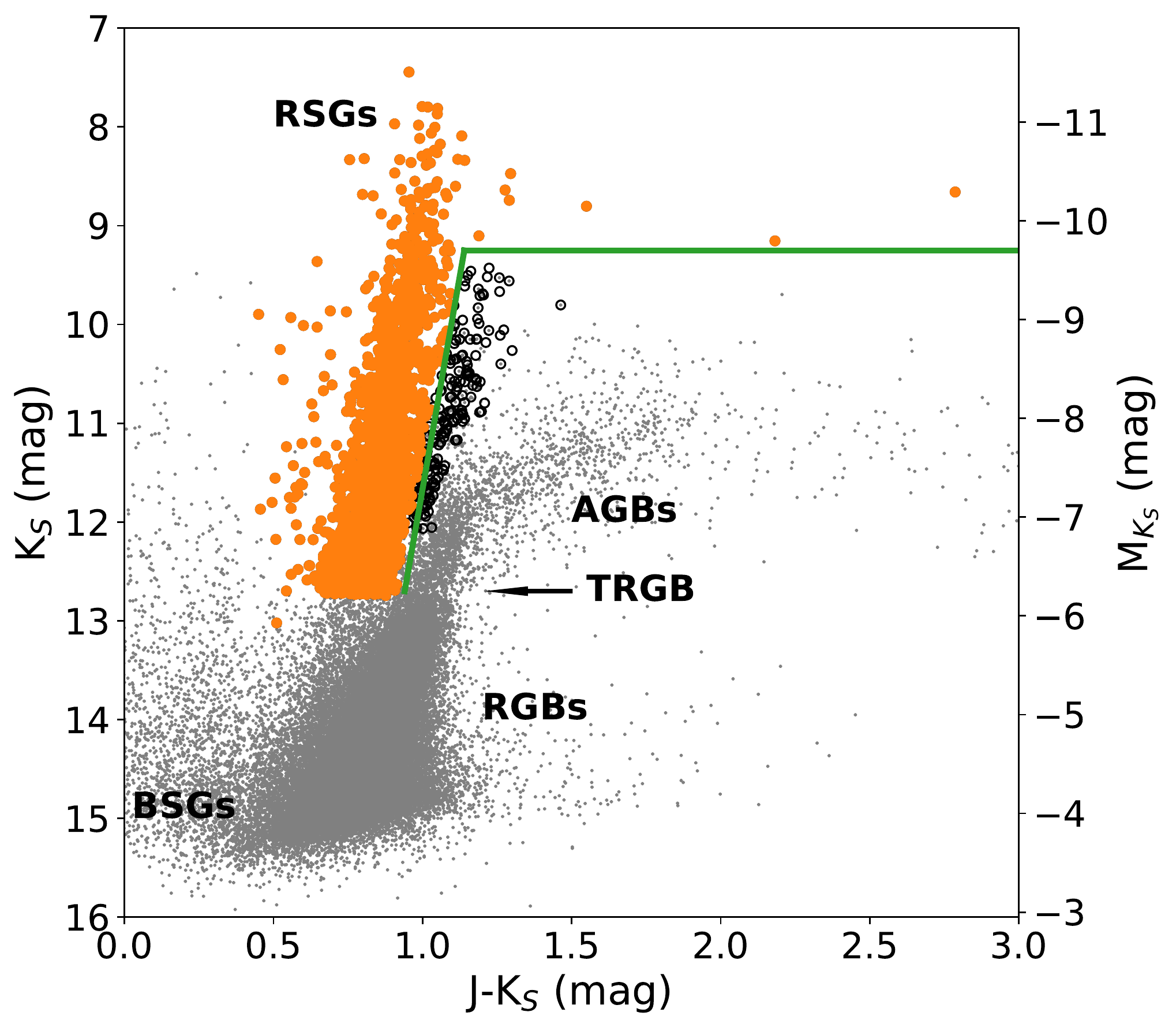}
\includegraphics[scale=0.43]{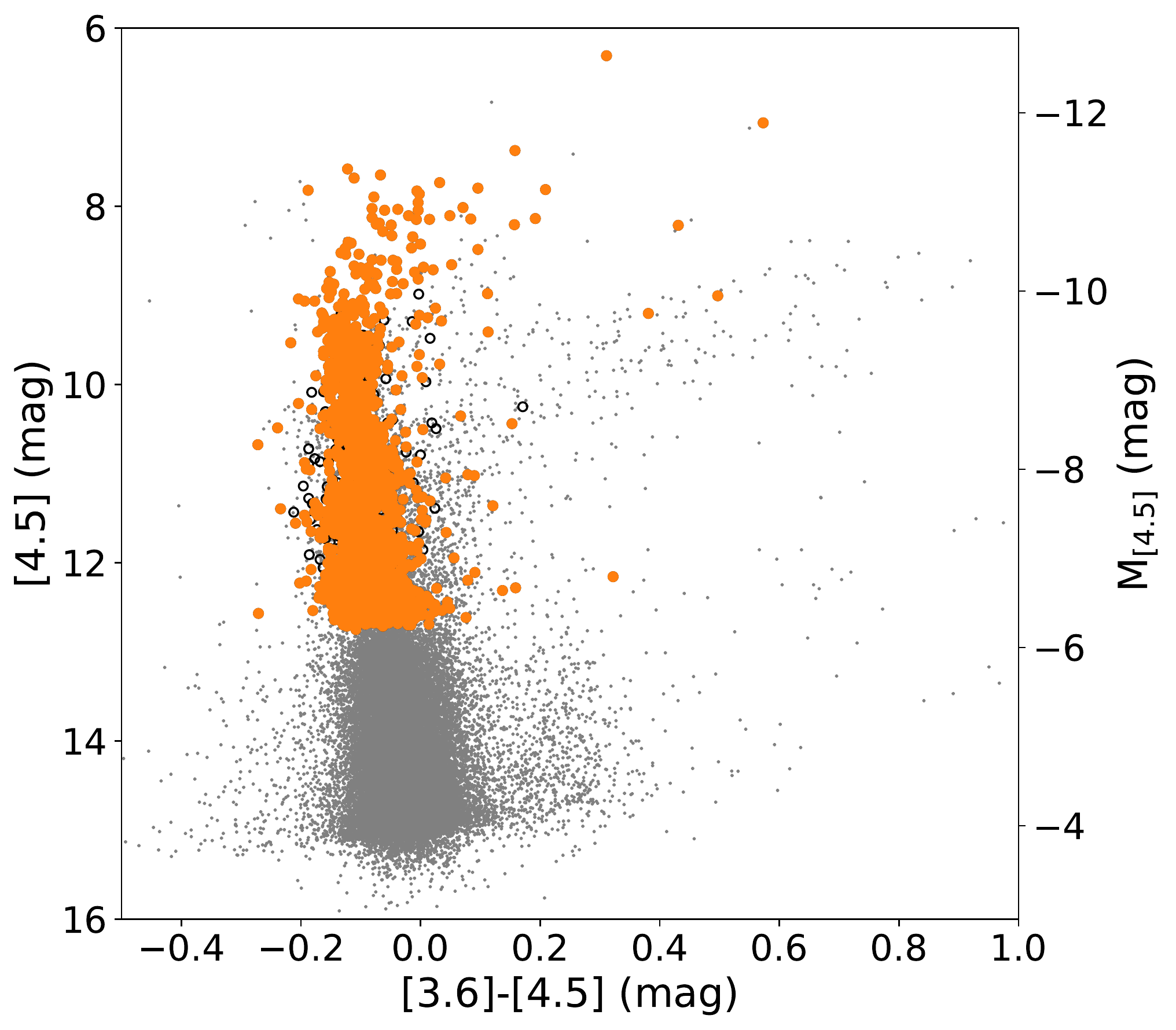}
\includegraphics[scale=0.43]{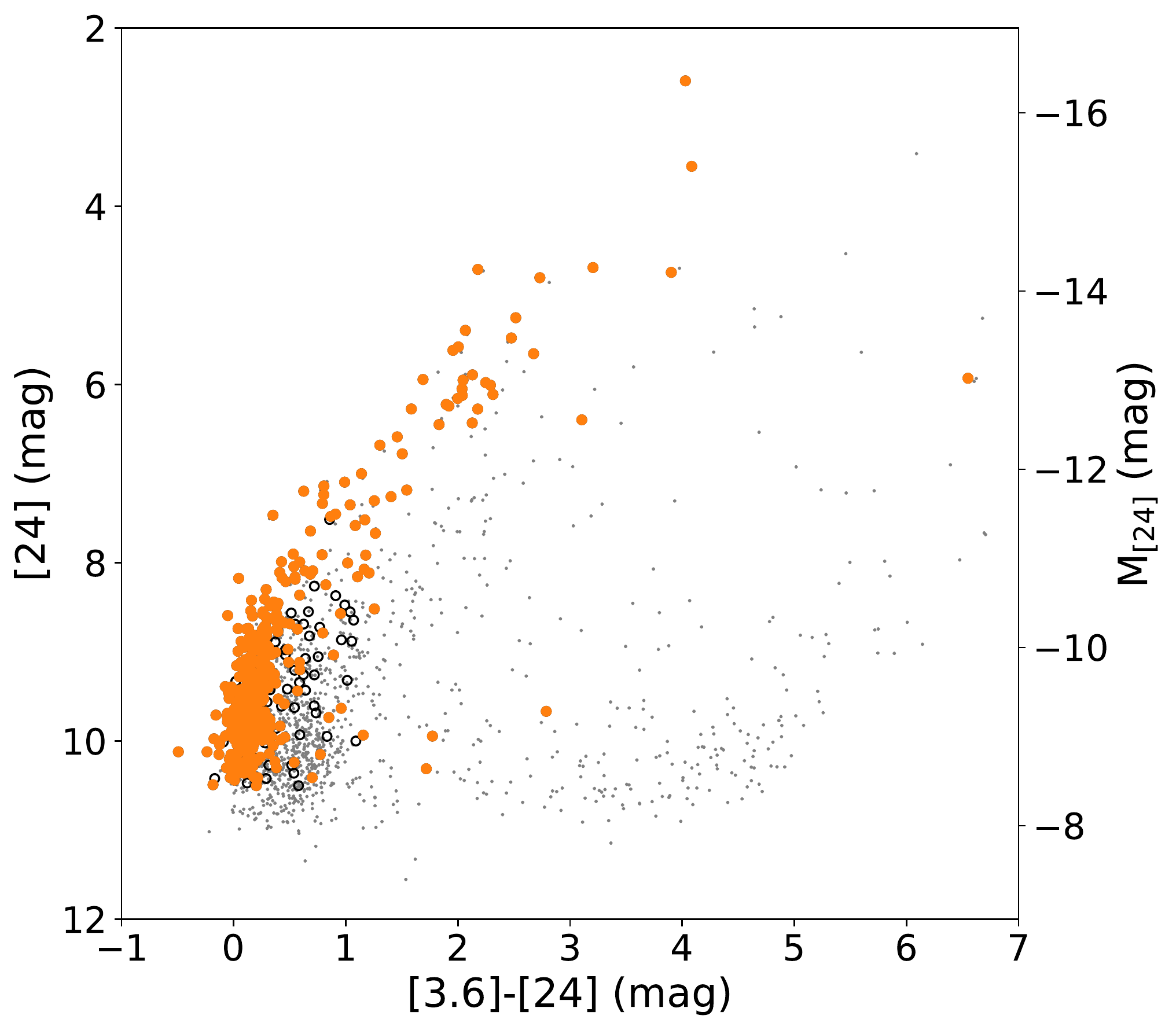}
\caption{Gaia (upper left), 2MASS (upper right), IRAC (bottom left), and MIPS (bottom right) color-magnitude diagrams of the final RSG sample in the SMC. Targets in the overlapping region (green lines) between AGBs and RSGs on the 2MASS CMD are removed (open black circles). Other general stellar populations, e.g. AGBs, red giant branch stars (RGBs), tip of the red giant branch (TRGB), blue supergiant stars (BSGs), are marked on the Gaia and 2MASS CMDs, but not on IRAC and MIPS CMDs due to the overlapping between different populations (see Figure 12 of \citealt{Yang2021a}). Background targets (gray dots) are from \citet{Yang2019}.
\label{sample_cmd}}
\end{figure*}

\begin{figure}
\center
\includegraphics[scale=0.47]{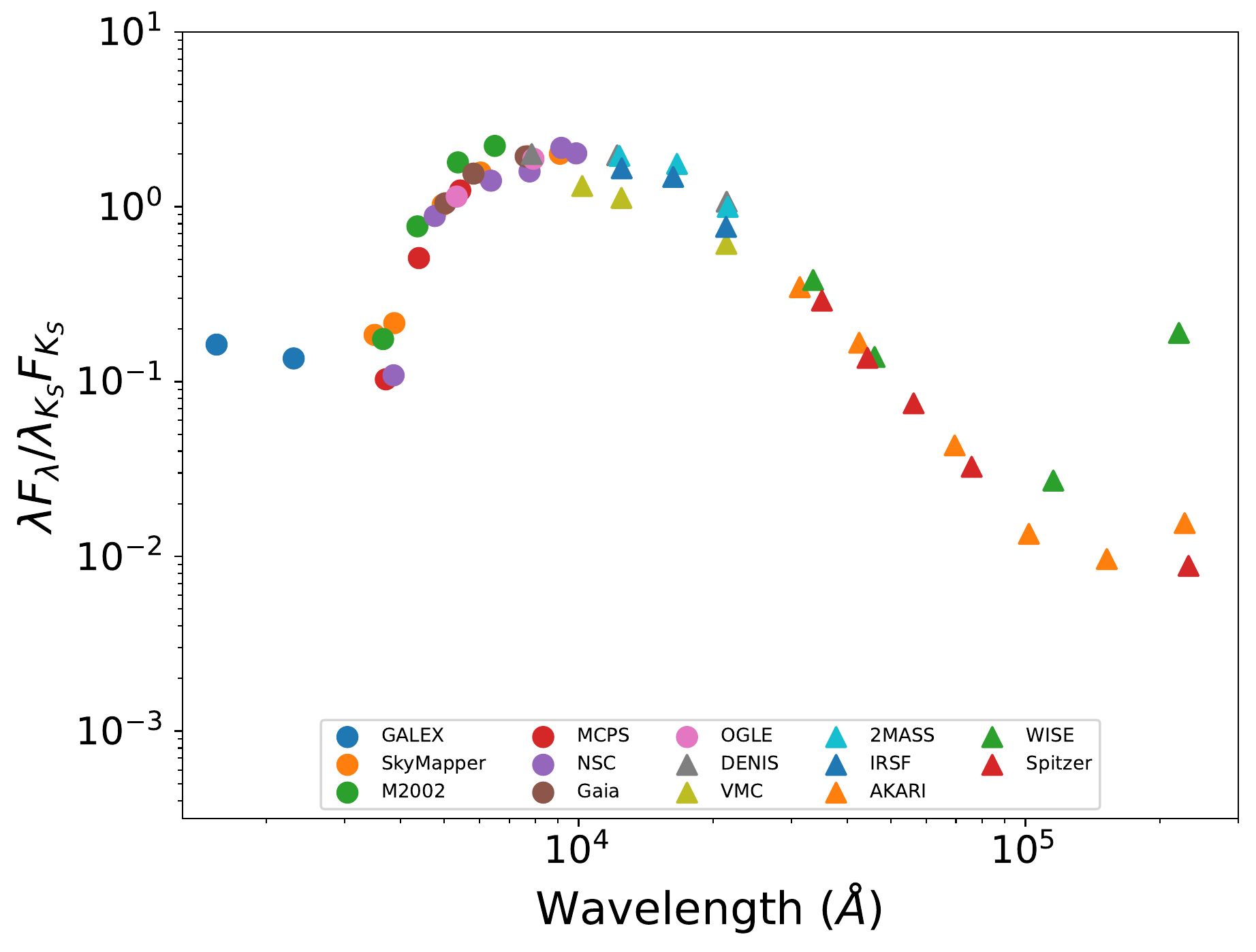}
\caption{The normalized (at 2MASS $\rm K_S$-band) median spectral energy distribution of the final RSG sample in the SMC. Due to the strict constraint of SNR, \textit{WISE} [12] and [22] bands data only include relatively bright targets with high fluxes.
\label{sed_median}}
\end{figure}

In the meantime, as indicated in \citet{Yang2020}, since there is a continuum with similarity and overlapping between RSGs and AGBs in photometry, spectra and lightcurves, the boundary between AGBs and RSGs is blurry with no perfect way to distinguish between them by using current data (e.g., see discussion in \textsection3.1 of \citealt{Yang2020}). Therefore, to be on the safe side, we removed the targets in the overlapping region between AGBs and RSGs on the 2MASS color-magnitude diagram (CMD; see Figure~\ref{sample_cmd}) by eye, which resulted in 2,121 targets for the final RSG sample. Figure~\ref{sample_spatial}, \ref{sample_cmd}, and \ref{sed_median} show the spatial distribution, CMDs, and the normalized (at 2MASS $\rm K_S$-band) median SED of the final RSG sample.

With $R_V=3.1$ and $E(B-V)=0.033$ \citep{Schlafly2011, Freedman2020}, the average Galactic foreground extinction of the SMC was corrected in the optical and IR bands by using the extinction law of \citet{Wang2019}, which obeyed the form of the model of \citet{Cardelli1989} with newly derived coefficients. Meanwhile, the foreground extinction in the far-UV (FUV) and near-UV (NUV) bands was corrected by using $A_{FUV}/E(B-V)=7.44$ and $A_{NUV}/E(B-V)=7.86$ \citep{Sun2021}. 

\begin{figure}
\center
\includegraphics[scale=0.47]{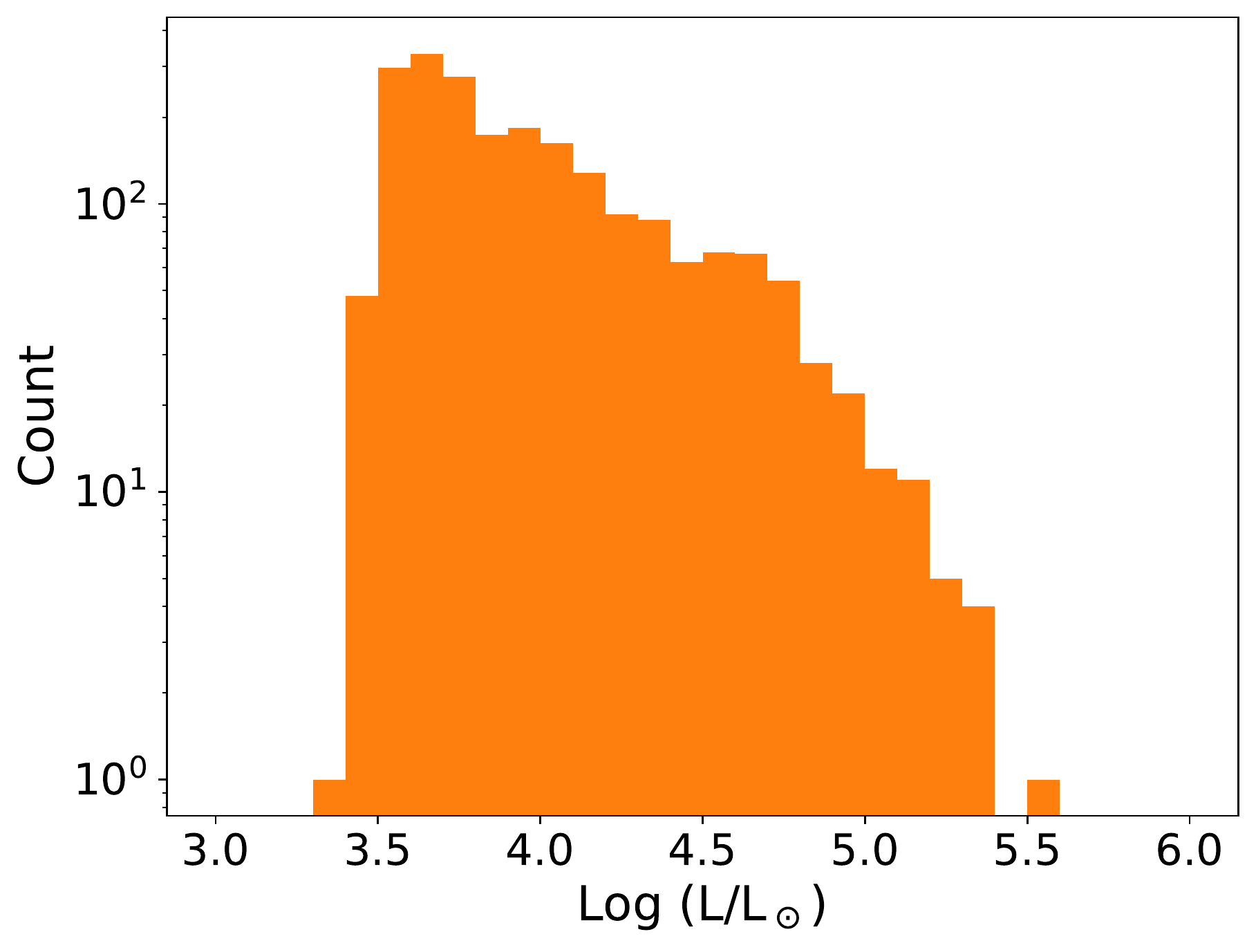}
\caption{Histogram of the luminosity of the final RSG sample. We note that the low-luminosity targets ($\log_{10}(L/L_\sun)\lesssim4.0$) may not be true RSGs, but $\sim6-8~M_\sun$ red helium-burning stars.
\label{hist_lum}}
\end{figure}

The luminosity of each individual target was calculated by integrating over the entire dereddened SED. Figure~\ref{hist_lum} shows the distribution of luminosity for the sample, spanning the range of $\log_{10}(L/L_\sun)\approx3.5-5.5$. In the majority of cases, this would yield an appropriate estimation of the luminosity with typical error less than 5\%. However, it might underestimate the luminosities of targets with high MLRs since there was no data at longer wavelength other than 24 $\mu$m, or overestimate the luminosities of targets with hot binary companions since there were additional UV detection (see more details in discussion). A further comparison between our results and the theoretical models (e.g., \citealt{Gilkis2021}) shows a general agreement of detected number of RSGs in the SMC, indicating the validity of our sample. Notice that, we followed the semi-empirical selection criteria (e.g., \citealt{Boyer2011, Dalcanton2012, Melbourne2012, Yang2019, Neugent2020, Tantalo2022}, etc.) to constrain our RSG sample as being brighter than the K$_S$-band TRGB (K$_S$-TRGB; see more details in \citealt{Yang2020} and \citealt{Ren2021}). However, the theoretical threshold for the massive stars is around 8 $M_\sun$ with $\log_{10}(L/L_\sun)\approx4.0$ (e.g., \citealt{Woosley1986}), which is in contradiction with the lower limit ($\log_{10}(L/L_\sun)\approx3.5$) of our sample. Therefore, we indicate that the low-luminosity targets ($\log_{10}(L/L_\sun)\lesssim4.0$) may not be true RSGs, but $\sim6-8~M_\sun$ red helium-burning stars (however, it does not impact our main result as shown in the latter sections). Moreover, inevitably, we may still have some small amount of contamination from the AGB population at the red and/or faint end of our sample (the Gaia and 2MASS CMDs in Figure~\ref{sample_cmd} show that, the contamination from the general population of AGBs is avoided), which is the main reason why we remove the targets in the overlapping region between AGBs and RSGs as mentioned before.

\begin{figure}
\center
\includegraphics[scale=0.47]{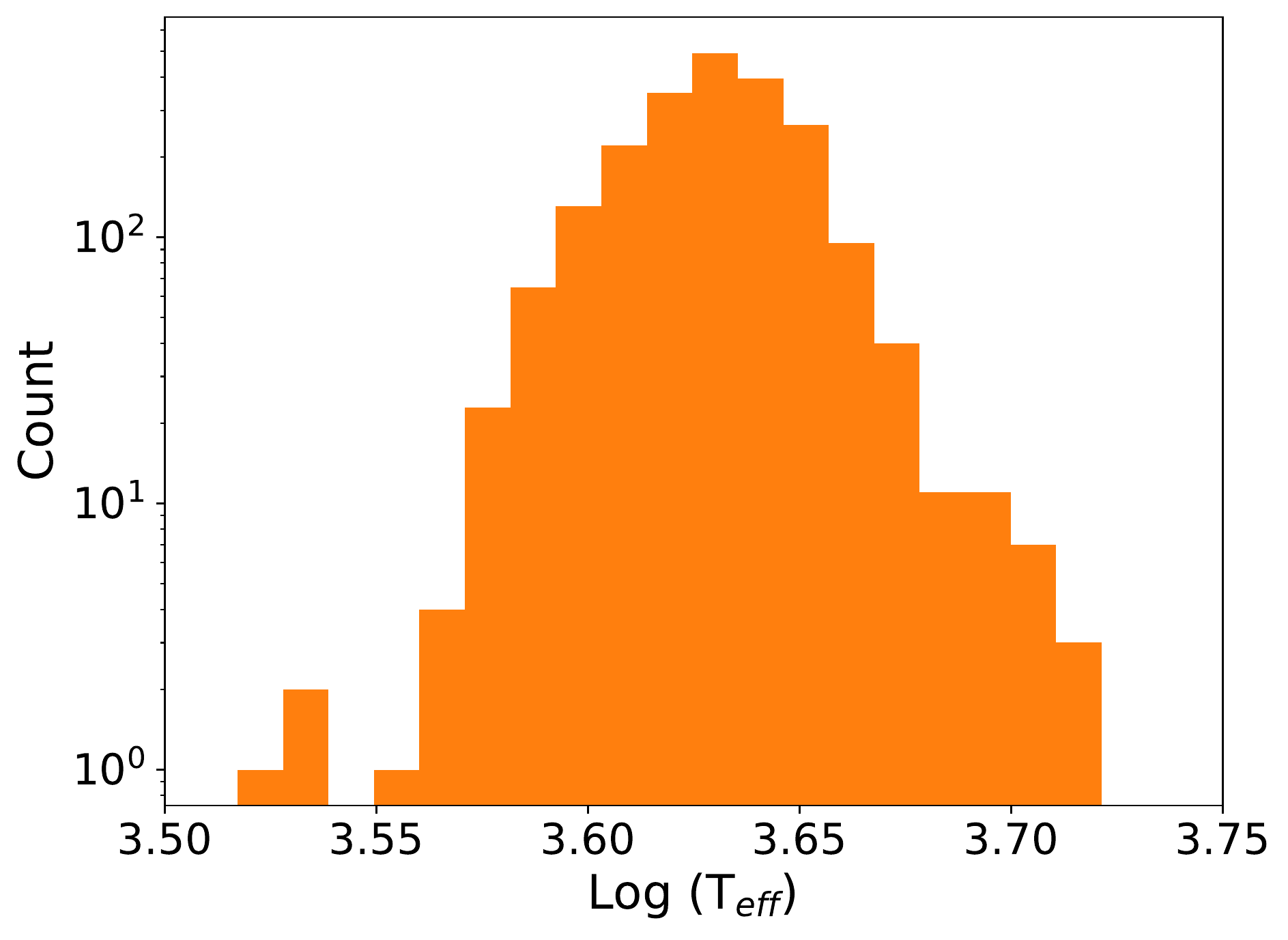}
\caption{Histogram of the $T_{\rm eff}$ of the final RSG sample.
\label{hist_teff}}
\end{figure}

The effective temperatures ($T_{\rm eff}$) of targets were calculated by using an algorithm from \citet{Yang2020} as,
\begin{equation}
log_{10}T_{\rm eff}=-0.23(J-K_S)_0+3.82,
\end{equation}
which converted the observed $\rm J-K_S$ color to $T_{\rm eff}$ with total extinction correction (including foreground extinction of the Milky Way, interstellar extinction of the SMC, and circumstellar extinction of the star) of $A_V=0.1$ mag (this value was justified by the fact that massive stars, like RSGs, are typically close to the star formation region due to their short evolutionary time scale, and RSGs produce dust by themselves). The typical uncertainty of $T_{\rm eff}$ is $\sim$0.017 dex \citep{Yang2020}. However, the applied reddening correction was an average value for the whole SMC, therefore targets within complicated environments or with high MLRs might be underestimated. Figure~\ref{hist_teff} shows the distribution of $T_{\rm eff}$ for the sample, spanning the range of $\log_{10}(T_{\rm eff})\approx3.5-3.75$.

\section{DUSTY grids and spectral energy distribution fitting}

The dust particles in a circumstellar dust shell absorb and re-emit the radiation from the central stars, which results in changes of the output stellar spectra. There are several dust radiative transfer models available ranging from 1D to 3D. For example, \citet{vanLoon2005}, \citet{Goldman2017}, \citet{Beasor2020}, \citet{Humphreys2020}, and \citet{Wang2021} used the 1D DUSTY code \citep{Ivezic1997} to estimate the MLRs of RSGs and AGBs in the Milky Way, LMC, M31, and M33. \citet{Groenewegen2012} and \citet{Groenewegen2018} also used modified DUSTY code of `More of DUSTY' (MoD) to calculate MLRs of carbon- and oxygen-rich evolved stars in several Local Group galaxies. \citet{Verhoelst2009} used the 1D MODUST code \citep{Bouwman2000, Bouwman2001} to assess the chemical composition and MLRs of a sample of Galactic RSGs. \citet{Sargent2010} and \citet{Riebel2012} used the Grid of Red supergiant and Asymptotic giant branch star ModelS (GRAMS; \citealt{Sargent2011, Srinivasan2011}) calculated by the 2D 2DUST code \citep{Ueta2003} to compute the MLRs of RSGs and AGBs in the LMC. \citet{Liu2017} also used the 2DUST to investigate the chemical composite and MLR of a sample of O-rich evolved Galactic stars. Recently, \citet{Cannon2021} used 3D MCMax3D code \citep{Min2009} to characterize a dust clump of Antares. Similarly, \citet{Montarges2021} used 3D RADMC3D to investigate the effect of dust on the `Great Dimming' of Betelgeuse.

\subsection{The DUSTY library and grids}

We used one of the most popular 1D code, DUSTY (V4\footnote{https://github.com/ivezic/dusty}), to construct a grid, covering a wide range of dust parameters, in order to match the observed SEDs. DUSTY solves the radiative transfer equation for a central source surrounded by a spherically symmetric dust shell at a certain optical depth ($\tau$, at 0.55 $\mu$m in our case), with a range of dust condensation temperature ($T_{in}$) changing at the inner boundary of the dust shell. The grid is produced by the radiatively-driven wind model of DUSTY (assuming the dust shell extends to a distance $10^4$ times the inner radius). Full dynamics calculations of radiatively-driven winds can be found in \citet{Ivezic1995} and \citet{Elitzur2001} and references therein, while an analytic solution takes the form of,
\begin{equation}
\eta(y)\propto\frac{1}{y^2}\left[\frac{y}{y-1+(v_1/v_e)^2} \right]^{1/2},
\end{equation}
shown here as a reference. The density distribution is described by the dimensionless profile $\eta(y)$, which DUSTY normalizes according to $\eta dy=1$. $v_1$ and $v_e$ represent the initial and final velocities of the stellar wind, respectively \citep{Elitzur2001}. 

The central source used in the DUSTY modelling was taken from MARCS stellar atmosphere model (15 M$_\sun$; \citealt{Gustafsson2008}) at the metallicity of $Z=-0.75$ and a typical surface gravity of log $g = 0$, with $T_{\rm eff}$ of 3300, 3400, 3500, 3600, 3700, 3800, 3900, 4000, 4250, and 4500 K. As the broad-band photometric data were not enough to distinguish subtle spectral features and in order to reduce the computational time, we resampled the surface fluxes of MARCS model (ranging from 1300 to 200,000 $\AA$ with a constant resolving power of $R=20,000$) to match the custom DUSTY wavelength grid in steps of 200 $\AA$ between 1000 and 10,000 $\AA$, 400 $\AA$ between 10,000 and 25,000 $\AA$, 1,000 $\AA$ between 25,000 and 350,000 $\AA$, and 150,000 $\AA$ between 350,000 and 3,000,000 $\AA$. Wavelength ranges longer than 200,000 $\AA$ that were not covered by MARCS model were extrapolated by using a blackbody. Using a MARCS model is better than a simple blackbody since it also represents the evolutionary effect on the stellar atmosphere such as the ``H-bump'' \citep{Yang2021b}.

The chemical composition of dust species for the RSG wind was mainly oxygen-rich (identified by the wide, smooth, and featureless Si-O stretching 9.7 $\mu$m and O-Si-O bending 18 $\mu$m silicate features) as indicated in many previous works \citep{vanLoon2005, Groenewegen2009, Verhoelst2009, Sargent2011, Goldman2017, Wang2021}, for which the optical constants of ``astronomical silicate'' from \citet{Draine1984} were adopted with a dust bulk density of $\rho_d=3.3$ g cm$^{-3}$. The selection of this dust species is somewhat arbitrary (as ``astronomical silicate'' is a popular or general choice for many previous studies), since as noted above, broad-band photometric data are not able to accurately distinguish different silicate features (see more details in the discussion section). Moreover, some RSGs have optically thin shells ($\tau\approx0.001$) or even no dust at all, for which no discrimination can be made on the dust species and the actual MLR is anyway very low.

A simple power law with an exponential decay of Kim-Martin-Hendry distribution (KMH; $n(a) \propto a^{-q}e^{-a/a_0}$ with $a > a_{min}$; \citealt{Kim1994}) was adopted for the dust grain size distribution with a power-law index of $q = 3.5$, a lower limit of $a_{min}=0.005$ $\mu$m, and the scale height of $a_{0} = 1.0$ $\mu$m (observationally, there was evidence indicating that the dust grain sizes of RSG wind might be between 0.1-1$\mu$m ; \citealt{Groenewegen2009, Smith2001, Scicluna2015}). Such a distribution is better than the simplistic homogeneous sphere models of the Mathis-Rumpl-Nordsieck distribution (MRN; \citealt{Mathis1977}), since it avoided the sudden cutoff at large dust grain size. Notice that, the derived MLR would still depend on the size of the dust and many studies used different size distributions \citep{vanLoon2005, Ohnaka2008, Verhoelst2009, Sargent2010, Liu2017, Beasor2018}. However, again, photometric data hardly distinguish the differences.

The $T_{in}$ at the inner boundary of dust shell varied from 400 to 1,200 K with step of 100 K. The condensation temperature of dust species may depend on the $L$, $T_{\rm eff}$, metallicity and other factors of the central source, or the chemical composition of dust itself. Previous studies found that dust might begin to form further from the photosphere with lower $T_{in}$, so that the condensation temperatures of silicate dust varied from around 400 K to 1,500 K with typical values of $\sim$1,000-1,200 K \citep{Gail1984, Gail1999, vanLoon2005, Ohnaka2008, Sargent2010, Gail2020}. For simplicity, we also set the sublimation temperature of 1,200 K for the O-rich dust.

The models were divided into two groups depending on the value of $\tau$ at 0.55 $\mu$m. The first group (``normal grid''; $\tau\leq1.0$) corresponded to the normal targets with $\tau$ varied from 0.001 (optically thin) to 1.0 in 99 piecewise equally spaced logarithmic steps\footnote{There is a bug in memory management in DUSTY V4 that prevents more than 100 steps of $\tau$; Zeljko Ivezic, private communication.}. In particular, there were 30 steps between 0.001 and 0.1, and 70 steps between 0.1 and 1.0. The second group (``dusty grid''; $\tau>1.0$) corresponded to the very dusty targets with $\tau$ varied from 1.0 to 10.0 in 99 equally spaced logarithmic steps. The choice of such steps will be addressed in the discussion section.

In total, there were 17,820 theoretical O-rich models generated by DUSTY, for which 8,910 models were for the ``normal grid'' and 8,910 models were for the ``dusty grid''. Each model was then convolved with the filter profiles\footnote{http://svo2.cab.inta-csic.es/theory/fps/} of the 53 bands to derive the model flux at each wavelength.

\subsection{Spectral energy distribution fitting}

\begin{figure}
\center
\includegraphics[scale=0.5]{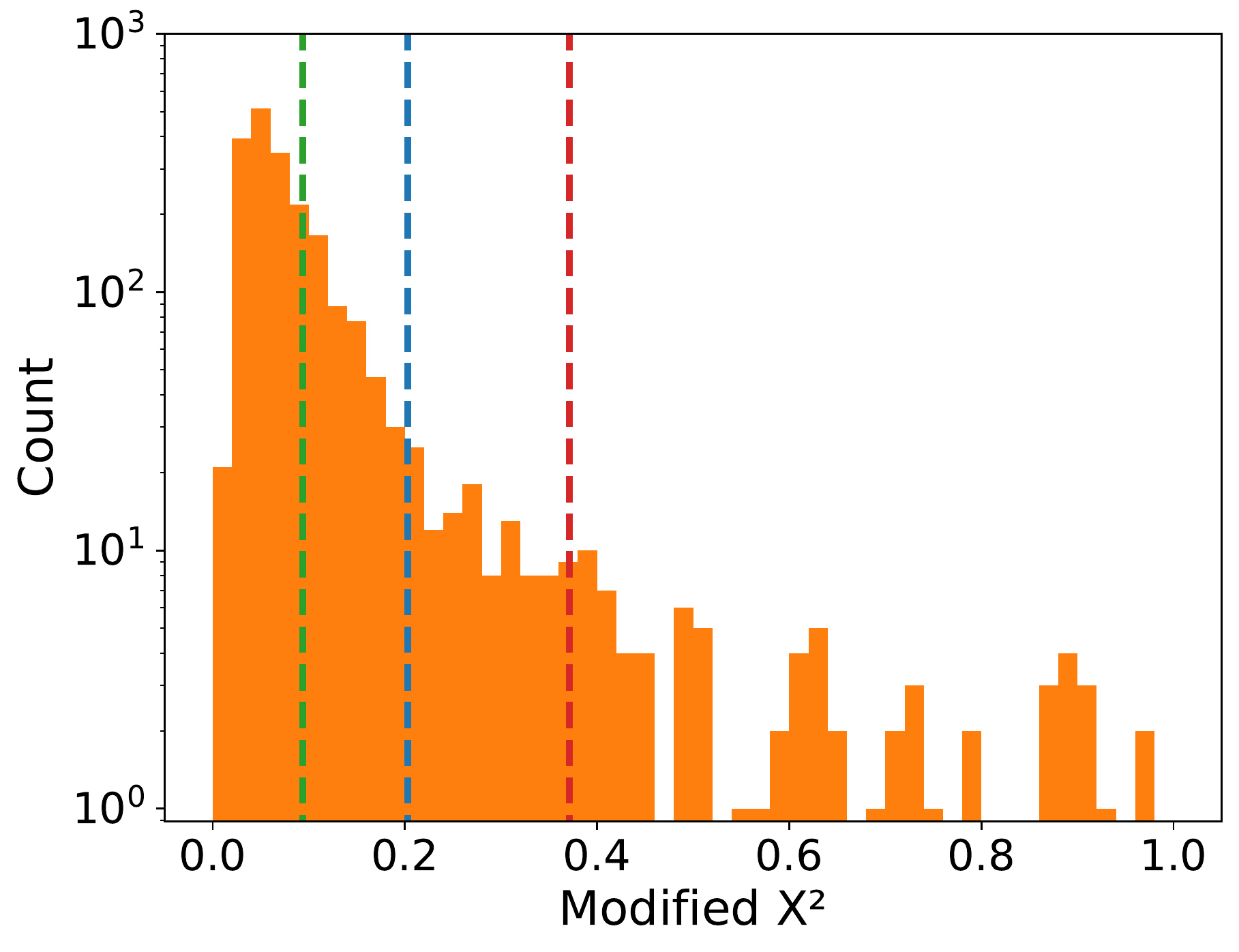}
\caption{Histogram of the $\chi_{m}^2$ distribution. The three vertical dashed lines indicate the $68^{th}$ ($\chi_{m}^2=0.094$), $90^{th}$ ($\chi_{m}^2=0.203$), and $95^{th}$ ($\chi_{m}^2=0.371$) percentiles, respectively.
\label{hist_chi2}}
\end{figure}

The best fitted model for each target was chosen by calculating a modified minimum chi-square ($\chi_{m}^2$), after normalizing the SED to the 2MASS K$_S$-band (this band was selected due to that it was much less affected by the extinction and had much smaller variability compared to the optical bands, e.g., the V-band). The $\chi_{m}^2$ for each source was calculated as,
\begin{equation}
\chi_{m}^2=\frac{1}{N-p-1}\sum^N\frac{[f(Obs, \lambda)-f(Model, \lambda)]^2}{f(Model, \lambda)},
\end{equation}
for which 
\begin{equation}
f(Obs, \lambda)=\frac{F(Obs, \lambda)}{F(Obs, K_S)}\times C(Obs, \lambda),
\end{equation}
\begin{equation}
f(Model, \lambda)=\frac{F(Model, \lambda)}{F(Model, K_S)}\times C(Obs, \lambda),
\end{equation}
\begin{equation}
C(Obs, \lambda)=\frac{F(Obs, K_S)}{F(Obs, \lambda)}.
\end{equation}
$N$ and $p$ are the numbers of data points and degrees of free parameters in the fitting, respectively. So the derived $\chi_{m}^2$ for each individual target can be directly compared ($N$ is different for each target). The formula also can be rewritten as,
\begin{equation}
\chi_{m}^2=\frac{1}{N-p-1}\sum^N\frac{[1-f_{norm.}(Model)/f_{norm.}(Obs)]^2}{f_{norm.}(Model)/f_{norm.}(Obs)},
\end{equation}
for which 
\begin{equation}
f_{norm.}=\frac{F(\lambda)}{F(K_S)}.
\end{equation}
It can be seen that, by multiplying with the constant $C$, the observed SED was converted into a flat spectrum with each wavelength having the same weight across the whole SED. The model with minimal $\chi_{m}^2$ then corresponded to the one closest resembling the flat spectrum. Meanwhile, the distribution of $\chi_{m}^2$ is more relevant to the goodness of fitting than its absolute values (see Figure~\ref{hist_chi2}). Notice that, this $\chi_{m}^2$ formula is a variant of the standard $\chi^2$ test \citep{Pearson1900}, 
\begin{equation}
\chi^2=\sum\frac{(O-E)^2}{E},
\end{equation}
where $O$ and $E$ represent the observed and expected (model) values, respectively. We didn't take the normal approach of using the reduced $\chi^2$ \citep{Andrae2010}, which is widely used in astronomy as,
\begin{equation}
\chi_r^2 = \frac{1}{p}\sum \left(\frac{O-E}{\sigma}\right)^2,
\end{equation}
where $p$ and $\sigma$ represent the the number of degrees of freedom and errors, respectively. In the discussion section we estimated the effect of our choice of fitting.

\begin{figure}
\center
\includegraphics[scale=0.5]{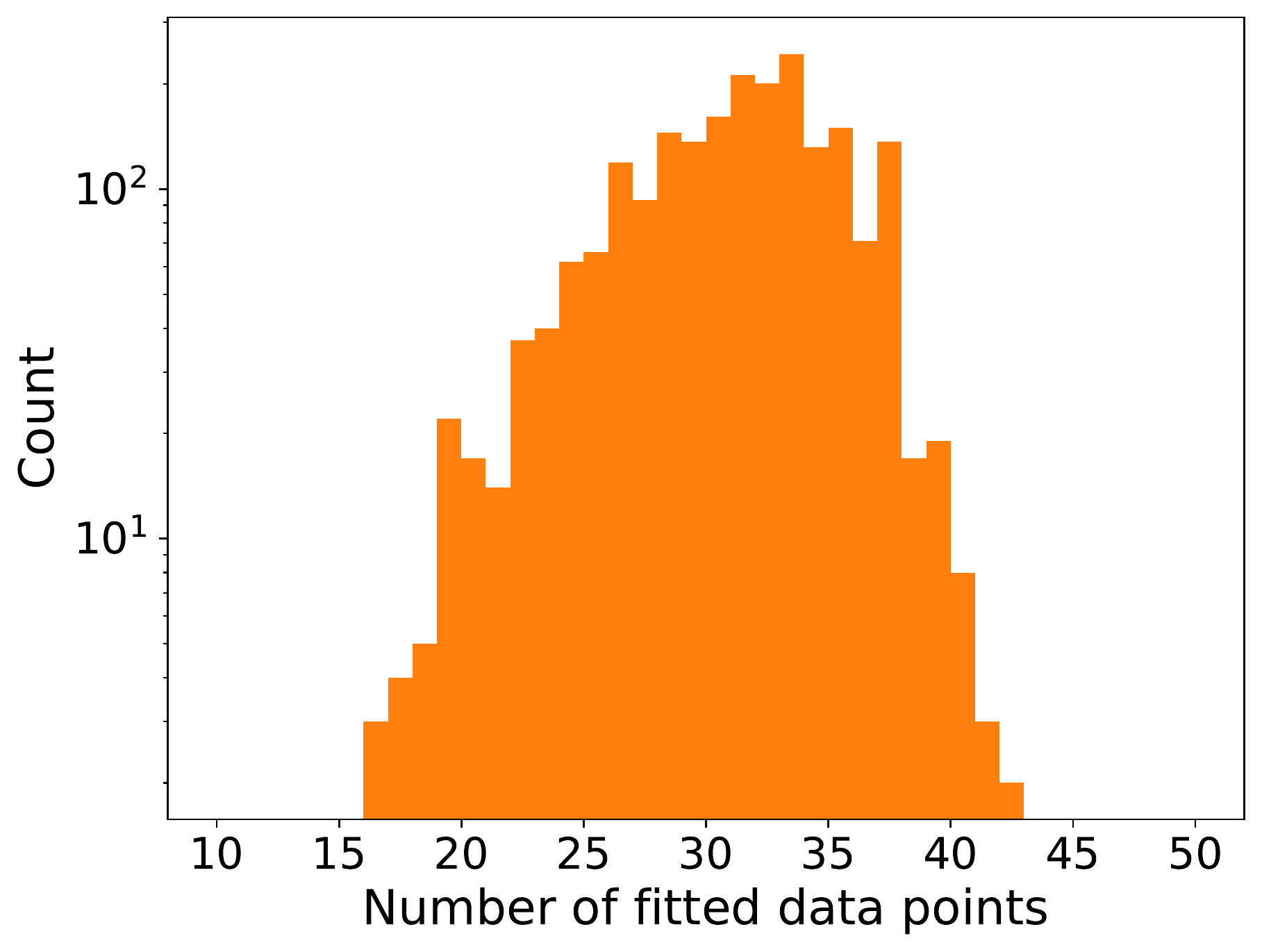}
\caption{Histogram of numbers of data points fitted for each targets. 
\label{hist_dpnum}}
\end{figure}

The final RSG sample was first fitted with the normal grids, then the dusty ones (six targets) were again fitted with the dusty grids. After the automated fitting process, we also visually inspected each target (simultaneous inspection of SED fitting, optical to mid-IR images, and locations on the CMDs) to evaluate the fitting result. Figure~\ref{hist_chi2} shows the histogram of the $\chi_{m}^2$ distribution from the normal grids. Three percentile values of $68^{th}$ ($\chi_{m}^2=0.094$), $90^{th}$ ($\chi_{m}^2=0.203$; presumably the beginning of relatively poor fitting), and $95^{th}$ ($\chi_{m}^2=0.371$) were shown in the diagram as references, respectively. Figure~\ref{hist_dpnum} shows the distribution of numbers of data points fitted for each target. On average, we had around 30 data points for each target so the shape of the SED was relatively well constrained. 

\begin{figure*}
\center
\includegraphics[scale=0.32]{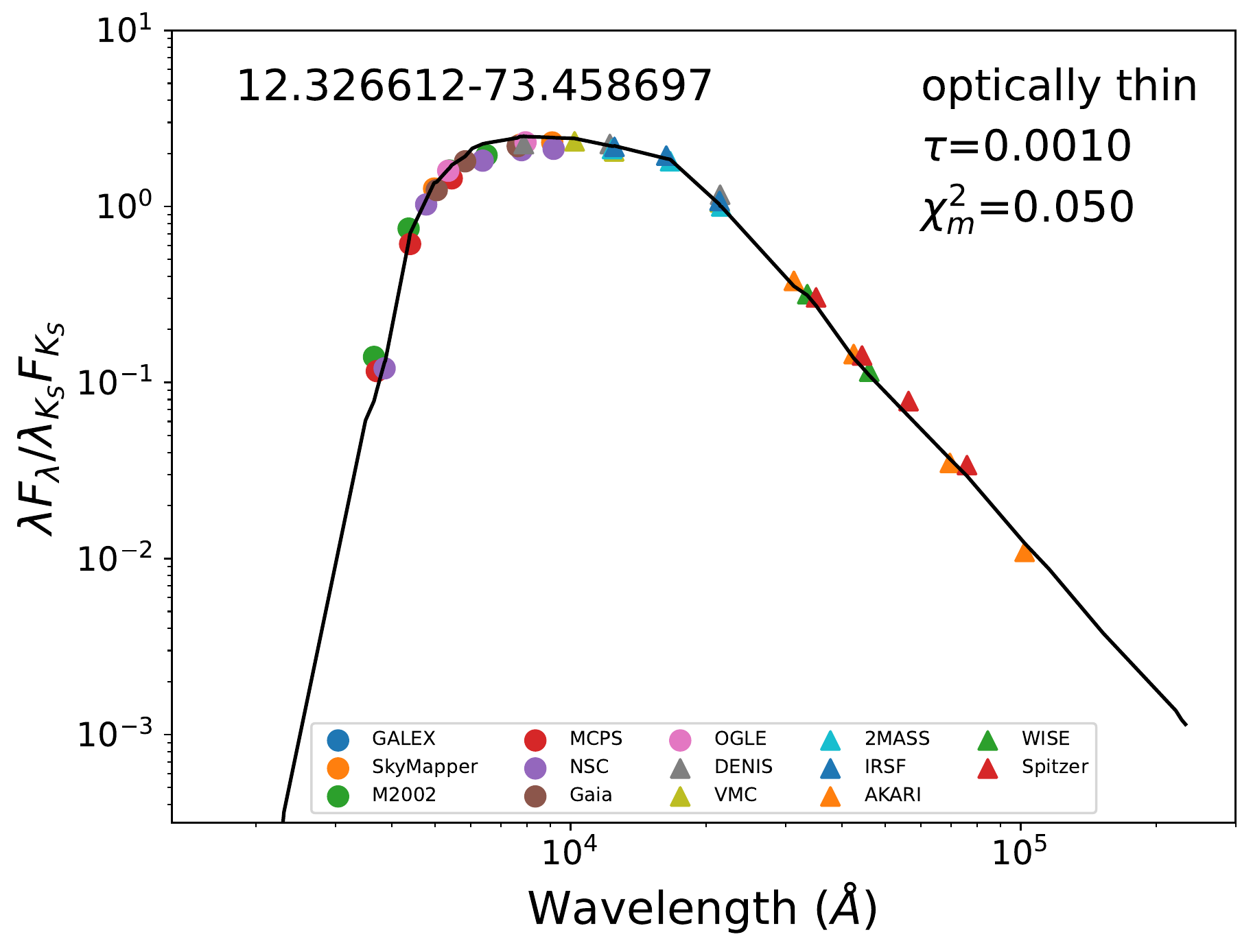}
\includegraphics[scale=0.32]{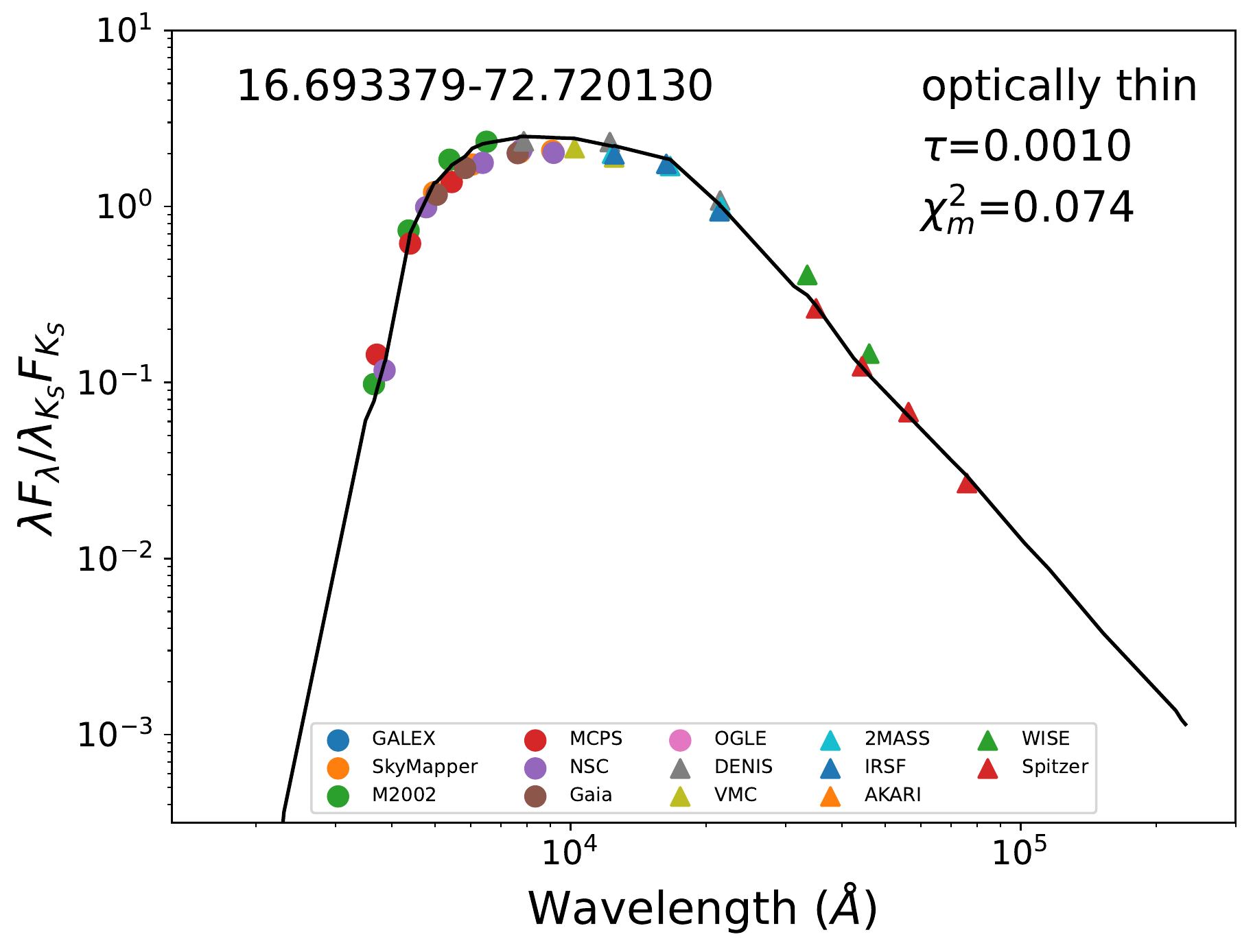}
\includegraphics[scale=0.32]{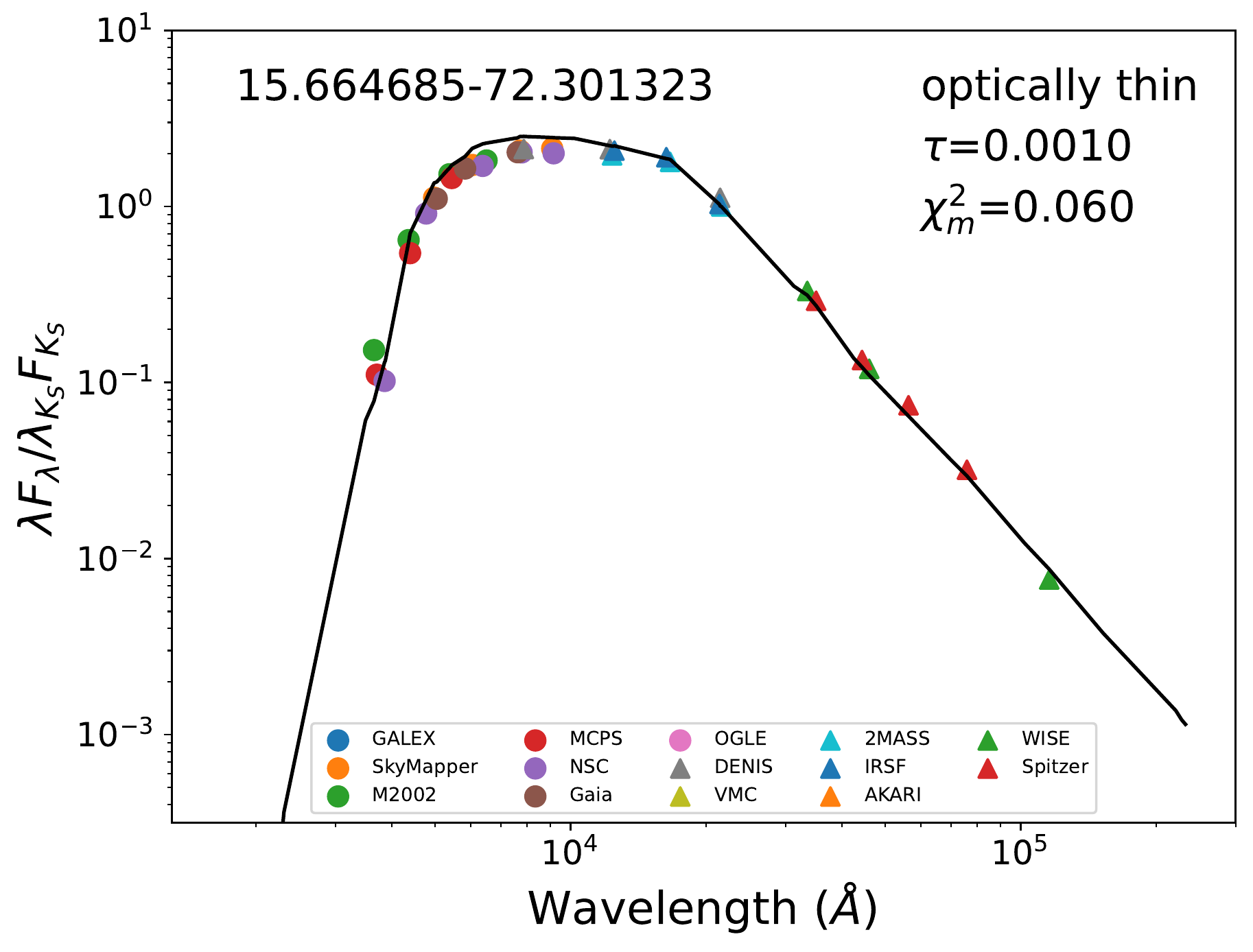}
\includegraphics[scale=0.32]{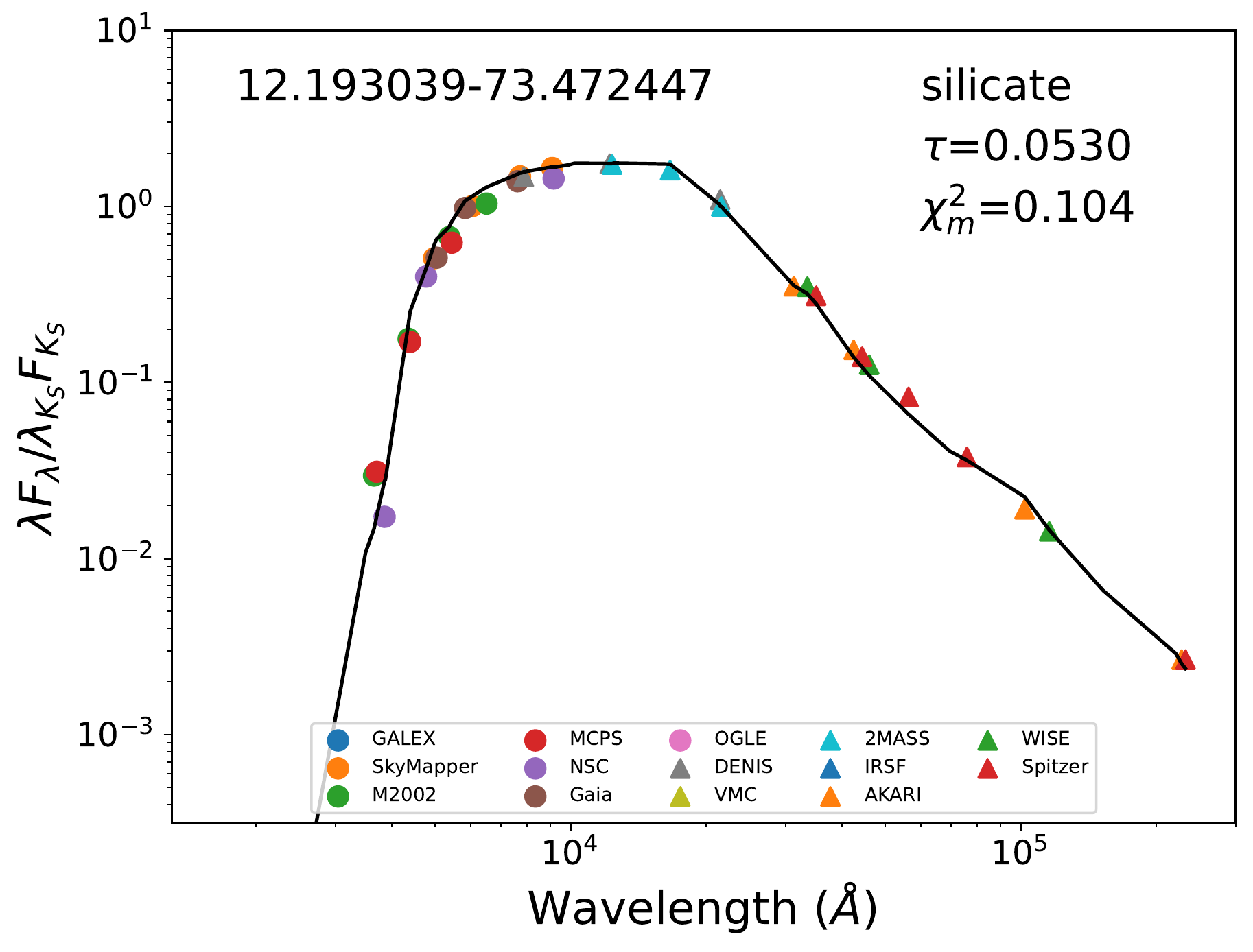}
\includegraphics[scale=0.32]{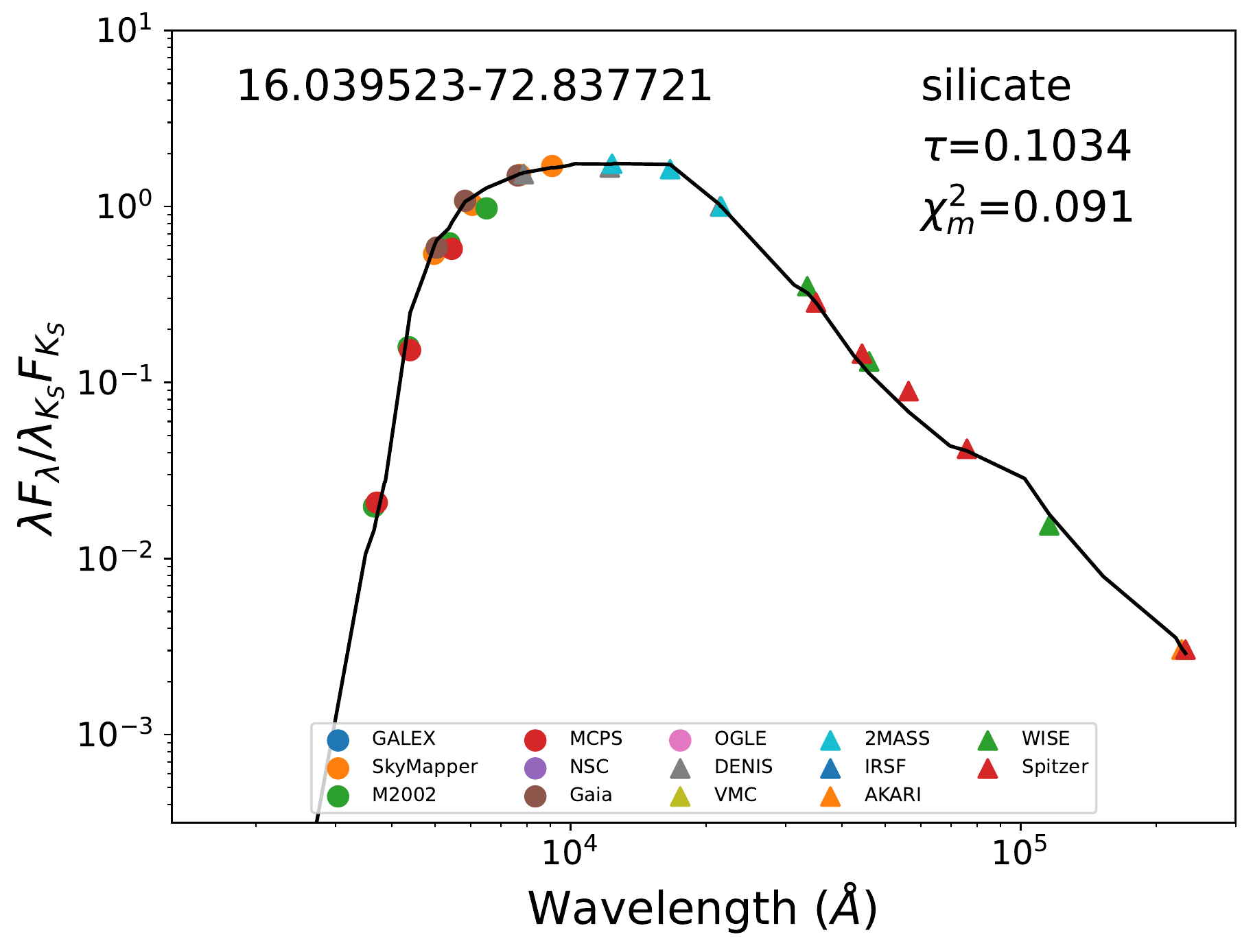}
\includegraphics[scale=0.32]{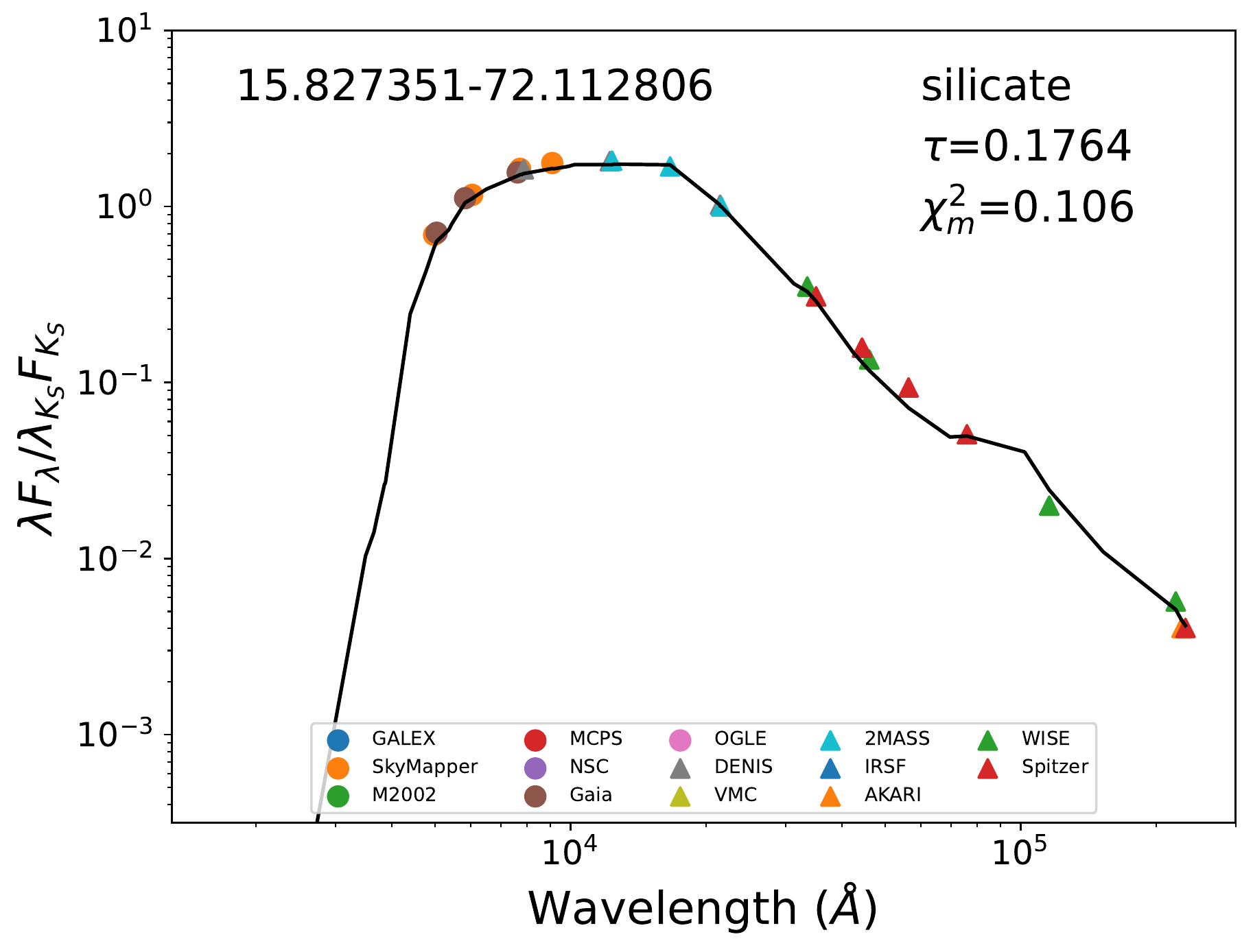}
\includegraphics[scale=0.32]{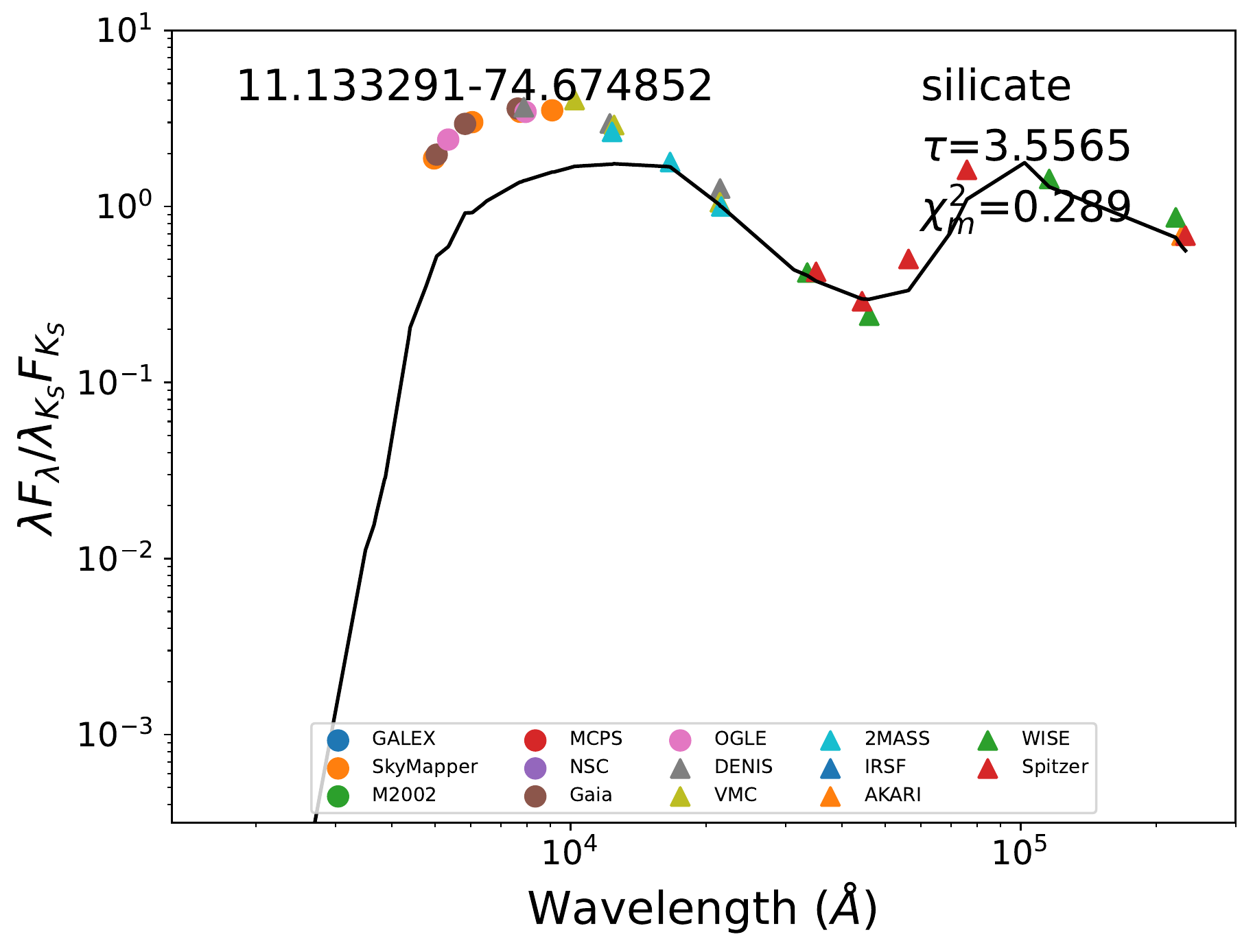}
\includegraphics[scale=0.32]{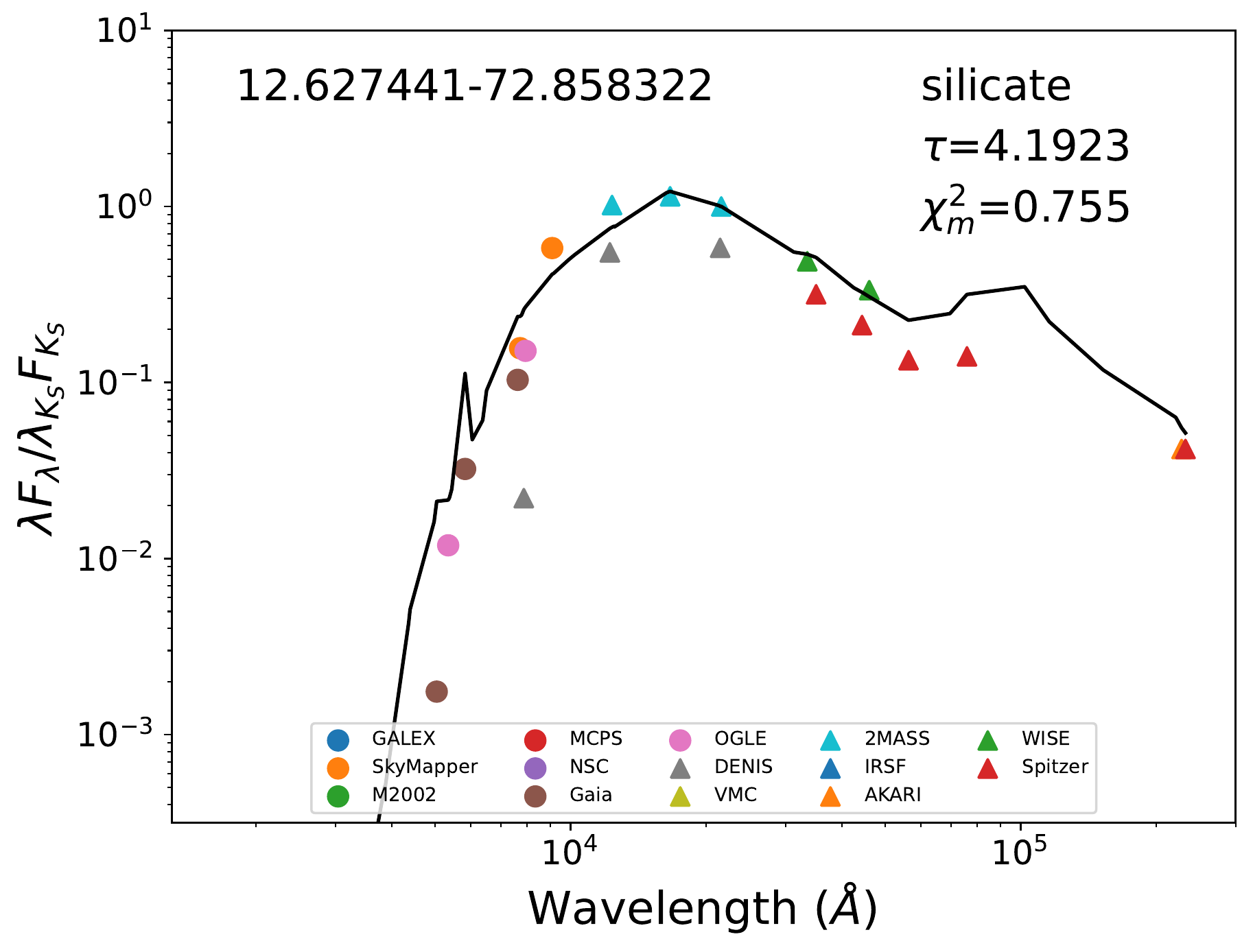}
\includegraphics[scale=0.32]{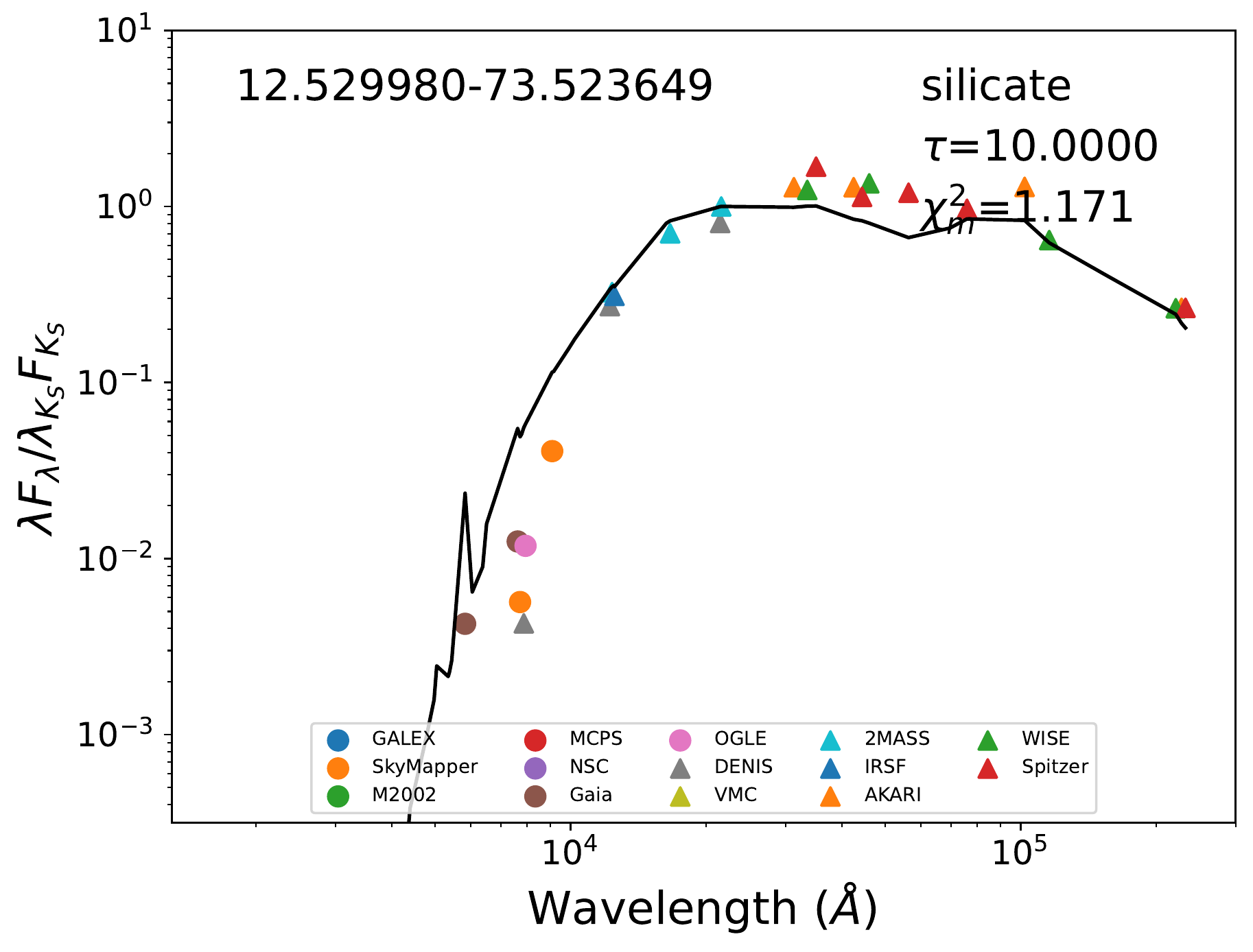}
\caption{Examples of the typical SED fitting for optically thin (upper row), normal (middle row), and very dusty targets (bottom row). In each panel, the coordinate is marked on the upper left and chemical composition, $\tau$, and $\chi_{m}^2$ are marked on the upper right.
\label{fitting_example1}}
\end{figure*}

\begin{figure*}
\center
\includegraphics[scale=0.32]{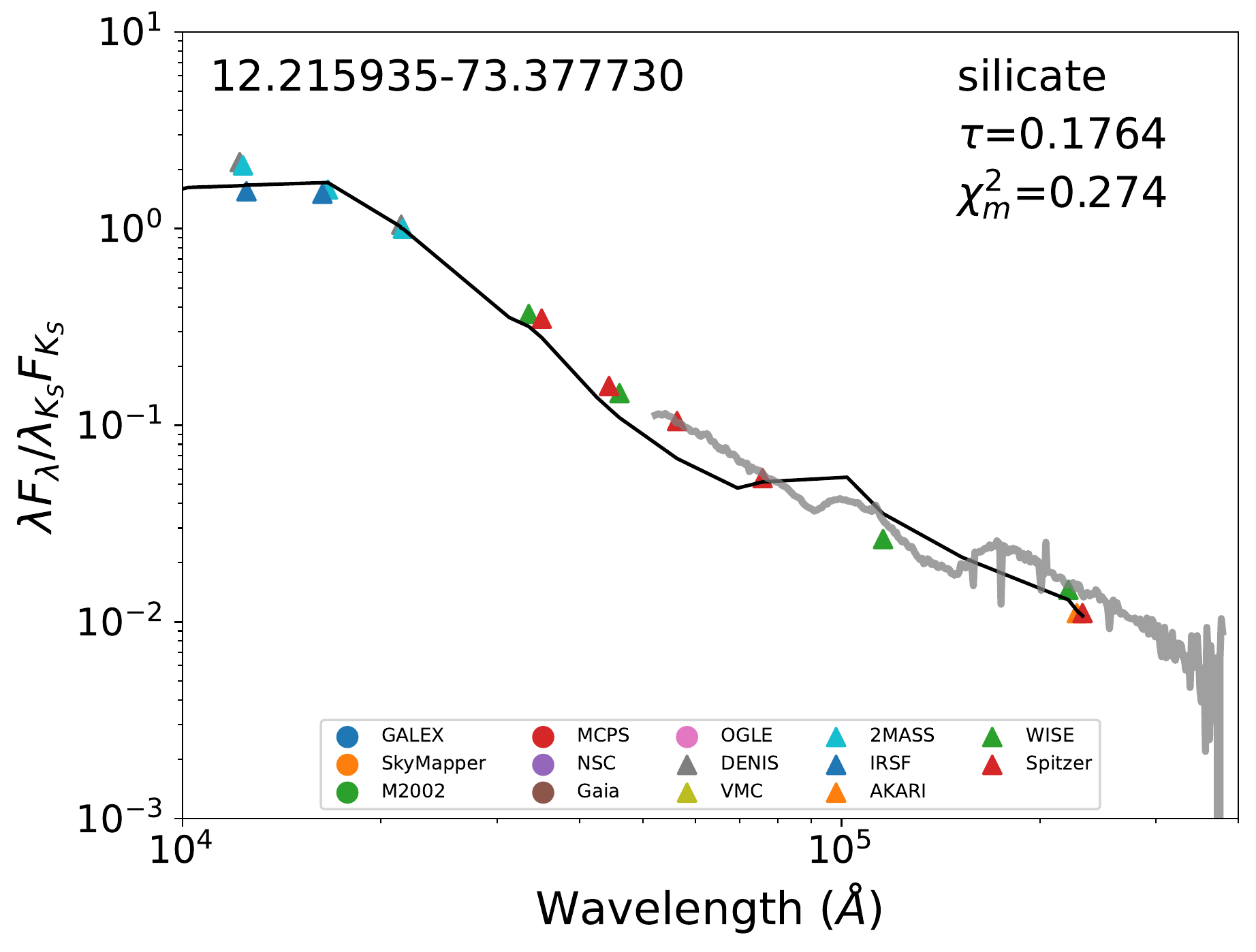}
\includegraphics[scale=0.32]{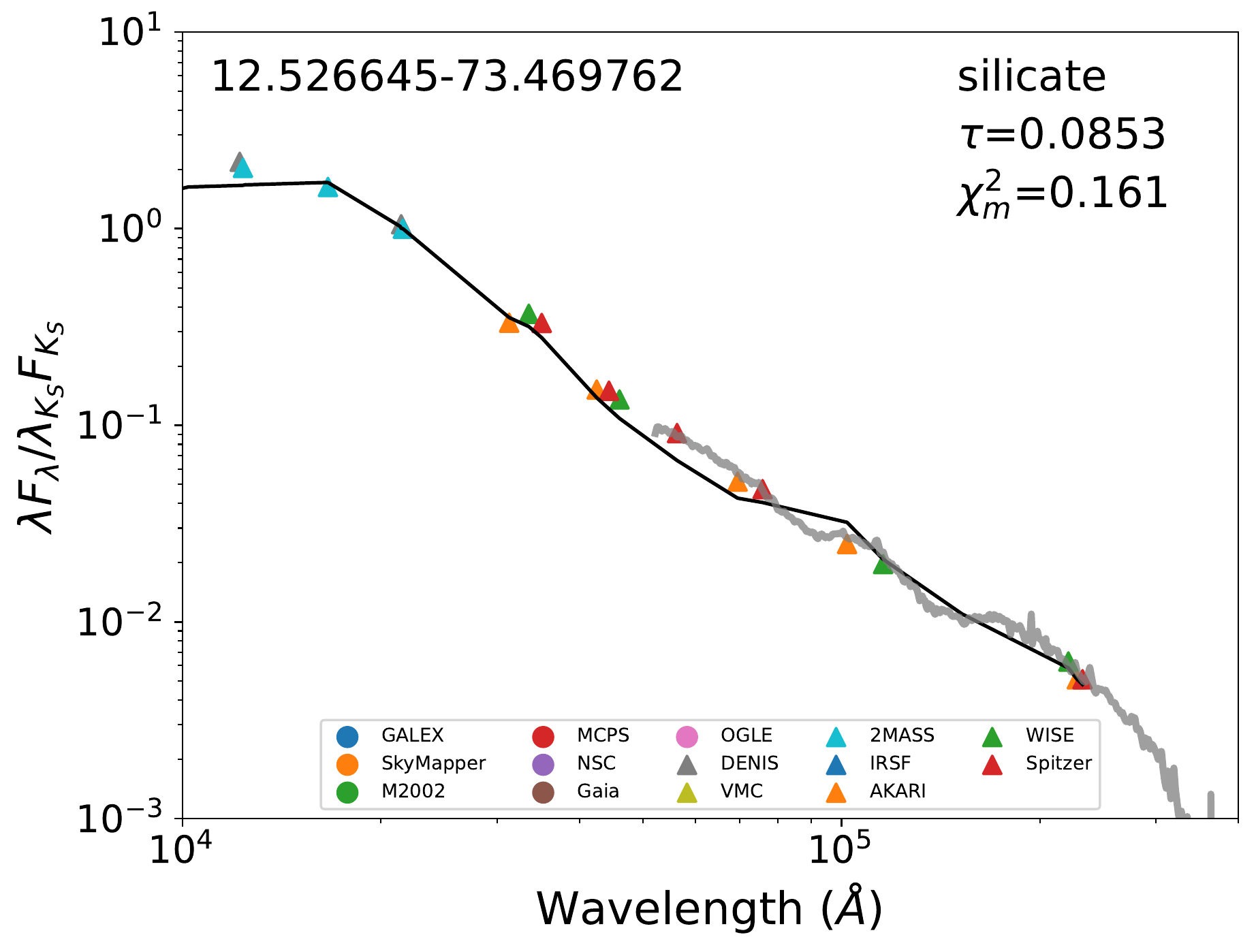}
\includegraphics[scale=0.32]{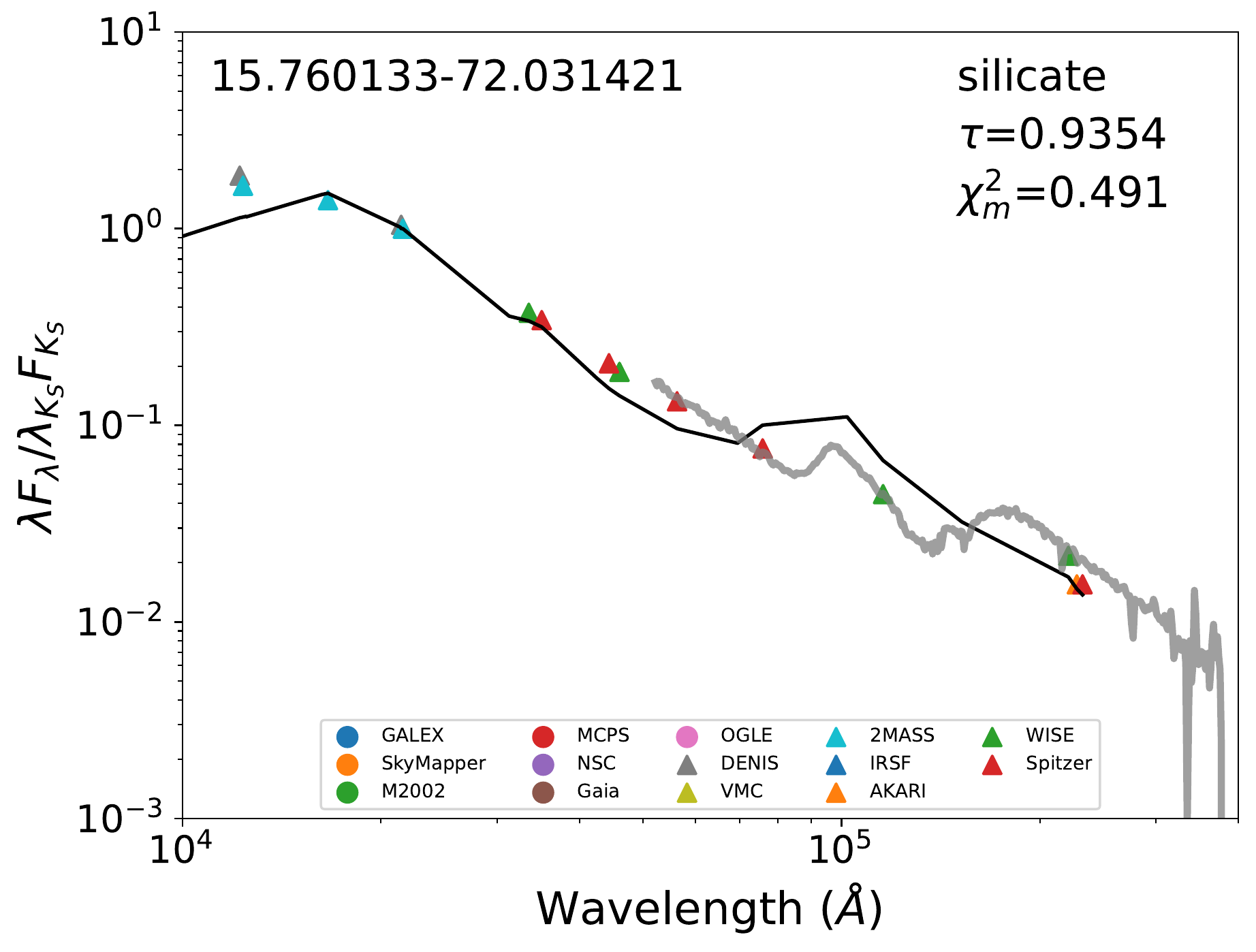}
\caption{Same as Figure~\ref{fitting_example1} but examples of zoomed-in region of SED fitting with \textit{Spitzer}/IRS spectra (grey) overlapped.
\label{irs_spec}}
\end{figure*}

Figure~\ref{fitting_example1} shows typical examples of the SED fitting for optically thin, normal, and very dusty targets (e.g., $\tau>1.0$) in the sample. In general, the very dusty targets are not well fitted, due to the lack of a dense grid at high optical depths, which would cost a large amount of computational time. Moreover, we also crossmatched our final sample with \textit{Spitzer}/IRS Enhanced Products which yielded 25 matches. After visual inspection, we found that almost all (96\%; 24/25) of the matched spectra in the sample showed identical chemical composition compared to the results of the SED fitting, indicating that our SED fitting was properly working. However, due to the complication of mid-IR spectra, these percentages could vary. A few examples of IRS spectra are shown in Figure~\ref{irs_spec}.

The resulting MLRs from DUSTY (for $L=10^4~L_\sun$, $\psi=200$, and $\rho_d = 3$ g cm$^{-1}$) were converted to real MLRs by using the following prescription \citep{Ivezic1999, vanLoon2000, Boyer2009, McDonald2011, Goldman2017}, based on the scaling relation as,
\begin{equation}
\dot{M}=\dot{M}_{DUSTY}\left(\frac{L}{10^4}\right)^{3/4}\left(\frac{\psi}{200}\right)^{-1/2}\left(\frac{\rho_d}{3}\right)^{1/2},
\end{equation}
where $\psi$ is the gas-to-dust ratio (assuming $\psi\approx1000$ in the SMC; e.g., \citealt{Bouchet1985, Gordon2009, RomanDuval2014}), $L$ is the stellar luminosity as derived in Section \textsection2, and $\rho_d$ is the bulk grain density mentioned in previous section. The errors of the MLRs could be calculated by repeating the SED fitting with adding and subtracting known variabilities (standard deviation) before and after the 2MASS K$_S$-band, and vice versa, which naturally mimic the effect of variability in different wavelengths (see more details in discussion). However, we would like to indicate that except for some specific wavelength bands, not all wavelengths contain variability information. Hence, the errors of MLRs may have large uncertainties. More discussion about the error of MLR can be found in Section \textsection5. Table~\ref{fsample} shows the full information of the sample with photometric datasets, number of data points, $\chi_m^2$, chemical composition, optical depth, $L$, $T_{\rm eff}$, derived MLR, etc.

\begin{table*}
\caption{Final sample of 2,121 red supergiant star candidates in the SMC} 
\label{fsample}
\centering
\begin{tabular}{ccccccccc}
\toprule\toprule
R.A.(J2000)& Dec.(J2000)  & 2MASS\_J & e\_2MASS\_J & ...... & $\log_{10}$(L)   & $\log_{10}$($T_{\rm eff}$) & $\log_{10}$(MLR)       & m$_{\rm bol}$\\
(deg)      & (deg)        & (mag)    & (mag)       & ...... & ($\rm L_\sun$)   & (K)                        & ($\rm M_\sun~yr^{-1}$) & (mag)        \\
\midrule 
 4.368634  &   -73.428555 &   12.497 &  0.022 &  ......  & 3.874 & 3.685 & -6.105 & 14.005 \\
 4.921237  &   -73.353052 &   11.675 &  0.022 &  ......  & 4.135 & 3.663 & -4.738 & 13.353 \\
 4.952093  &   -73.588549 &   12.672 &  0.024 &  ......  & 3.834 & 3.699 & -6.063 & 14.106 \\
 5.335794  &   -73.407555 &   13.358 &  0.026 &  ......  & 3.519 & 3.699 & -6.464 & 14.892 \\
 5.800023  &   -72.390737 &   12.920 &  0.029 &  ......  & 3.731 & 3.661 & -5.769 & 14.363 \\
 ...       &   ...        &   ...    &  ...   &  ......  & ...   & ...   &  ...   & ...    \\
\midrule 
\end{tabular}
\tablefoot{
This table is available in its entirety at CDS. A portion is shown here for guidance regarding its form and content.
}
\end{table*}

\section{Mass-loss rate and physical properties of red supergiants}

\begin{figure}
\center
\includegraphics[scale=0.47]{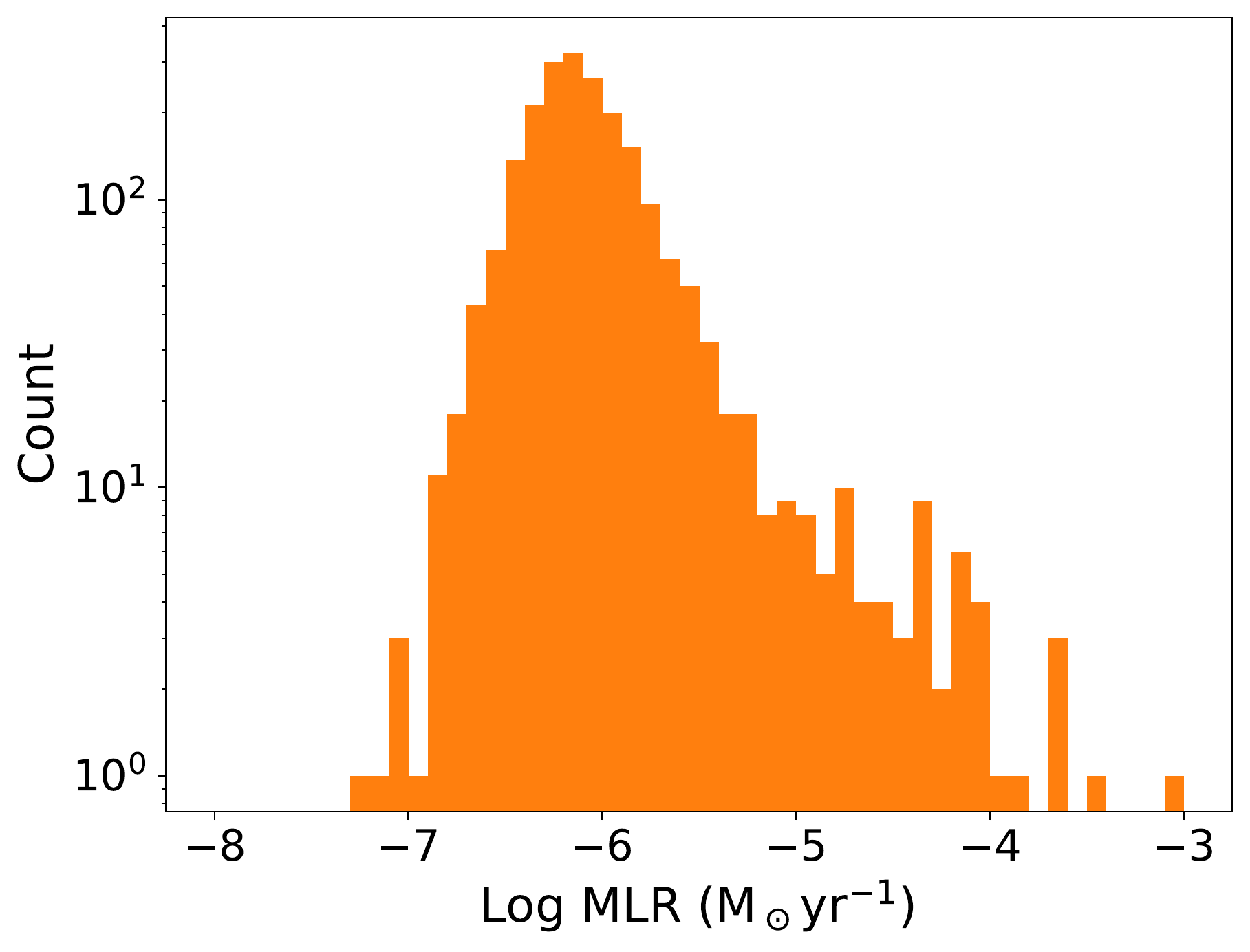}
\caption{Histogram of the derived MLRs of the final RSG sample. 
\label{hist_mlr}}
\end{figure}

\begin{figure}
\center
\includegraphics[scale=0.47]{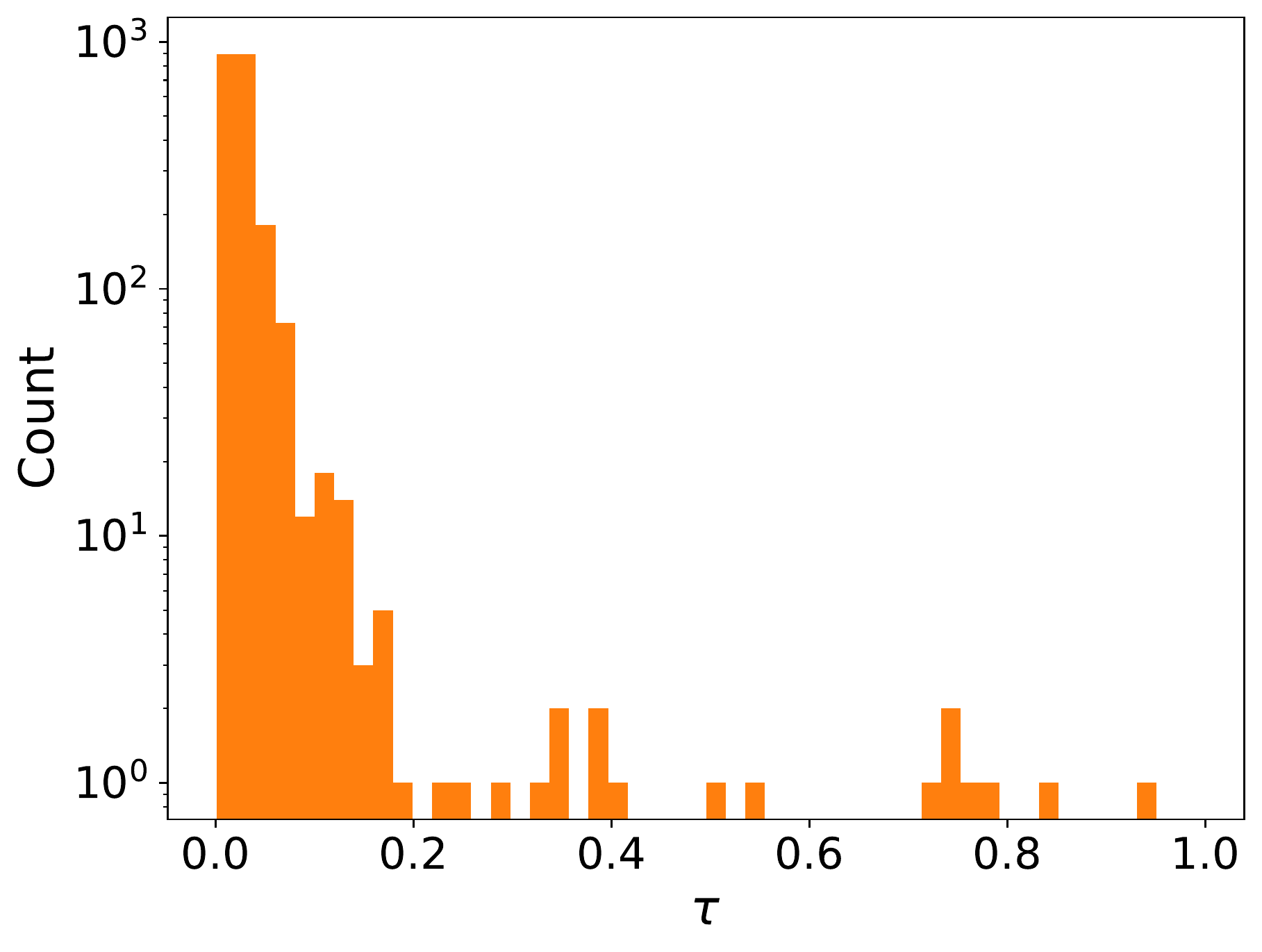}
\caption{Histogram of the optical depth of the final RSG sample. For clarity, targets with large $\tau$ (e.g., $>1.0$) are not shown in the diagram.
\label{hist_odepth}}
\end{figure}

Figure~\ref{hist_mlr} and Figure~\ref{hist_odepth} show the histograms of derived MLRs and $\tau$ of the final RSG sample. The majority of the targets has a typical MLR of $\sim10^{-6}$ $M_\sun$ yr$^{-1}$ ($\tau\lesssim0.1$), with a few outliers at the high MLR end (about $10^{-4}\sim10^{-3}$ $M_\sun$ yr$^{-1}$ with $\tau\gtrsim1.0$). A total MLRs of $6.16\times10^{-3}$ $M_\sun$ yr$^{-1}$ is measured for the sample. Notice that, this result is a lower limit, as we may still miss some RSGs in the SMC. However, we believe that our sample is quite close to complete compared to all the previous studies. Taking $\psi\approx1000$ as indicated before, this value can be converted to a dust-production rate (DPR) of $\sim6\times10^{-6}$ $M_\sun$ yr$^{-1}$. Previously, the dust ejection rate by AGBs/RSGs in the SMC was estimated by several studies, e.g., \citet{Boyer2012}, \citet{Matsuura2013}, and \citet{Srinivasan2016}, etc. \citet{Boyer2012} calculated a total (AGBs and RSGs) DPR of $(8.6-9.5)\times10^{-7}$ $M_\sun$ yr$^{-1}$, with RSGs contributed the least ($<4\%$; $\sim0.04\times10^{-6}$ $M_\sun$ yr$^{-1}$). \citet{Matsuura2013} found a higher global DPR of $\sim7\times10^{-6}$ $M_\sun$ yr$^{-1}$, for which $\sim3\times10^{-6}$ $M_\sun$ yr$^{-1}$ came from O-rich AGBs and RSGs. \citet{Srinivasan2016} also derived a similar global AGBs/RSGs dust-injection rate of $1.3\times10^{-6}$ $M_\sun$ yr$^{-1}$. Given the above, there is a general agreement between our result and those studies within a few factors, except the one from \citet{Boyer2012}. The discrepancies could be due to differences in sample sizes, the choice of $\rho_d$, $\psi$ and $v_{exp}$, adopted optical constants, model assumptions, etc., for which the MLR/DPR can largely vary. In addition, we also would like to indicate that, previous studies derived the DPR for RSGs and AGBs together, in which AGBs were dominant. That is to say, the contribution of RSGs is even smaller, while the difference between our result and those studies can be larger (e.g., may up to one order of magnitude).

Moreover, a significant part of the MLR actually comes from a few dusty ones. About 34\% of the MLR, which is $\sim2.11\times10^{-3}$ $M_\sun$ yr$^{-1}$, originates from seven targets with MLR larger than $10^{-4}$ $M_\sun$ yr$^{-1}$. About 31\% of the MLR, which is $\sim1.93\times10^{-3}$ $M_\sun$ yr$^{-1}$, originates from 55 targets with $10^{-5}<MLR\leq10^{-4}$ $M_\sun$ yr$^{-1}$. Overall, about 3\% of the relatively dusty targets has contributed about 65\% of the MLR. Meanwhile, as mentioned in Section \textsection3.2, there are six very dusty targets in our sample which are not well fitted with the models. Their uncertainties are likely to be around $10^{-4}$ $M_\sun$ yr$^{-1}$, so the total MLR and the contribution from the dusty ones may be even larger than our estimation.

\begin{figure*}
\center
\includegraphics[scale=0.46]{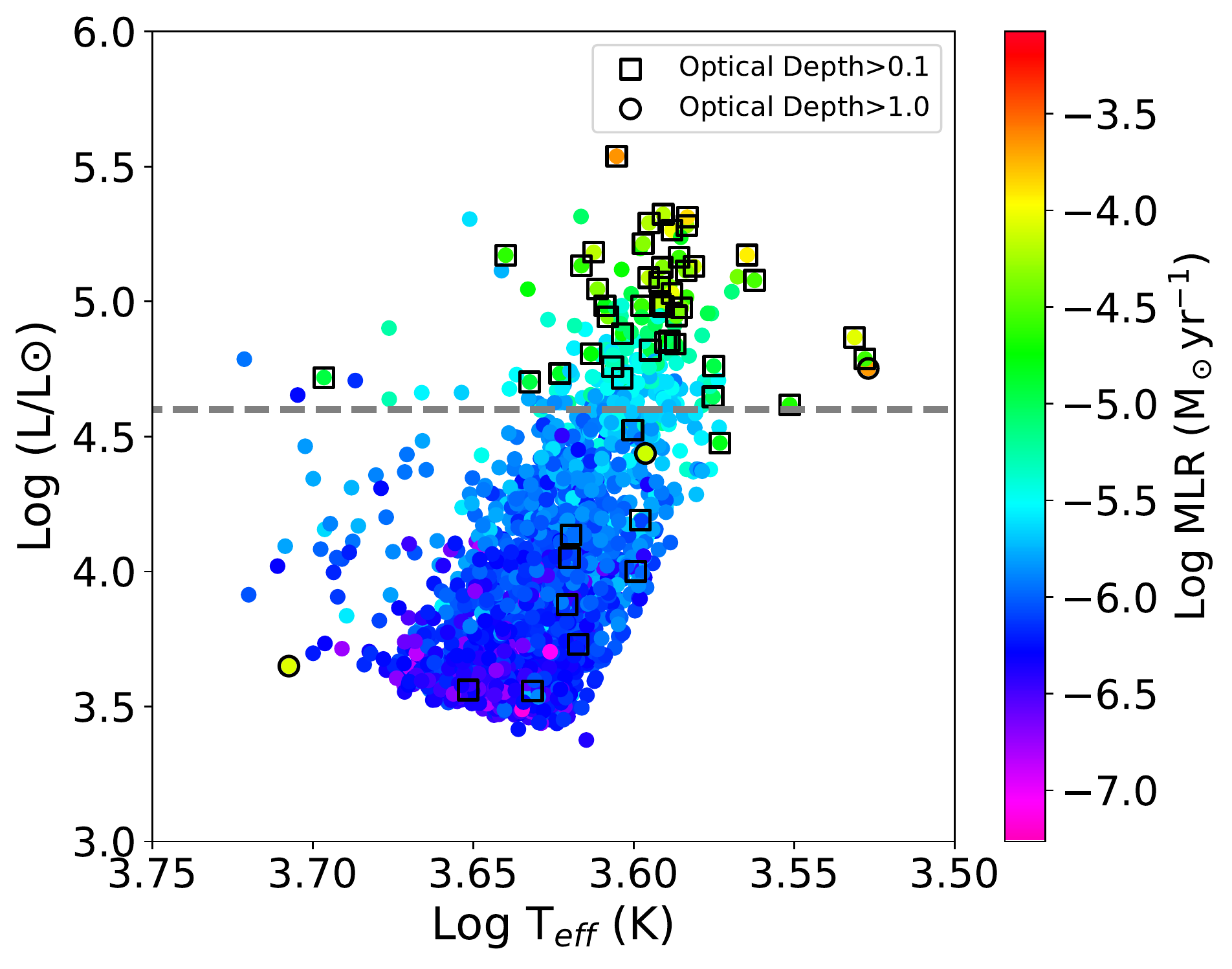}
\includegraphics[scale=0.46]{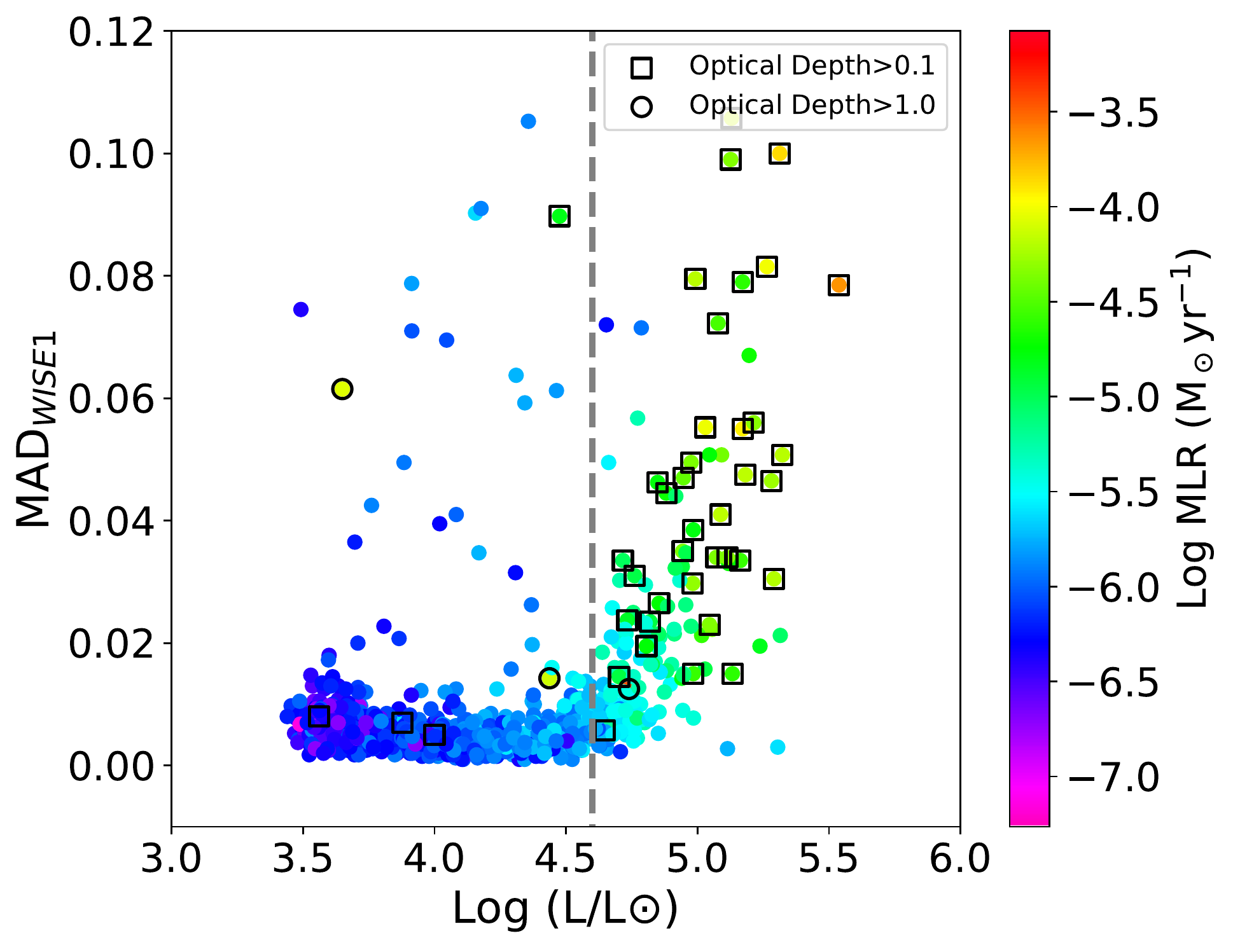}
\caption{Hertzsprung-Russell (left) and luminosity versus MAD (right) diagrams for the final RSG sample. Targets are color-coded with the MLR, while targets with relatively large ($>0.1$) and very large ($>1.0$) $\tau$ are marked by the open squares and open circles in the diagram, respectively. The dashed lines indicate the luminosity of $\log_{10}(L/L_\sun)=4.6$. For clarity, only the general population (without the extreme outliers) of the sample is shown in the diagram. In the luminosity versus MAD diagram, there are a few targets showing large MAD with low luminosity and MLR, which are mostly targets crossing the instability strip (because of their large variability, they are changing their $T_{\rm eff}$, so that sometimes they are RSGs, sometimes they are yellow supergiants).
\label{teff_lum_var_mlr}}
\end{figure*}

Figure~\ref{teff_lum_var_mlr} shows the H-R diagram and luminosity versus median absolute deviation (MAD) diagram of the final RSG sample, respectively. For the H-R diagrams, one prominent feature can be seen from the diagrams that, most of the relatively dusty targets ($\tau>0.1$) appear at the bright end, confirming the correlation between luminosity and MLR \citep{Yang2018, Yang2020}. The luminosity versus MAD diagram shows the similar trend that the MLR increases along with the increasing of luminosity and variability (we derived MAD based on the WISE/NEOWISE time-series data in [3.4] band, for which MAD is a variability index that measures the robust variability of the object; see more details in \citealt{Yang2018, Yang2019}). Moreover, as there is a positive correlation between variability and luminosity for the RSGs \citep{Kiss2006, Yang2011, Yang2012, Soraisam2018, Yang2018, Chatys2019}, the MLR can be considered as a monotonous function of the luminosity. 

\begin{figure}
\center
\includegraphics[scale=0.48]{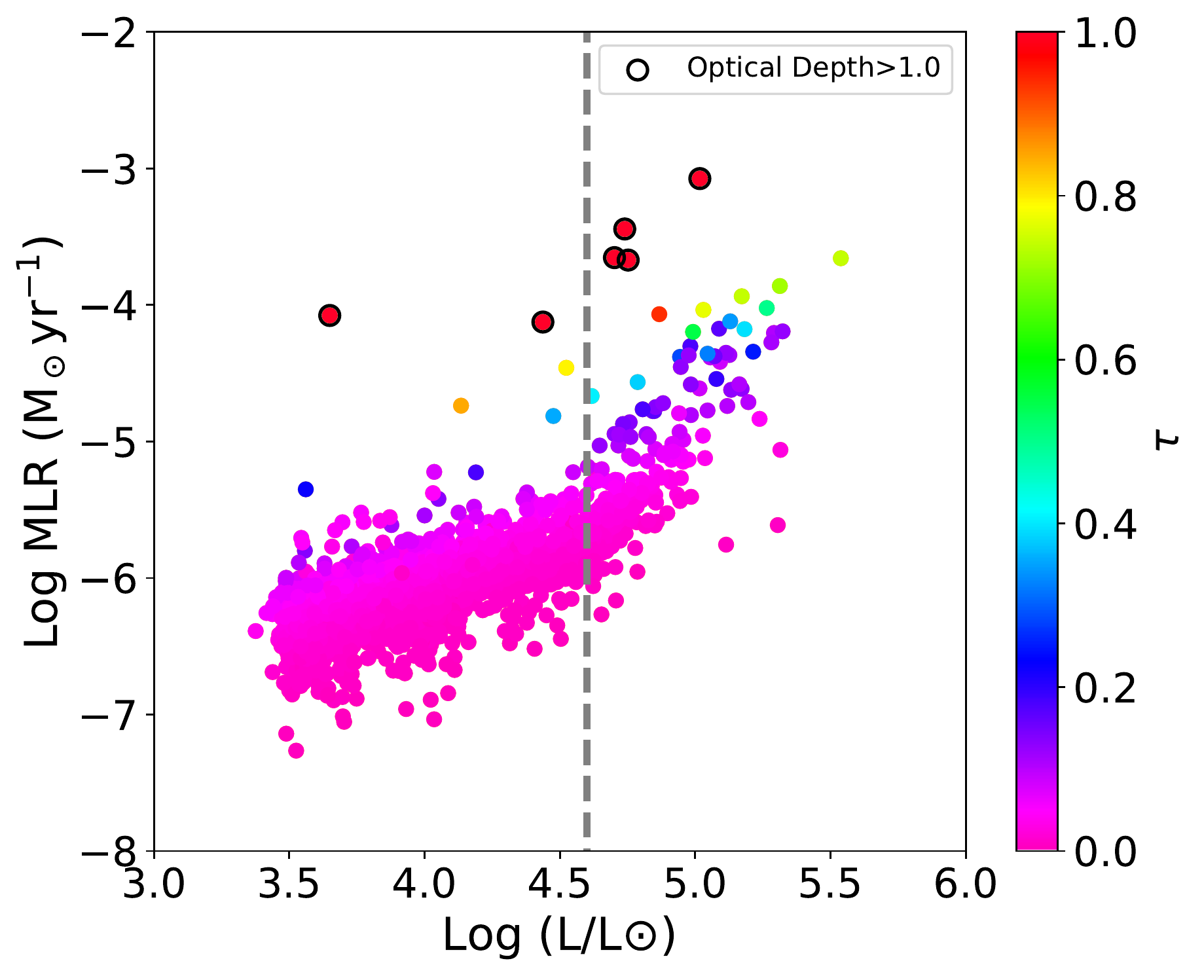}
\caption{Luminosity versus MLR diagrams for the final RSG sample. Targets are color-coded with $\tau$, while targets with very large $\tau$ ($>1.0$) are marked by the open circles in the diagram. The dashed line indicates the luminosity of $\log_{10}(L/L_\sun)=4.6$. We note that the low-luminosity targets ($\log_{10}(L/L_\sun)\lesssim4.0$) may not be true RSGs, but $\sim6-8~M_\sun$ red helium-burning stars.
\label{lum_mlr}}
\end{figure}

Figure~\ref{lum_mlr} shows the luminosity versus MLR diagram. The sample shows a ``knee-like'' shape with a notable turning point around $\log_{10}(L/L_\sun)\approx4.6$, where the enhanced mass-loss seems to take place. This scenario can be understood straightforwardly by the radiatively-driven wind mechanism of the RSGs (see also \textsection1), i.e., the pulsation may ``throw'' the materials of outer layers of the stars to larger radii (well past the stellar photosphere) where the dust can be condensed. The dust envelope is then expanded by the radiation force on the dust grains, which collides with the surrounding gas and drags it along to form the stellar wind. As indicated by the luminosity versus MAD diagram, evident variability can be seen for the bright RSGs ($\gtrsim \log_{10}(L/L_\sun)\approx4.6$), for which the pulsation effectively develops (along with the increased radiation pressure at higher luminosity) to enhance the mass-loss and further create the ``knee-like'' point. However, the exact mechanism of this turning point is still unknown. The degeneracy of luminosity, pulsation, low surface gravity, convection, and other factors is inevitable and may result in the appearance of such catastrophic effect \citep{Yang2018, Yang2020}.

\begin{figure*}
\center
\includegraphics[scale=0.31]{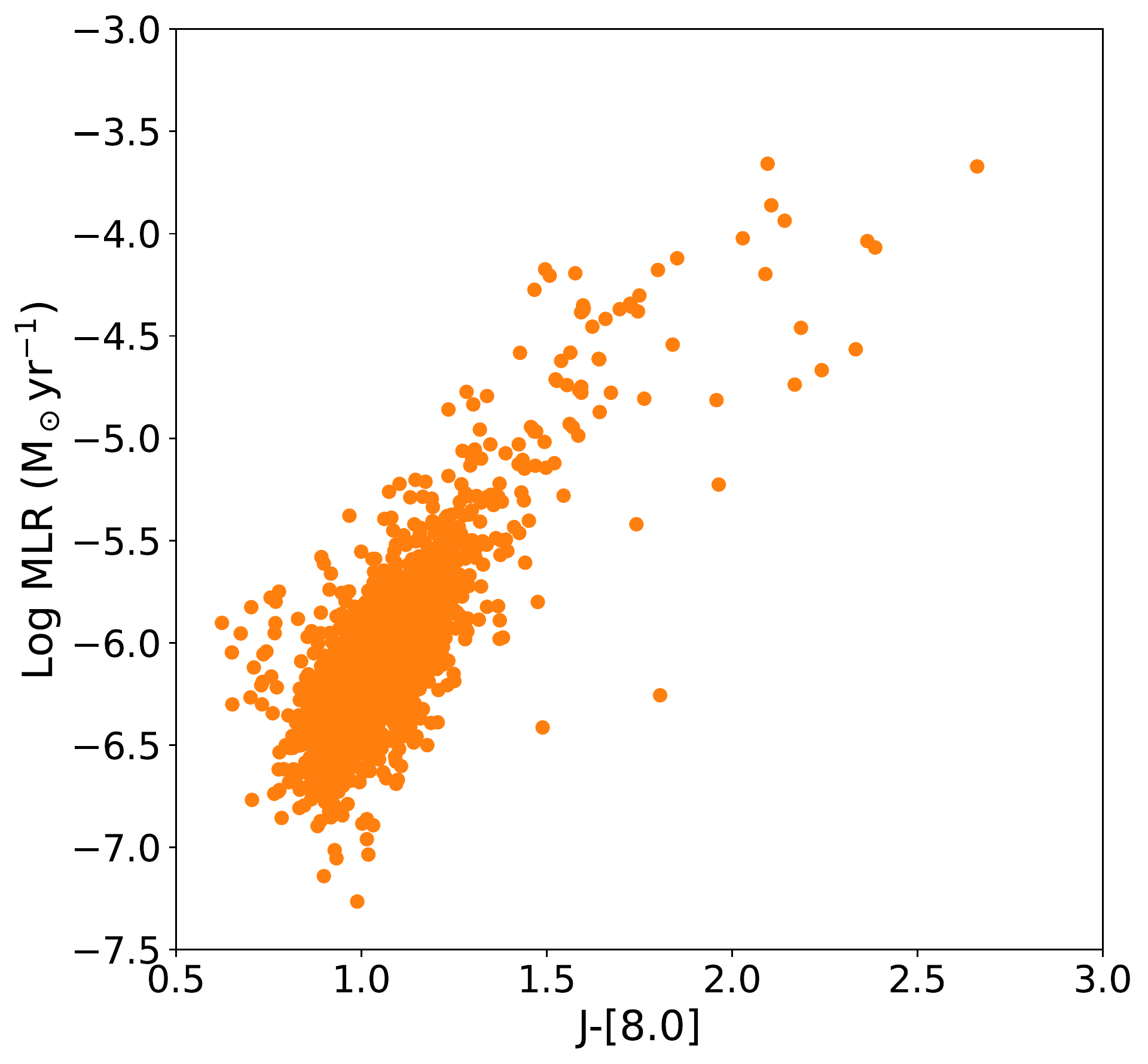}
\includegraphics[scale=0.31]{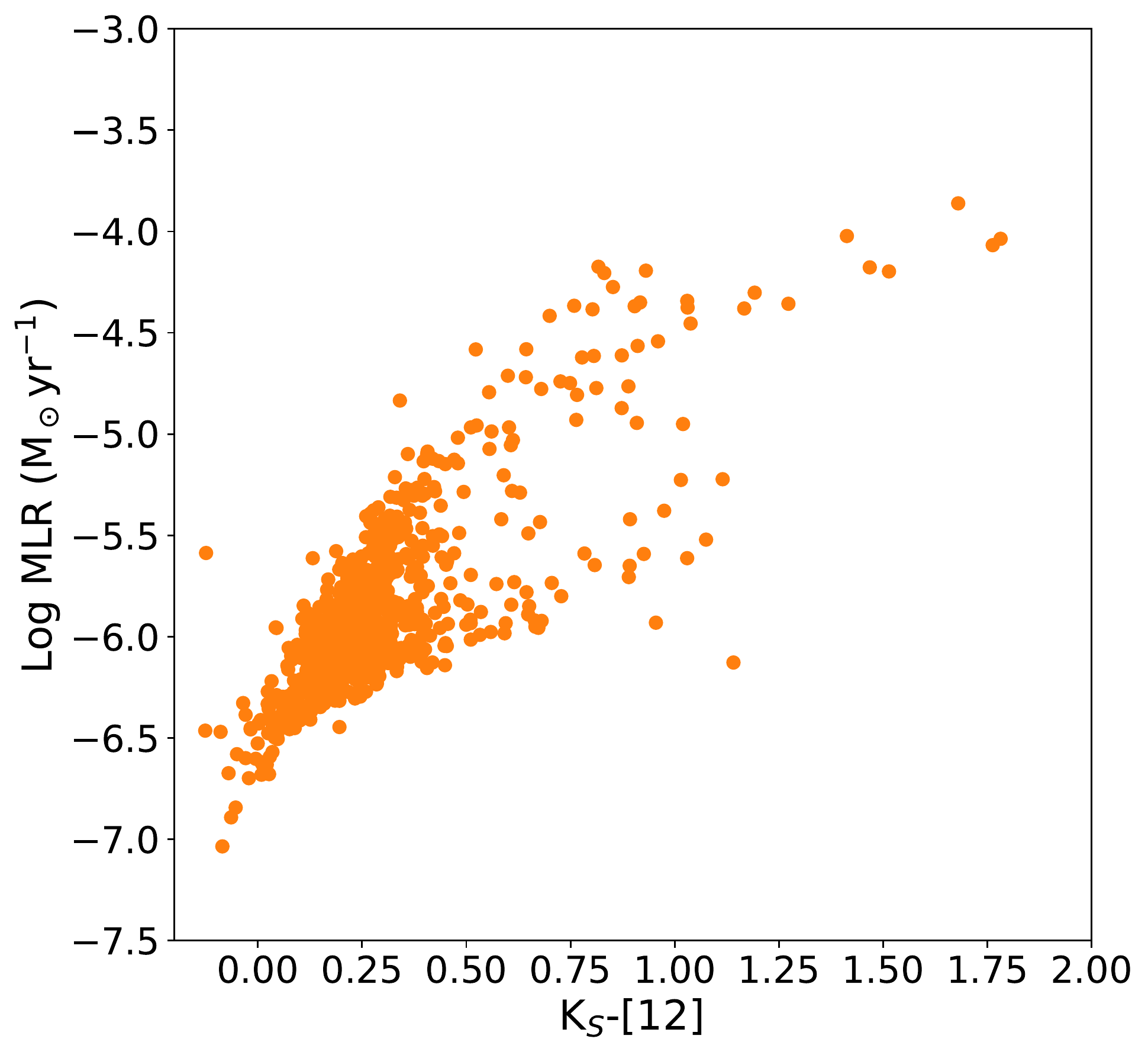}
\includegraphics[scale=0.31]{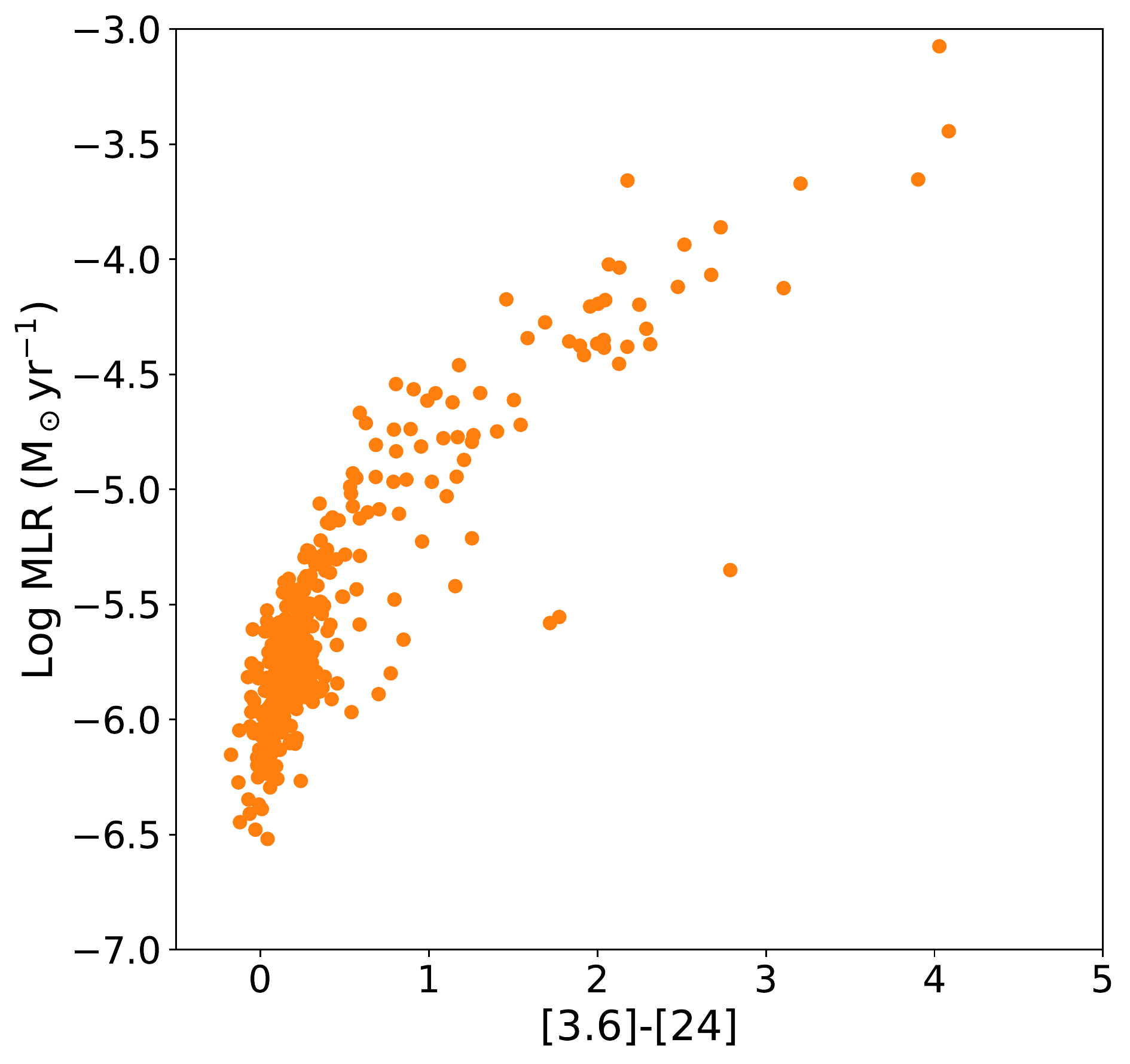}
\caption{$J-[8.0]$ (left), K$_S-[12]$ (middle), and $[3.6]-[24]$ (right) versus MLR diagrams for the final RSG sample.
\label{color_mlr}}
\end{figure*}

Figure~\ref{color_mlr} shows the color versus MLR diagrams with three typical MLR indicators of $J-[8.0]$, $K_S-[12]$, and $[3.6]-[24]$. All of them show the positive correlation between the redder color and larger MLR, ranging from 1 to 2.5 in $J-[8.0]$, 0 to 1.75 in $K_S-[12]$, and 0 to 4 in $[3.6]-[24]$, respectively.


The left panel of Figure~\ref{lum_mlr_ref} shows the final luminosity versus MLR diagram, where a third-order polynomial fitting is adopted as,
\begin{equation}
    \begin{split}
log_{10}(\dot{M}) =~~&0.45\times[log_{10}(L/L_\sun)]^3-5.26\times[log_{10}(L/L_\sun)]^2\\
              &+20.93\times[log_{10}(L/L_\sun)]-34.56.
    \end{split}
\end{equation}
Overall, except for the very dusty targets ($\tau>1.0$), the general population of the sample is well fitted with the third-order polynomial but may underestimate the MLR at the extended faint end (e.g., $\log_{10}(L/L_\sun)\approx3.0$). However, since the MLR will be indeed very small at such luminosity as can be predicted from the diagram, this effect should have no big impact. Moreover, as indicated in previous sections, both luminosities and MLRs have large uncertainties at the very bright end where a significant part of the contribution comes from, therefore the relation may be less precisely determined.

\section{Discussion}

\subsection{The complexity of mass-loss estimation based on the spectral energy distribution}

First of all, we would like to indicate that, the derived MLRs of RSGs based on current photometric datasets have large uncertainties. The reason for this is simple and straightforward: the complexity of the error budgets and parameter spaces in both methodology and physics. 

In methodology, for example, the adopted magnitudes are derived based on the total system throughput (e.g., filters convolve with camera, telescope, and atmosphere for ground-based telescopes), which may have small uncertainties and change with time or external conditions (e.g., J-band is sensitive to the amount of precipitable water in the atmosphere\footnote{https://old.ipac.caltech.edu/2mass/releases/allsky/doc/\\sec3\_1b1.html}). In that sense, it may result in a few percent difference in flux (Philip Massey, private communication). The difference between maximum and minimum flux for each individual target spans around 2-3 orders of magnitude (e.g., see Figure~\ref{fitting_example1}), for which the weights are largely different (the typical solution for this problem is to fit the SED in logarithmic space, while we adopt a flat spectrum to solve it as described in Section \textsection3.2). The photometric data used here are all derived from broad-band filters, so that some prominent spectral features may be smoothed out. For example, the 9.7 $\mu$m silicate feature, which covers about 8-13 $\mu$m, is not easily distinguished, except for a few targets ($\sim$4.4\%) detected by \textit{AKARI} [S11] ($\lambda_{\rm eff}\approx10.2$ $\mu$m with a range of $\sim$8.3-15.3 $\mu$m; see also Figure~\ref{fitting_example1}), since it is right between the \textit{Spitzer} [8.0] ($\lambda_{\rm eff}\approx7.6$ $\mu$m with a range of $\sim$6.3-9.6 $\mu$m) and \textit{WISE} [12] ($\lambda_{\rm eff}\approx11.6$ $\mu$m with a range of $\sim$7.4-17.3 $\mu$m) bands (notice that, the quantum efficiency drops fast at the edges of each filter). The photometric error is inhomogeneous across the SED, as it is usually about a few thousandths of a magnitude in the optical bands, but increases to $\sim$0.03-0.05 mag when wavelengths are longer than 10 $\mu$m (an important reason that we have not used reduced $\chi^2$, since it takes error as the weight and therefore the most important wavelength range of mid-IR will have relatively low weight; more discussion about the pros and cons of reduced $\chi^2$ can be found in \citealt{Andrae2010}). Similarly, the angular resolutions also have large inconsistencies between different wavelengths, ranging from subarcsec to a dozen of arcsecs (e.g., the typical angular resolutions are about 4.5'', 2'', 0.9'', 0.9'', 5'', 6'', 2'', 12'', and 6'' for GALEX, \textit{Gaia}, NSC, VMC, 2MASS, \textit{WISE} [3.6], \textit{Spitzer} [8.0], \textit{WISE} [22], and \textit{Spitzer} [24], respectively). Hence, neighboring targets/environment may contaminate the measurement in low-resolution data, which is quite common in nearby and distant galaxies. Moreover, for wavelengths with available time-series information, the sampling of lightcurves is often sparse without the full coverage of the period (the typical periods of bright RSGs are about 300-800 d; \citealt{Yang2011, Yang2012, Chatys2019, Ren2019}). In our case, only the \textit{WISE} [3.4] and [4.6] data have long enough coverage to derive relatively accurate median magnitudes. Thus, the single-epoch/incomplete time-series measurements in some wavelengths may not represent the appropriate luminosities. Meanwhile, different datasets are taken at different times, ranging from the late 1990 to the present. Moreover, problems also come from the poor data quality as mentioned in Section \textsection2, i.e. that inaccurate measurements will influence the derived MLR (especially for the mid-IR data that dominate the MLR modelling; we have tried our best to filter out the low quality data but still some cases may be present). Different fitting methods or normalized wavelengths may also result in about 10-30\% differences, while the majority of them appear in the low optical depth region ($\tau<0.1$). Differences in the sample sizes (incomplete or biased samples) will also lead to different results. Finally, for the most important factor, the degeneracy of different models is inevitable and complex due to many parameters in the modelling. For examples, the typically used consecutive equally spaced logarithmic steps of optical depths, e.g., 0.001 to 10.0 in 99 intervals, is too dense at the low optical depth (half of the steps will be less than 0.1), which cannot really distinguish adjacent models by using only photometric data; this is also the reason why we set up our own steps for $\tau$. Moreover, one may always find a best model with low $\chi^2$ for the target, as long as the numbers of parameters and generated models are large enough, so the solution may fall in a local minimum instead of the global minimum. In such cases, the choice of the correct solution cannot be justified, unless each parameter (or the dominant parameters) is precisely constrained (to say, a few percent; this is indeed very difficult for almost every case of SED fitting, not just for our study of MLRs).

In physics, for example, the emerged theoretical spectra are slightly different if we use different central illuminating sources, like blackbody or stellar atmosphere models (e.g., PHONEIX, ATLAS, or MARCS). The variability of RSGs may also cause some uncertainties. RSGs are semi-regular variables (considering also the contribution of convection). Typically, as the wavelength increases, the amplitude will gradually decrease, the lightcurve will progress towards a more symmetrical variation, and the phase of the maximum will shift toward later phases. In that sense, the phase/time lags in variability will change the SED shape accordingly (along with the pulsation, the convection may also play an important role in the mass-loss of RSGs as they may help the release of materials from the outer layer of the star). Besides that, it is also possible to have small contamination from the AGB population at the red and/or faint end of our sample as mentioned in Section \textsection2. The extinction is another issue, as we have only corrected the average foreground extinction towards the SMC. However, since the 3D structure and internal extinction of the SMC, and the distances of the targets are not accurately determined, the internal extinction of the SMC may largely vary from star to star, especially for RSGs that are supposedly close to the star formation regions. Moreover, the physical properties and driving mechanisms of the mass-loss of RSGs are still poorly understood, not to mention the dependence of MLR on many factors. For example, for the chemical composition, one may use silicate from \citet{Ossenkopf1992} instead of \citet{Draine1984}. Additionally, there is evidence for Polycyclic Aromatic Hydrocarbons (PAHs) or SiC existing in RSGs. Such occurrence of carbon species in an O-rich environment is possible, due to the dissociation of CO by the strong ultraviolet emission from chromosphere \citep{Beck1992, Hofner2007}. The produced free carbon atoms may form carbon dust along with the expected O-rich silicate dust. Meanwhile, the UV photons from chromospheric emission may also excite the IR emission from PAHs or similar carbonaceous species \citep{Allamandola1989}. However, on the other hand, the diffuse ISM is also known to exhibit PAH emission features, which can be confused with the observed target. Therefore, the exact scenario of carbonaceous dust formation in the RSGs is still unclear (as well as the solid definition of ``C-rich'' RSG, since in almost all cases, the 9.7 $\mu$m silicate bump and PAH emission coexist) and we exclude carbon dust in our modelling. For the dust density and size distribution, one may use a steady-state density distribution of $r^{-q}$ with power-law index $q\leq2$ instead of a radiatively-driven wind, or choosing an average grain size instead of size distribution. For $\psi$, traditionally, one may assume an average value of 100-200 for the Milky Way, 400-500 for the LMC, and $\sim$1000 for the SMC. However, it may largely vary for each individual source. For the wind-expansion velocities ($v_{exp}$), it highly depends on the luminosity of the central source, metallicity, and other factors \citep{Goldman2017}. For $\rho_d$, it may change upon the chemical composition and structure of the dust grain. For the metallicity, many studies often include both Milky Way and MC RSGs, spanning a range of metallicities. There is also an inherent uncertainty of $\sim$30\% for the DUSTY radiatively-driven wind model \citep{Ivezic1999}. Finally, some external factors, like binarity, may also play an important role. More discussions about modelling parameters can be found in \citet{Ivezic1997}.

One may argue that each of the mentioned issues just create a small error (e.g., a few percent). However, the errors are accumulated and propagated, so that the final error of MLR becomes large and complicated (potentially up to one order of magnitude or more; see also below). Thus, in brief, the MLRs derived in this work (or any other works based on photometric data) can be only considered as a mixture of ``snap shot'' with large uncertainties. This effect is especially true for the bright RSGs with larger variabilities and MLRs, but is much mitigated for the faint RSGs with smaller variabilities and MLRs. While it is difficult to assess the MLR with a few percentage accuracy, our result help to understand the underlying relation between MLR and other physical parameters and putting a relatively stringent constraints on it. More details about the analysis will be presented in our future paper (Wen et al., in preparation).


\subsection{Comparison with previous mass-loss rate prescriptions}

We also compared our work with previous studies. Historically, many empirical relations have been used in the stellar evolutionary codes in order to estimate the MLR as a function of basic stellar parameters, such as $L$, $T_{\rm eff}$, mass ($M$), and radius ($R$). However, in the majority of cases, the dominant factor is the $L$ (e.g., the $T_{\rm eff}$ range of RSGs is relatively narrow, while $R$ and $M$ are $L$-dependent as shown below; \citealt{Mauron2011}), which is also relatively easy to be measured. Thus, following previous studies, we simplified the MLR prescriptions. 

The first widely used relation was the Reimers law (hereafter R75; \citealt{Reimers1975, Kudritzki1978}) as, 
\begin{equation}
\dot{M}=5.5\times10^{-13}LR/M,
\end{equation}
which was based on a small sample of stars including both red giants and RSGs ($L$, $R$, and $M$ in solar units, same below). For RSGs, this relation can be simplified for an average temperature of $T_{\rm eff}=3750$ K (same below when taking $T_{\rm eff}$ into account) as, 
\begin{equation}
\dot{M}=3.26\times10^{-12}(L)^{1.17},
\end{equation}
since $R/R_\sun=(L/L_\sun)^{0.5}\times(T_{\rm eff}/5770)^{-2}$ and $L/L_\sun =f(M/M_\sun)^3$ with $f\approx15.5$ (same below for the simplification; \citealt{Mauron2011, Kee2021}). Notice that, different studies might use different definitions of $M$ term (as well as the corresponding $L$ term), e.g., one might use initial mass, current mass, most expected mass, final mass, etc. Using the same $M$-$L$ relation for all of the MLR prescriptions may be slightly misleading, but good enough for the simplification, considering the large uncertainties mentioned in previous section.

About a decade later, \citet{deJager1988} developed the famous MLR prescription for stars located over the whole H-R diagram, including 15 Galactic RSGs, which was presented as a sum of Chebyshev polynomials (hereafter dJ88). Commonly used in stellar evolution codes, it is expressed as the first-order approximation,
\begin{equation}
\dot{M}=10^{-8.158}(L)^{1.769}(T_{\rm eff})^{-1.676}.
\end{equation}
Meanwhile, a second formula was published later by also taking into account the stellar mass (hereafter NJ90; \citealt{Nieuwenhuijzen1990}), as
\begin{equation}
\dot{M}=10^{-14.02}(L)^{1.24}(M)^{0.16}(R)^{0.81}.
\end{equation}

Later on, \citet{Feast1992} found that there was a fairly well relation between pulsation period ($P$), $L$, and MLR for 15 RSGs in the LMC based on the data from \citet{Reid1990} as,
\begin{equation}
log_{10}(\dot{M})=1.32 \times log_{10}P-8.17, 
\end{equation}
and 
\begin{equation}
M_{bol}=-2.38\times log_{10}P-1.46.
\end{equation}
In that sense, the relation between MLR and $L$ (Feast law; hereafter F92) can be written as,
\begin{equation}
\dot{M}=10^{-11.609}(L)^{1.387}.
\end{equation}
Meanwhile, a MLR prescription from \citet{Vanbeveren1998} was also based on the work of \citet{Reid1990} as (hereafter V98), 
\begin{equation}
\dot{M}=10^{-8.7}(L)^{0.8}.
\end{equation}

Another relation was proposed by \citet{Salasnich1999}, who adopted the Feast law but took also into account the possibility that the $\psi$ varied with the stellar luminosity. The Salasnich relation (hereafter S99) is,
\begin{equation}
\dot{M}=10^{-14.5}(L)^{2.1}.
\end{equation}

Afterwards, \citet{vanLoon2005} derived a well-known mass-loss law for oxygen-rich dust-enshrouded AGBs and RSGs in the LMC (hereafter vL05). The van Loon law is written as,
\begin{equation}
\dot{M}=10^{-5.56}\left(\frac{L}{10^4}\right)^{1.05}\left(\frac{T_{\rm eff}}{3500}\right)^{-6.3}.
\end{equation}

More recently, \citet{Goldman2017} (hereafter G17) and \citet{Beasor2020} (hereafter B20) developed new MLR prescriptions for RSGs, based on samples of AGBs and RSGs with OH masers in the LMC and the Milky Way, and RSGs in the clusters, respectively. The G17 relation is written as,
\begin{equation}
\dot{M}=1.06\times10^{-5}\left(\frac{L}{10^4}\right)^{0.9}\left(\frac{P}{500d}\right)^{0.75}\left(\frac{\psi}{200}\right)^{-0.03}, 
\end{equation}
where $P$ is the pulsation period. The B20 relation is written as,
\begin{equation}
\dot{M}=10^{-26.4-0.23M_{ini}}(L)^{4.8},
\end{equation}
where $M_{ini}$ is the initial mass.

\citet{Wang2021} also developed a MLR prescription of RSGs in relatively metal-rich galaxies of M31 and M33 by using DUSTY (hereafter W21). The prescription is applied for O-rich and C-rich RSGs, respectively, as,
\begin{equation}
\begin{split}
\dot{M}=10^{-8.31}(L)^{0.83} \ (\rm M31, silicate), \\
\dot{M}=10^{-10.03}(L)^{0.91} \ (\rm M31, carbon), \\
\dot{M}=10^{-8.64}(L)^{0.92} \ (\rm M33, silicate), \\
\dot{M}=10^{-11.13}(L)^{1.14} \ (\rm M33, carbon).
\end{split}
\end{equation}

\begin{figure*}
\center
\includegraphics[scale=0.47]{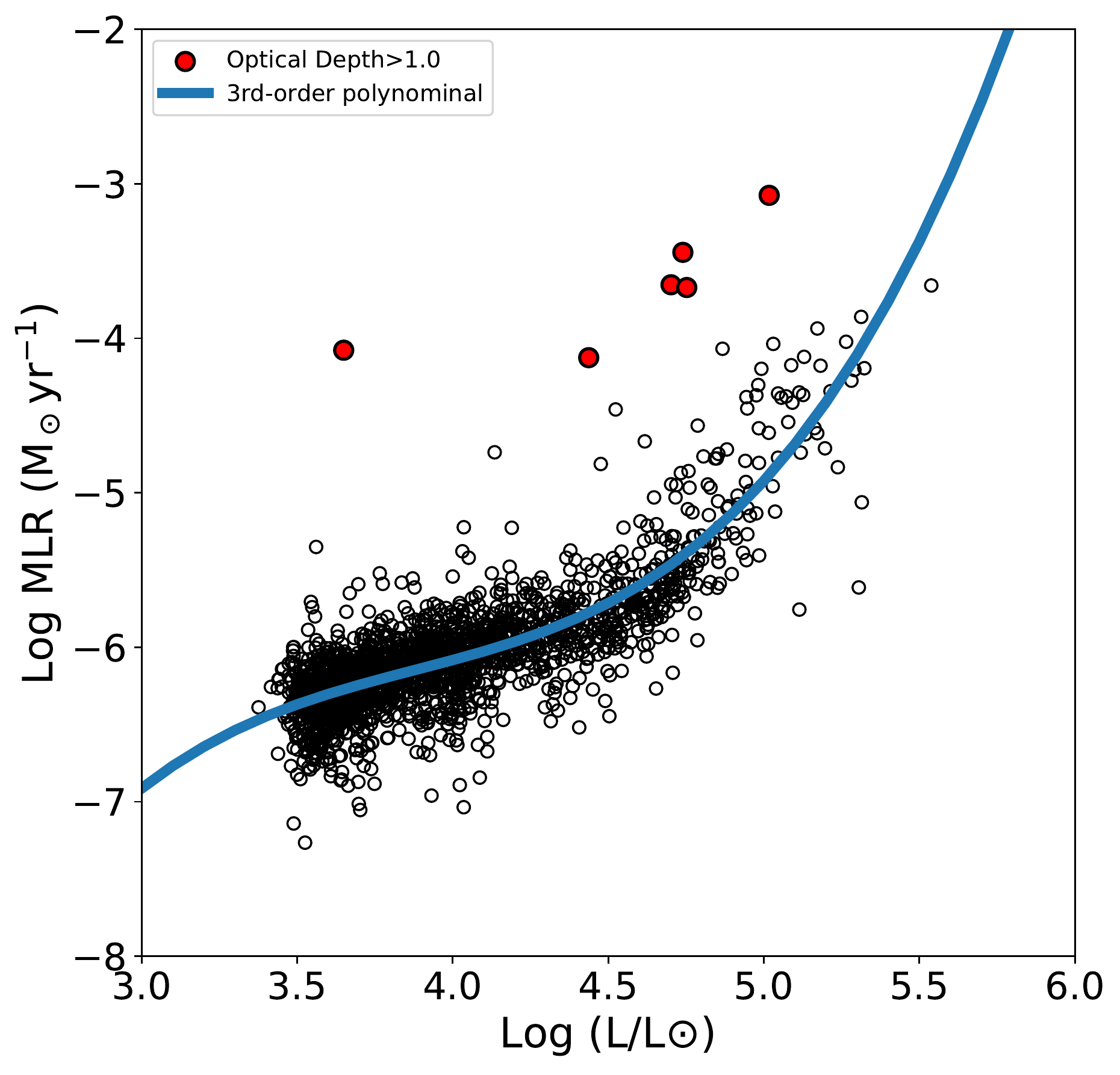}
\includegraphics[scale=0.47]{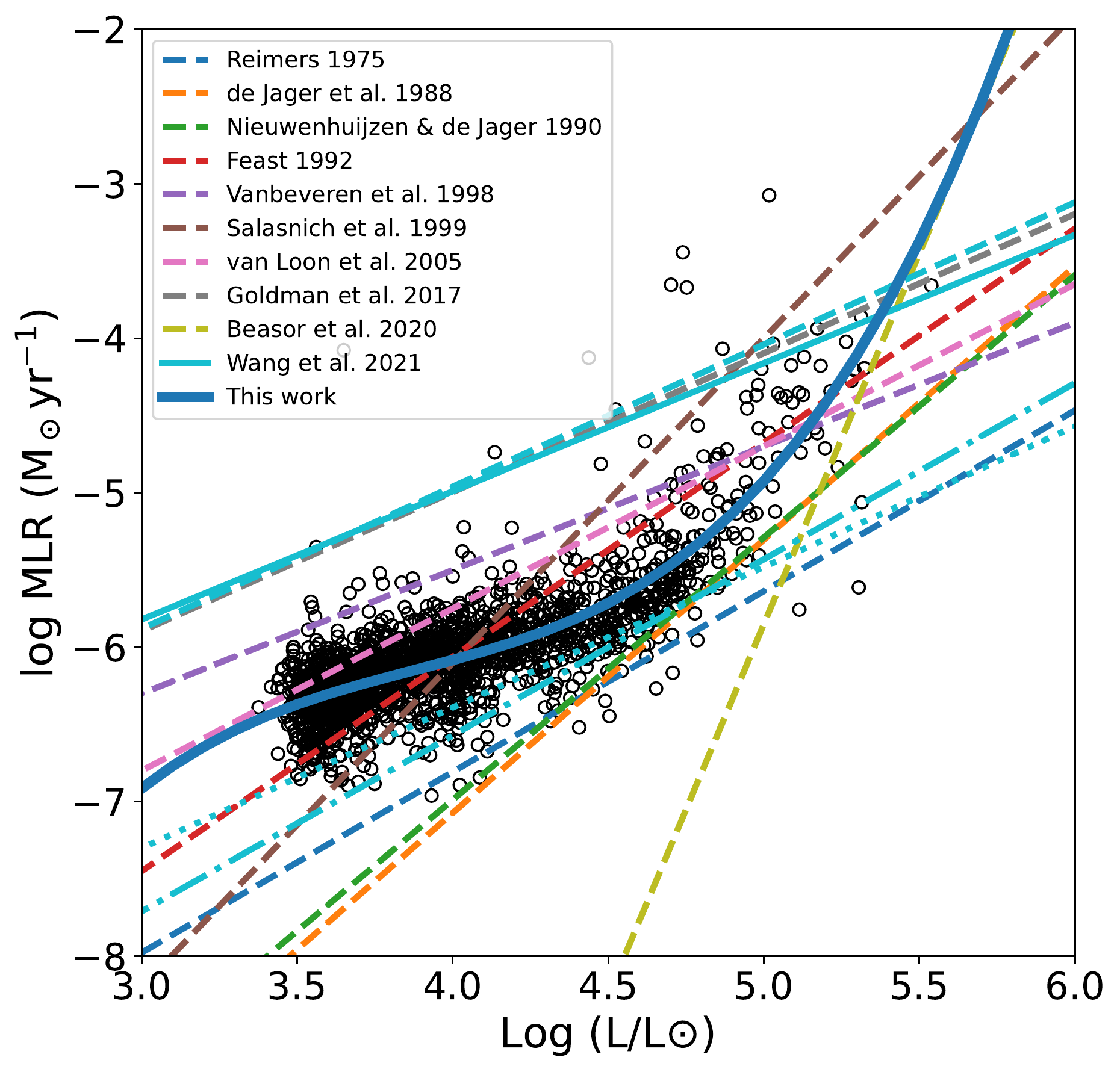}
\caption{Derived MLR-$L$ relation from this work (left) and comparison of the same relation between this and previous works (right). In the left panel, the very dusty targets ($\tau>1.0$) are marked with red colors. In the right panel, lines of the same color are variations of the same relation (as shown in Figure~\ref{lum_mlr_ref_each}).
\label{lum_mlr_ref}}
\end{figure*}

\begin{figure*}
\center
\includegraphics[scale=0.24]{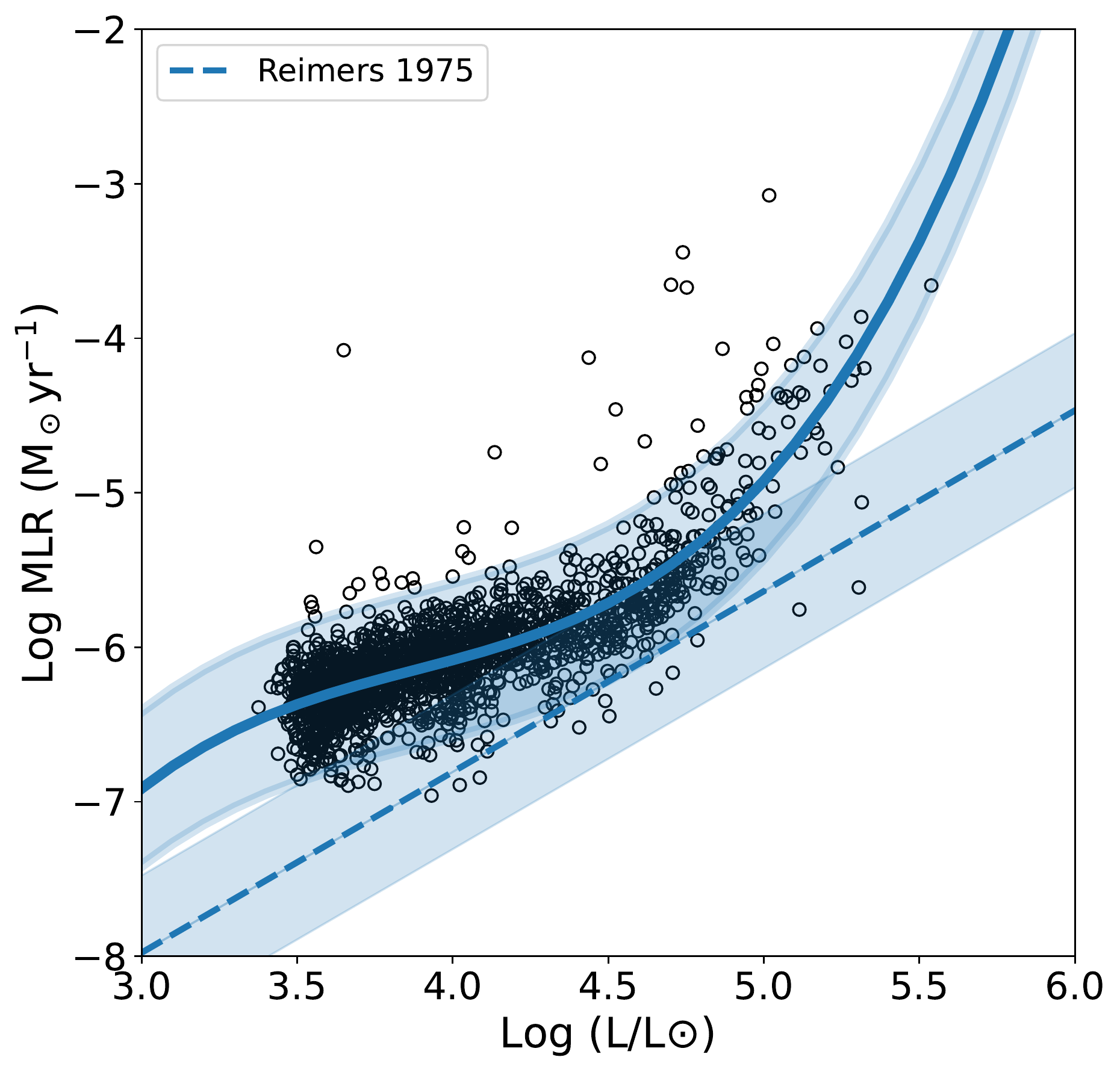}
\includegraphics[scale=0.24]{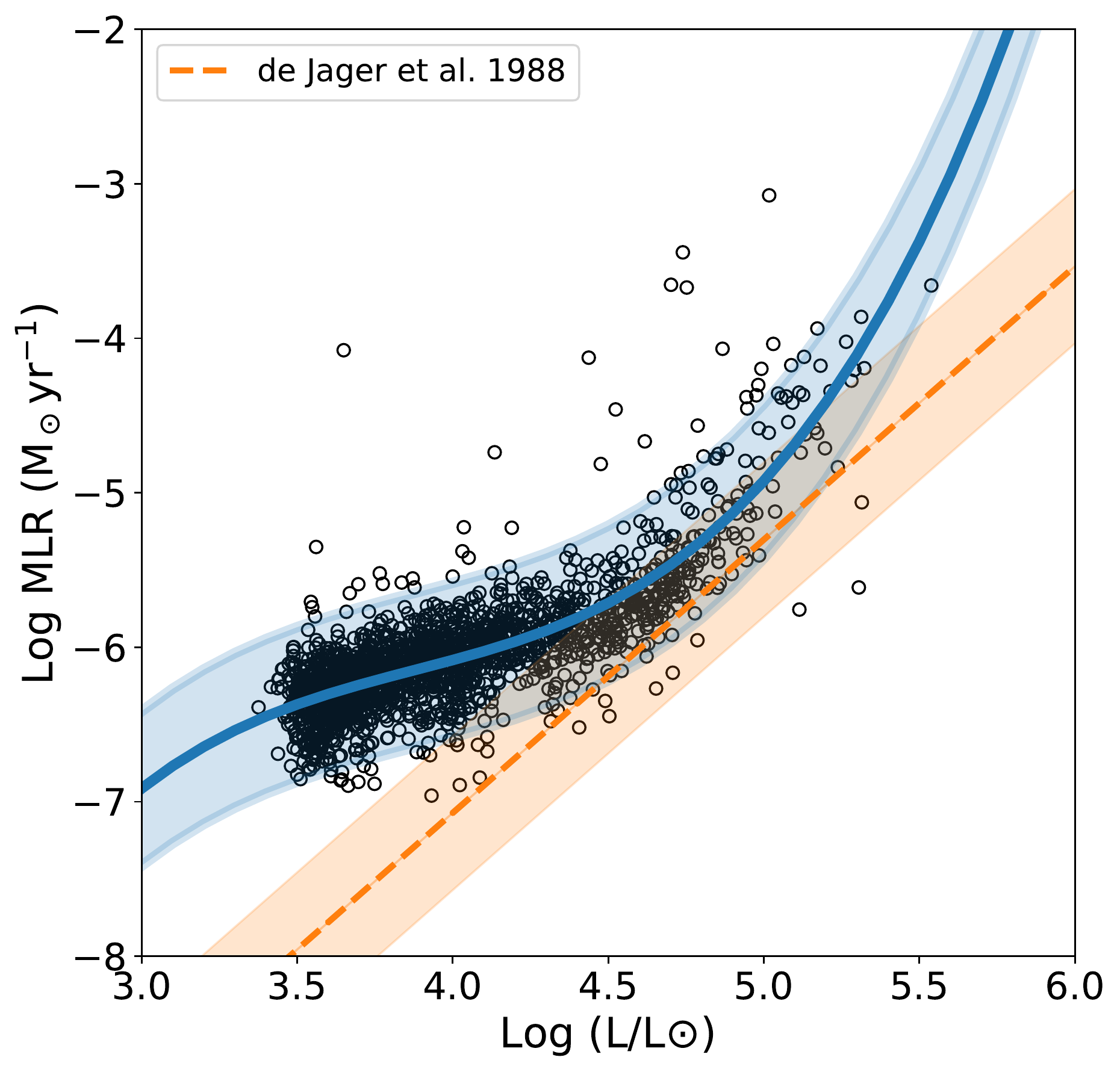}
\includegraphics[scale=0.24]{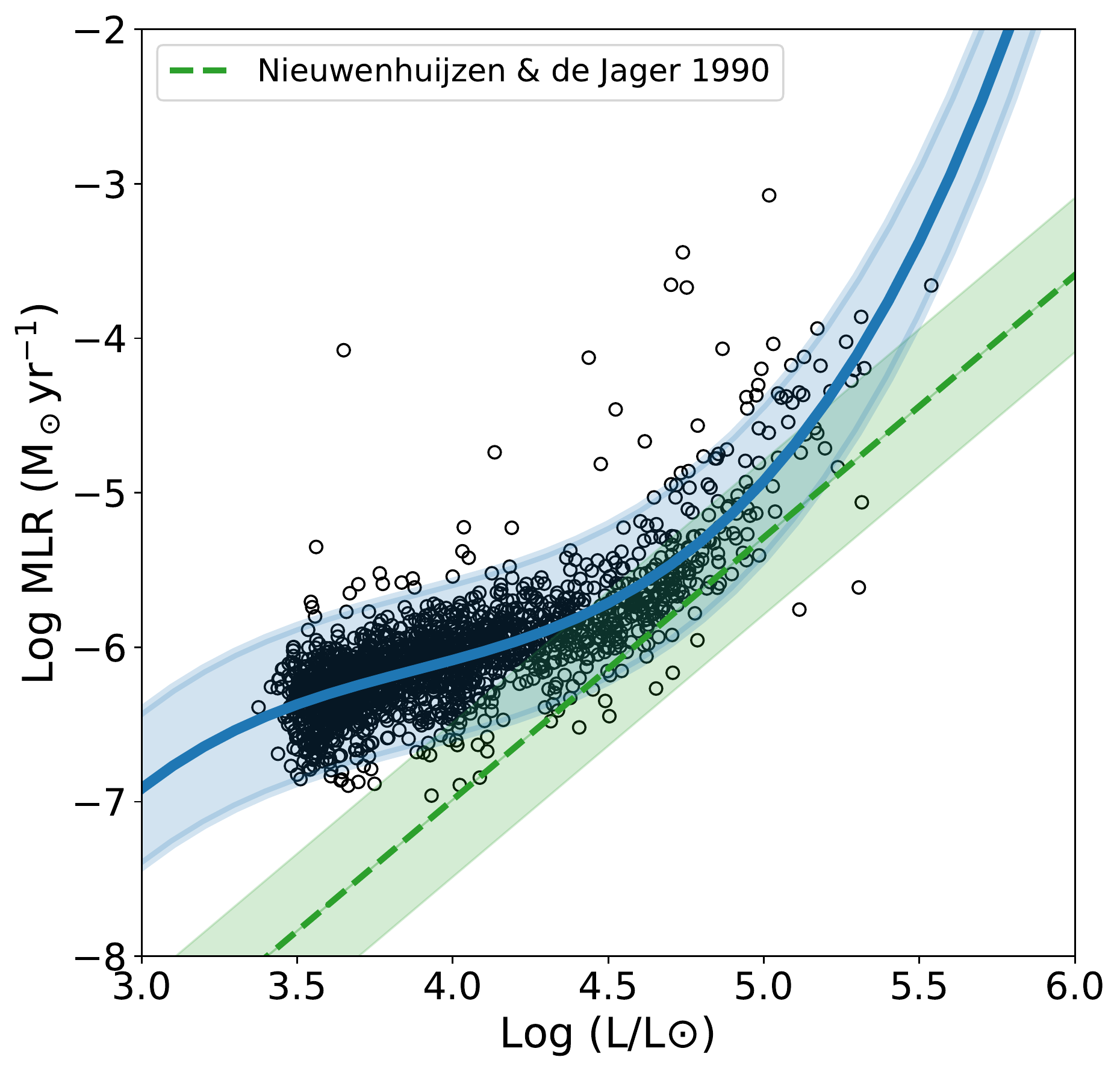}
\includegraphics[scale=0.24]{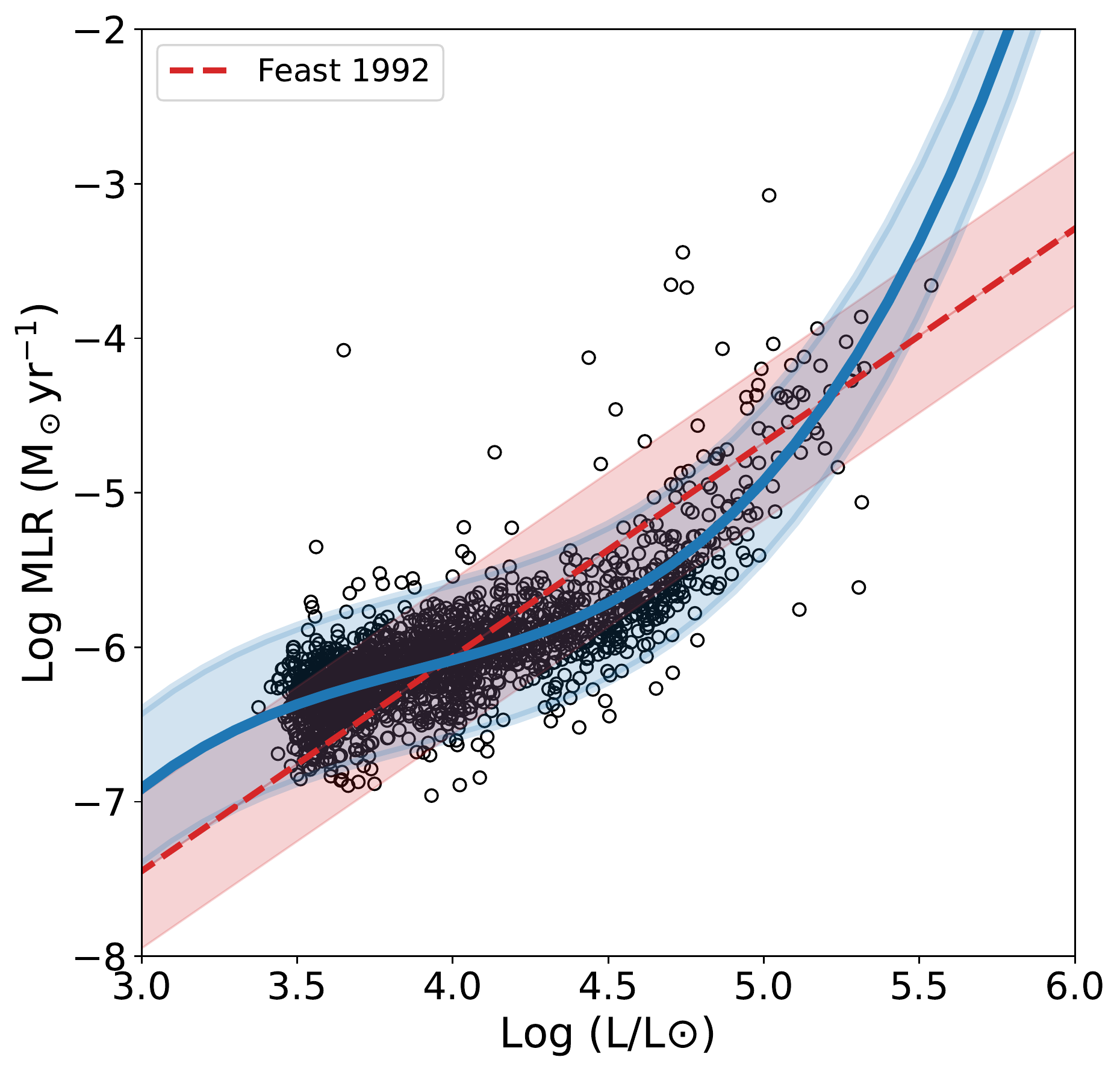}
\includegraphics[scale=0.24]{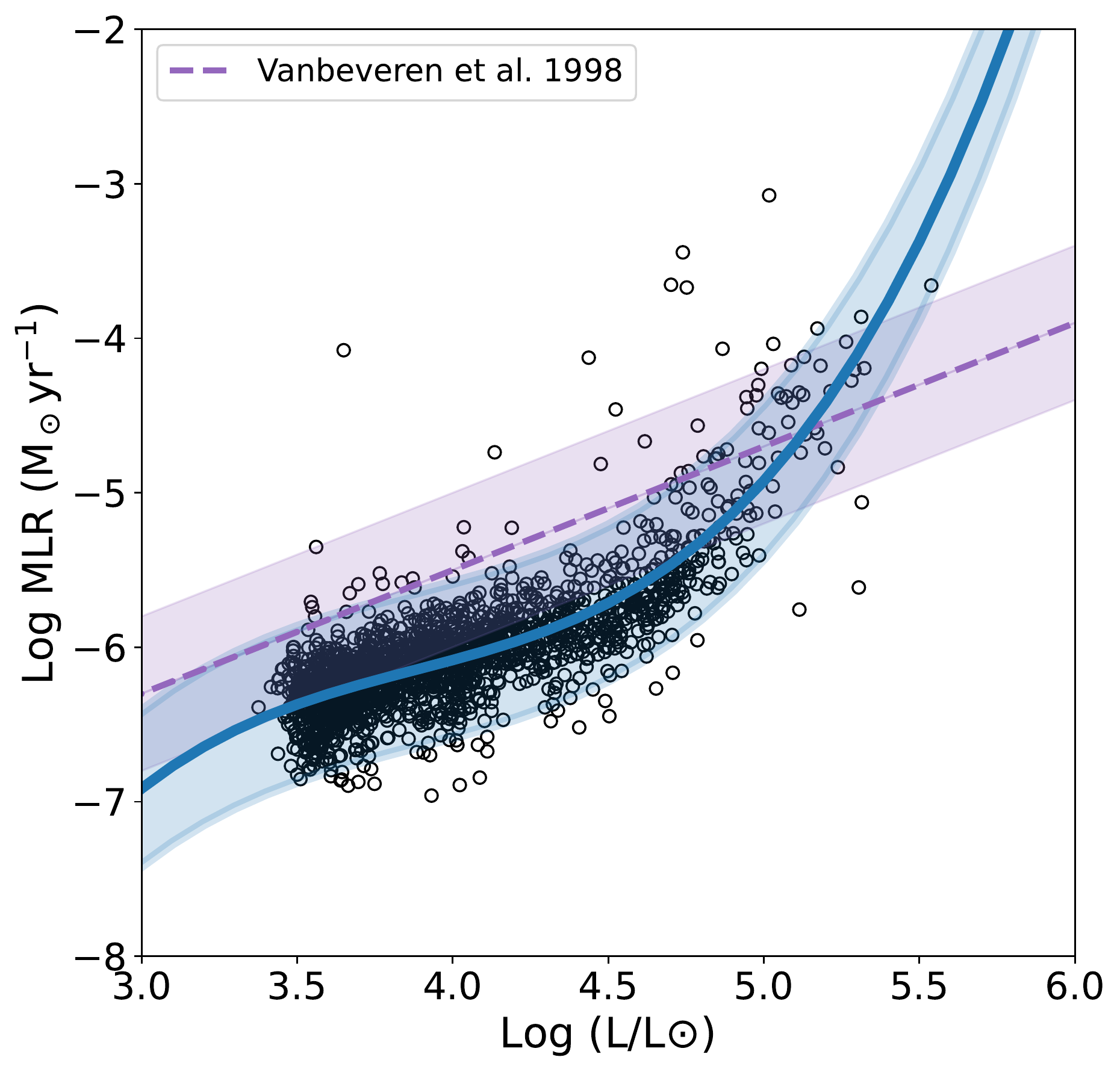}
\includegraphics[scale=0.24]{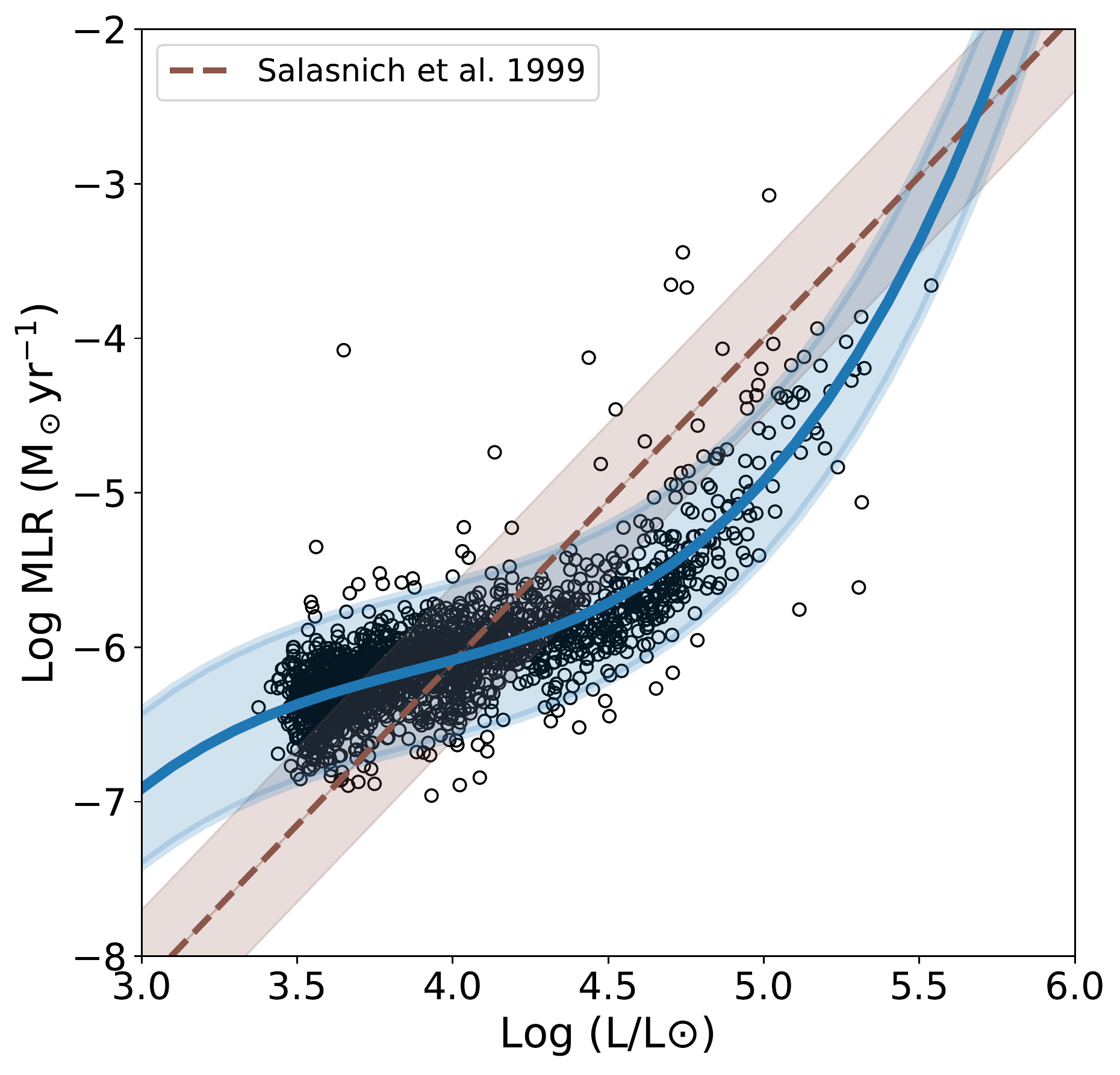}
\includegraphics[scale=0.24]{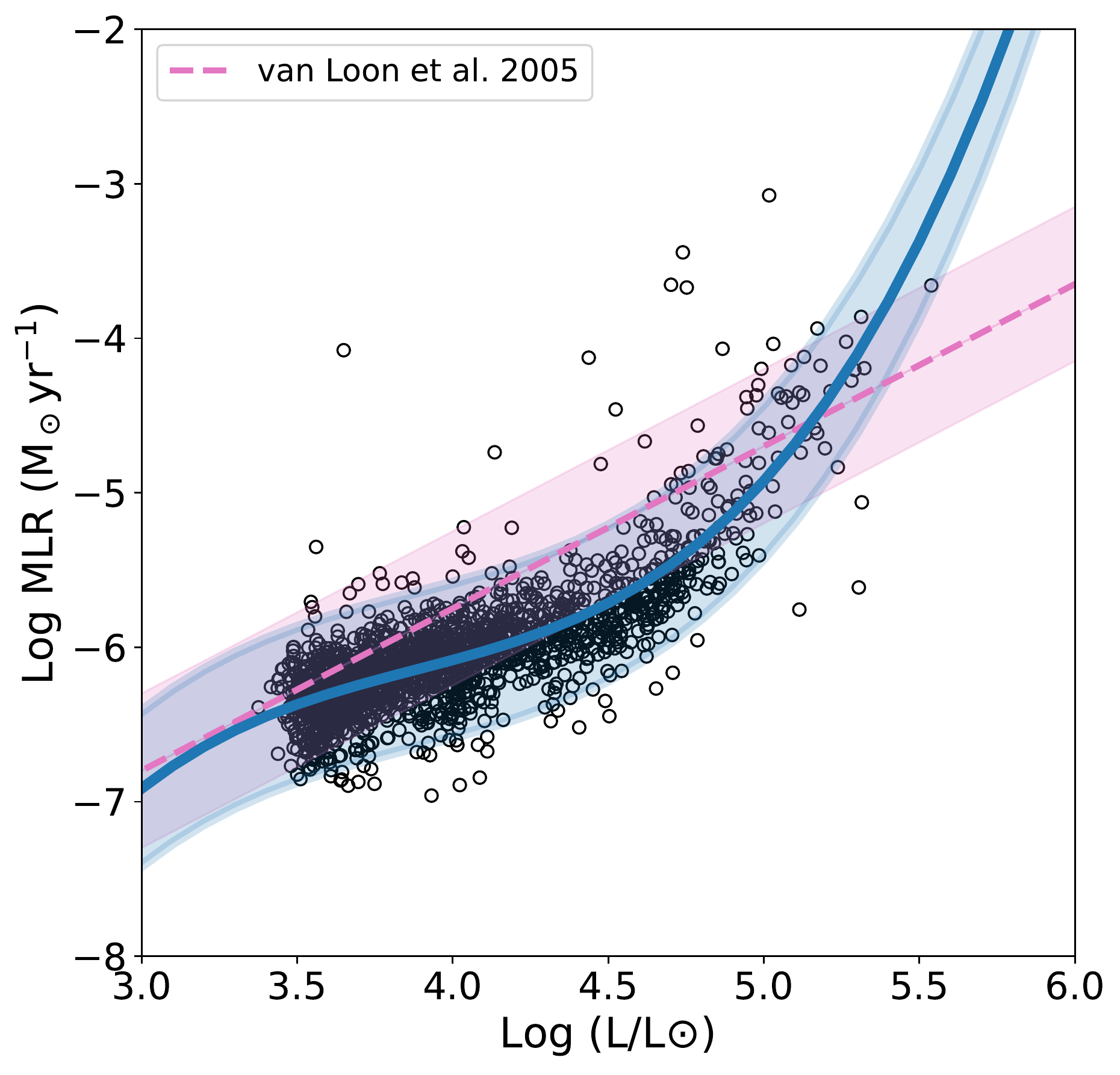}
\includegraphics[scale=0.24]{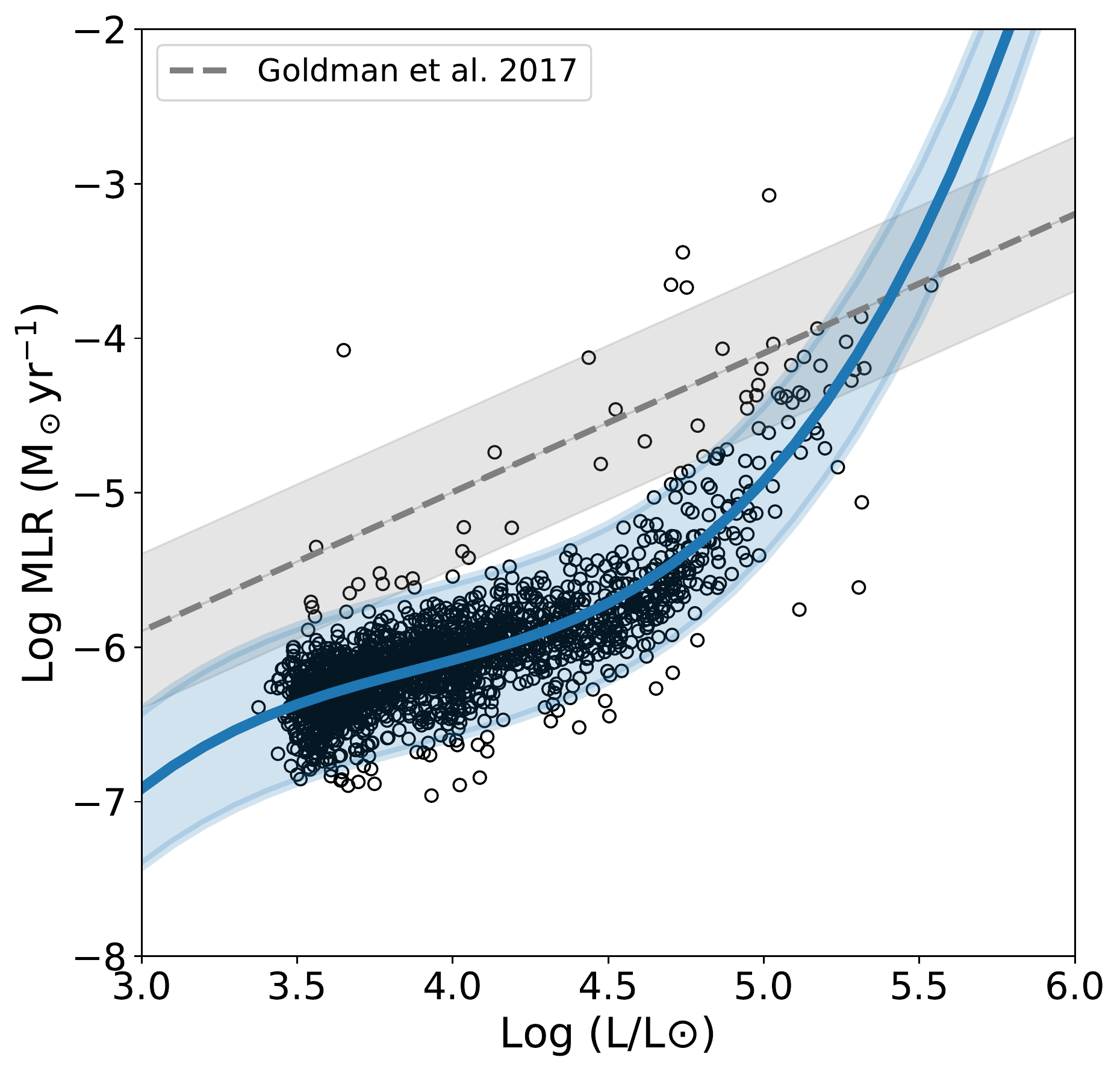}
\includegraphics[scale=0.24]{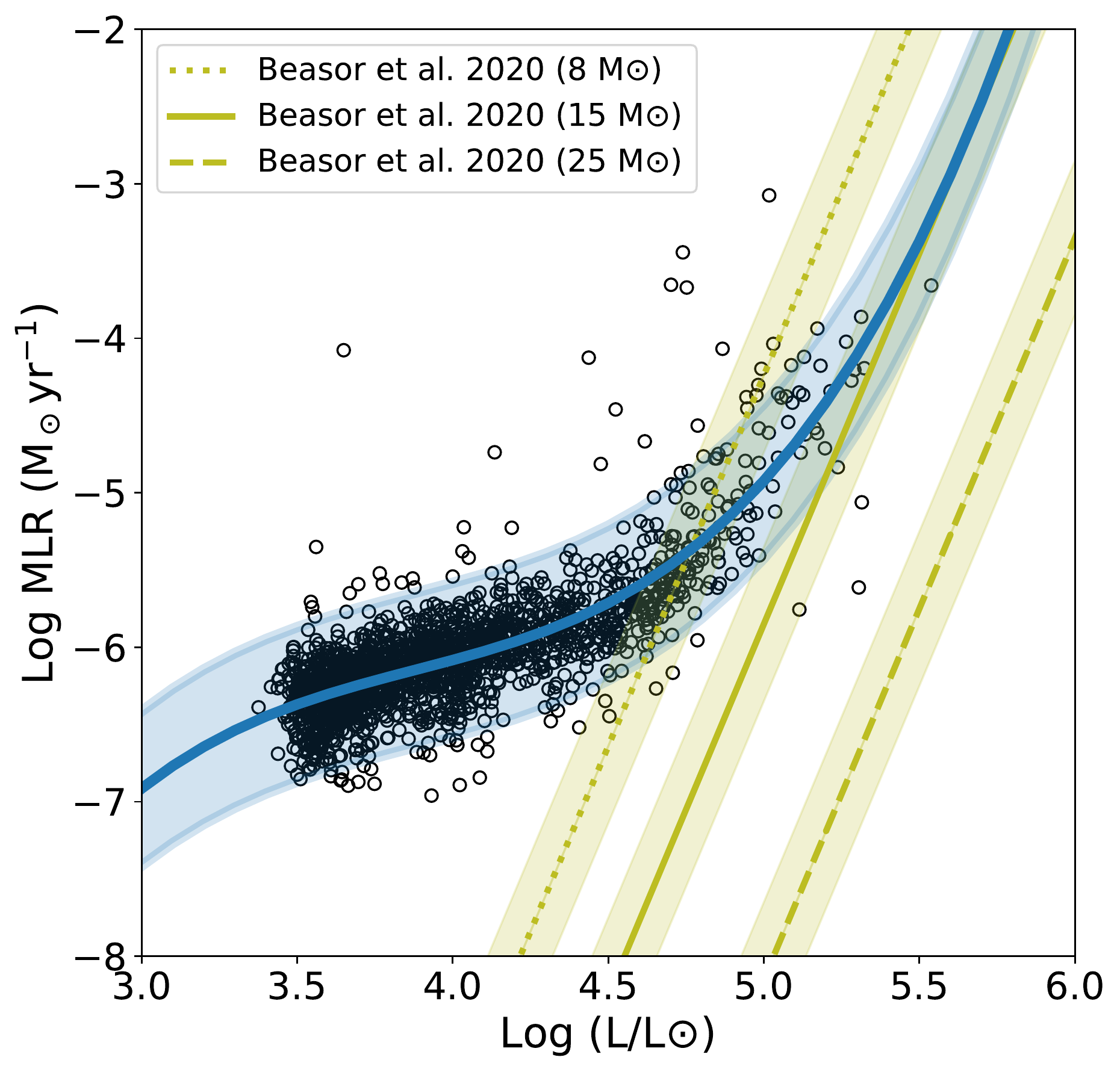}
\includegraphics[scale=0.24]{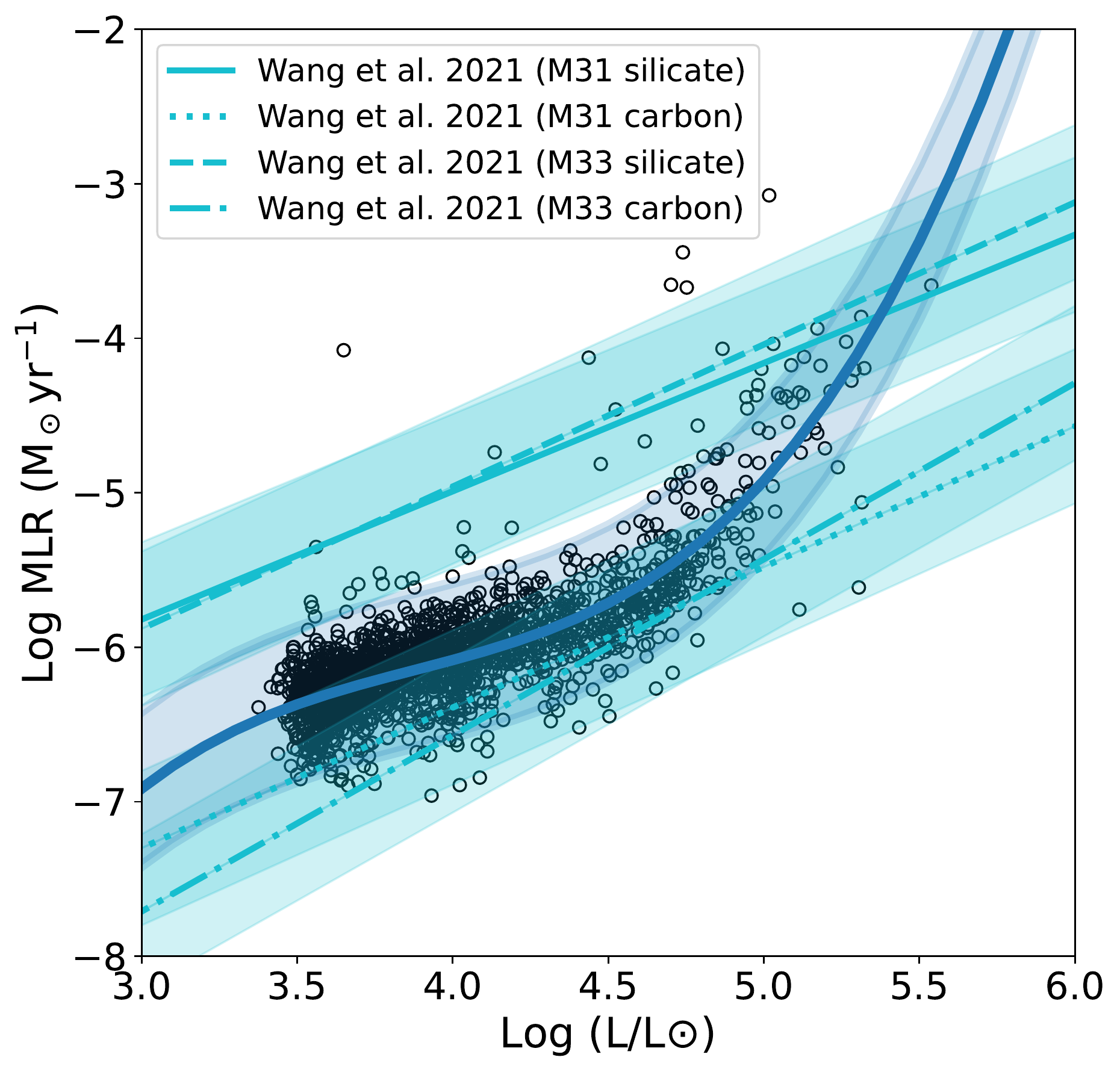}
\caption{Comparison of derived MLR-$L$ relation between this and each previous individual study. A 0.5 dex error (the shaded region) is shown as a reference for all prescriptions (including ours). 
\label{lum_mlr_ref_each}}
\end{figure*}

The right panel of Figure~\ref{lum_mlr_ref} shows the comparison between our work and previous studies. It can be seen that the majority of the MLR prescriptions have a similar trend but the scatter is large (within two orders of magnitude), except for a few ones. For clarity, we also showed the comparison between our work and each previous individual study in Figure~\ref{lum_mlr_ref_each}. In general, the errors of MLR prescriptions are estimated based on some parameters like $T_{\rm eff}$, $M$, $R$, or $P$, etc. However, in almost all cases, such kinds of errors are underestimated and inhomogeneous, which makes the comparison between different prescriptions hard. Thus, we adopted a 0.5 dex error for all prescriptions (including ours) as a reference shown in the diagrams. Moreover, as discussed in the previous section, a metallicity dependence and sample size may also contribute to the large scatter, as the comparison includes studies from both Milky Way and MC RSGs, which span a range of metallicities and sample sizes.

As can be seen from the diagrams, the ones most resembling our relation are vL05 and F92, which almost overlap with our sample and relation over the whole luminosity range ($3.5\lesssim log_{10}(L/L_\sun) \lesssim5.3$) with a slight overestimation around $\log_{10}(L/L_\sun)\approx4.6$. Meanwhile, at higher luminosity ($\log_{10}(L/L_\sun)>5.3$), our prescription seems more than a order of magnitude different than vL05 and F92. V98 and S99 are also very similar to our relation, for which V98 is more flattened with a slight overestimation at the medium and faint end of luminosity, while S99 is steeper with overestimation at the medium luminosity and underestimation at the faint end of luminosity. The classic R75, dJ88 and NJ90 are similar to each other within a few factors, but are all underestimating at both faint and bright ends, though more impact may be found in the bright end with large MLRs. G17 and W21 (silicate) are largely overestimating at the medium and faint end of luminosity. However, G17 used OH/IR stars from LMC, Galactic Centre and Galactic bulge to derived MLRs of O-rich AGBs and RSGs, for which the OH/IR stars were targets with large MLRs and thick cirumstellar envelopes. Meanwhile, W21 derived the MLRs of RSGs in the M31 and M33, where much higher metallicities might play an important role. Thus, the overestimation of G17 and W21 are relatively reasonable. Finally, B20 is in good agreement with our relation at the very bright end ($\log_{10}(L/L_\sun)\gtrsim5.2$ with initial mass of $M=15~M_\sun$; we have indicated three typical initial masses of $M=8,15,25~M_\sun$ in the diagram) but largely underestimate MLR at the medium and faint end of luminosity, which may be due to their sparse sampling of initial masses of RSGs.

Overall, our prescription captured the change of slope in MLR that the rest could not, because they all had a more limited sample. Therefore, our result may provide a better solution at the cool and luminous region (e.g., about $3500 \leq T_{\rm eff} \leq 5500~\rm K$ and $3.5\leq log_{10}(L/L_\sun) \leq5.5$) of the H-R diagram at low-metallicity (the SMC level) compared to previous works, given that is based on a much larger sample. In the meantime, we believe that our prescription could be extrapolated to a wider range (possibly covering a large part of the RHeBS), e.g., $\sim3000 \leq T_{\rm eff} \leq 6000~\rm K$ and $\sim3.0\leq log_{10}(L/L_\sun) \leq6.0$, with caution (above $log_{10}(L/L_\sun)=5.5$ the extrapolation is more uncertain). We also did some preliminary tests of our new prescriptions in MESA stellar evolution models \citep{Paxton2011, Paxton2013, Paxton2015, Paxton2018}. The high MLR of massive luminous RSGs leaves them with only a thin envelope at the end of their life. This could in principle be a (partial) solution to the ``RSG problem'' \citep{Smartt2009}, although further investigation is needed. Meanwhile, the widely used vL05 law and F92 law are two prescriptions that are most similar to ours. Still, we would like to emphasize again that there are many uncertainties in the estimation of MLR for RSGs.


\subsection{Other interesting factors}

\begin{figure}
\center
\includegraphics[scale=0.47]{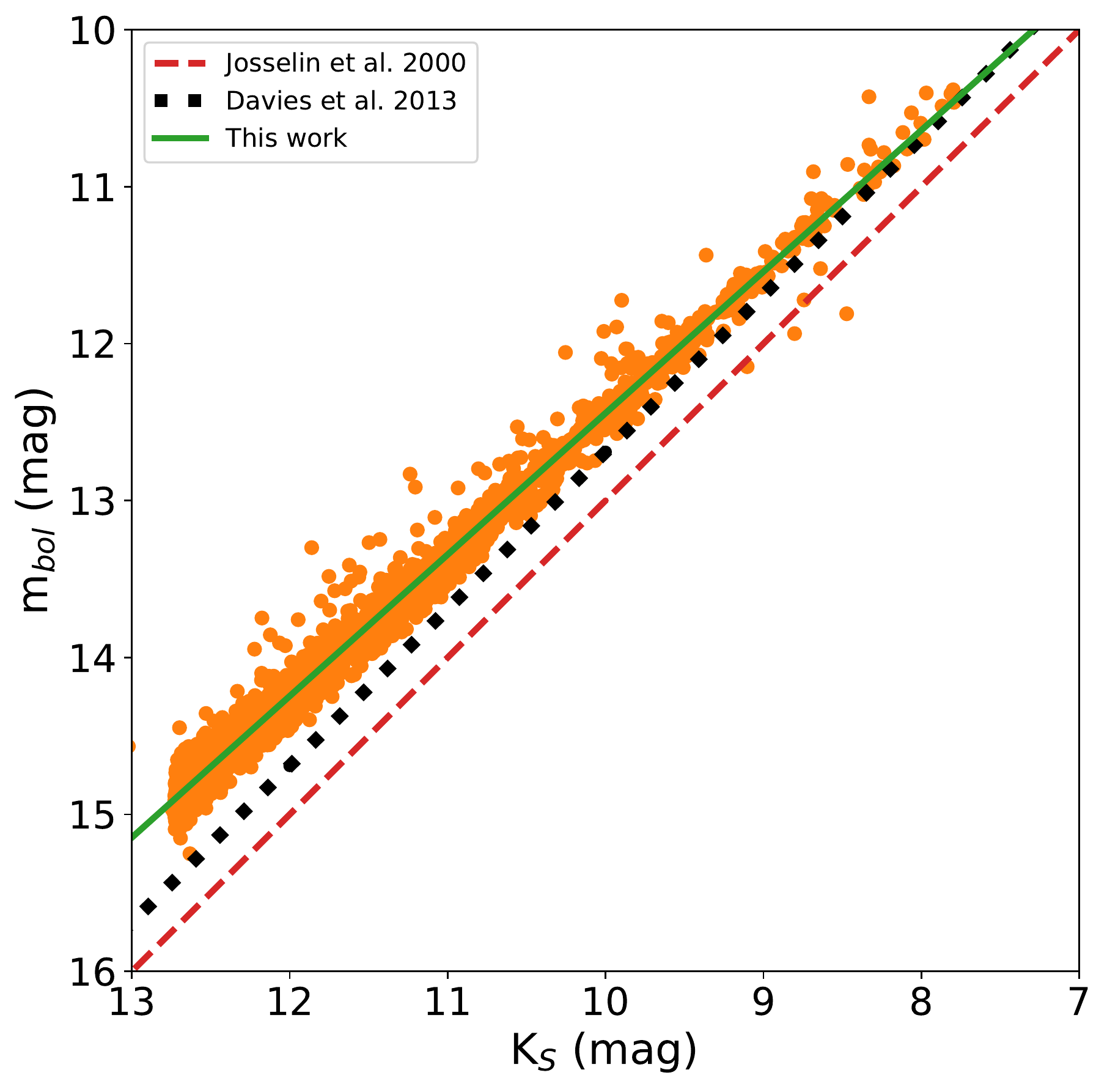}
\caption{K$_S$ versus bolometric magnitudes diagram. Our work indicates an improved relation based on a larger sample compared to previous studies.
\label{k_mbol}}
\end{figure}

Additionally, we derived the relation between the K$_S$-band and bolometric magnitudes as, 
\begin{equation}
m_{bol}=0.90\times K_S+3.43,
\end{equation}
shown in Figure~\ref{k_mbol}. Compared to previous studies, our relation is flatter at the faint end, but very similar to \citet{Davies2013} at the bright end. Meanwhile, both ours and that of \citet{Davies2013} are different from \citet{Josselin2000}. All the differences are most likely due to the sample sizes and/or metallicities, for which our work indicates an improved relation based on a much larger sample.

\begin{figure*}
\center
\includegraphics[scale=0.32]{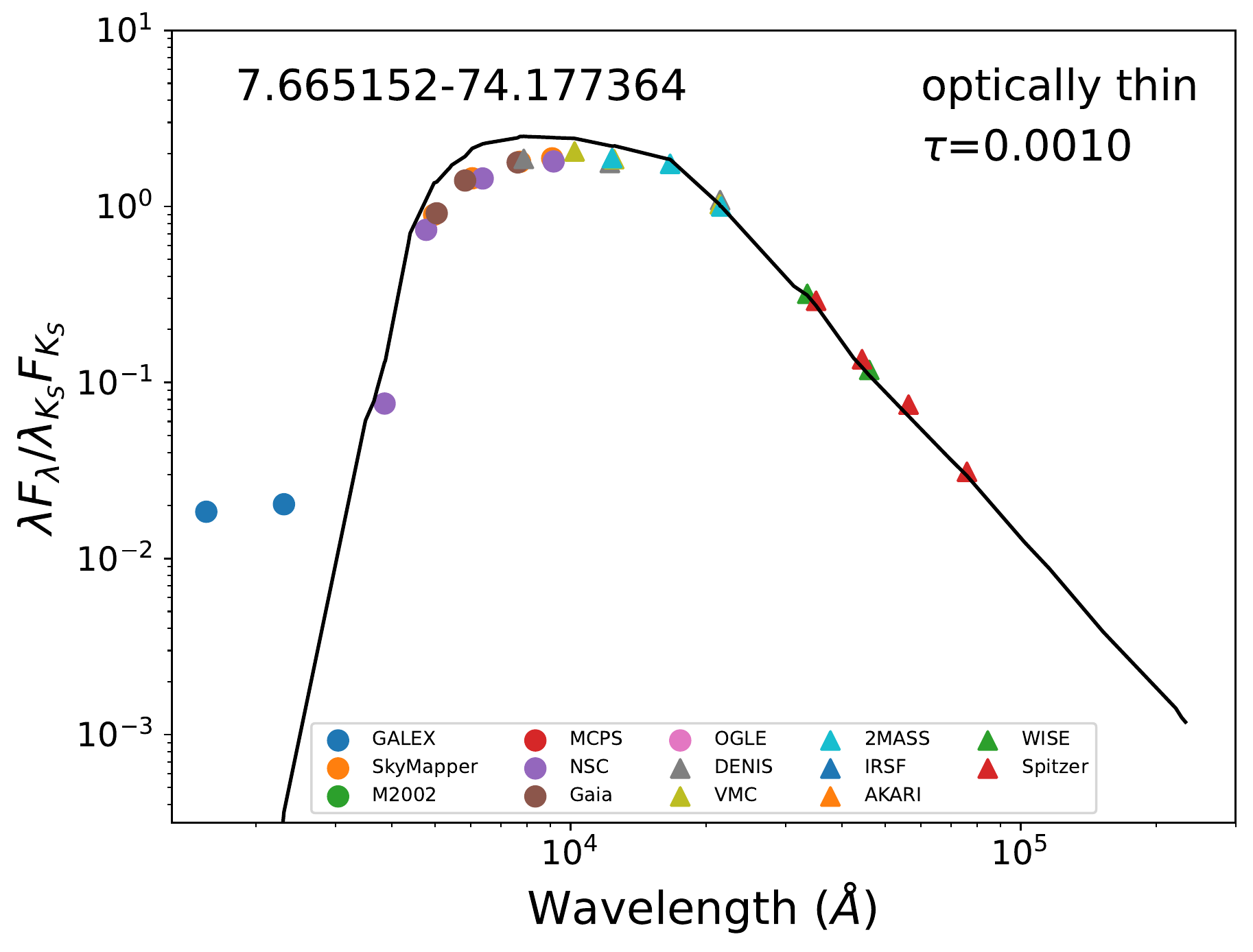}
\includegraphics[scale=0.32]{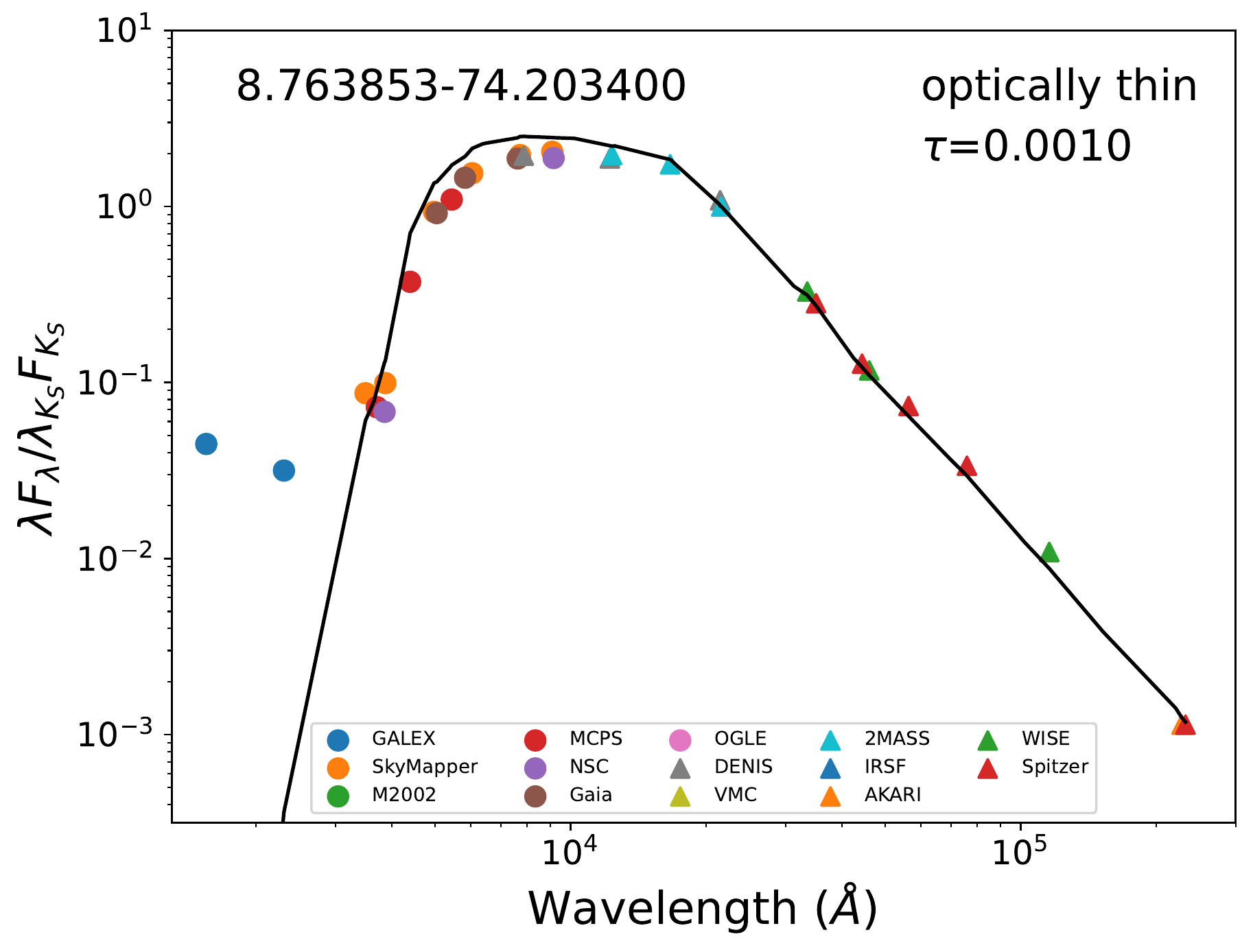}
\includegraphics[scale=0.32]{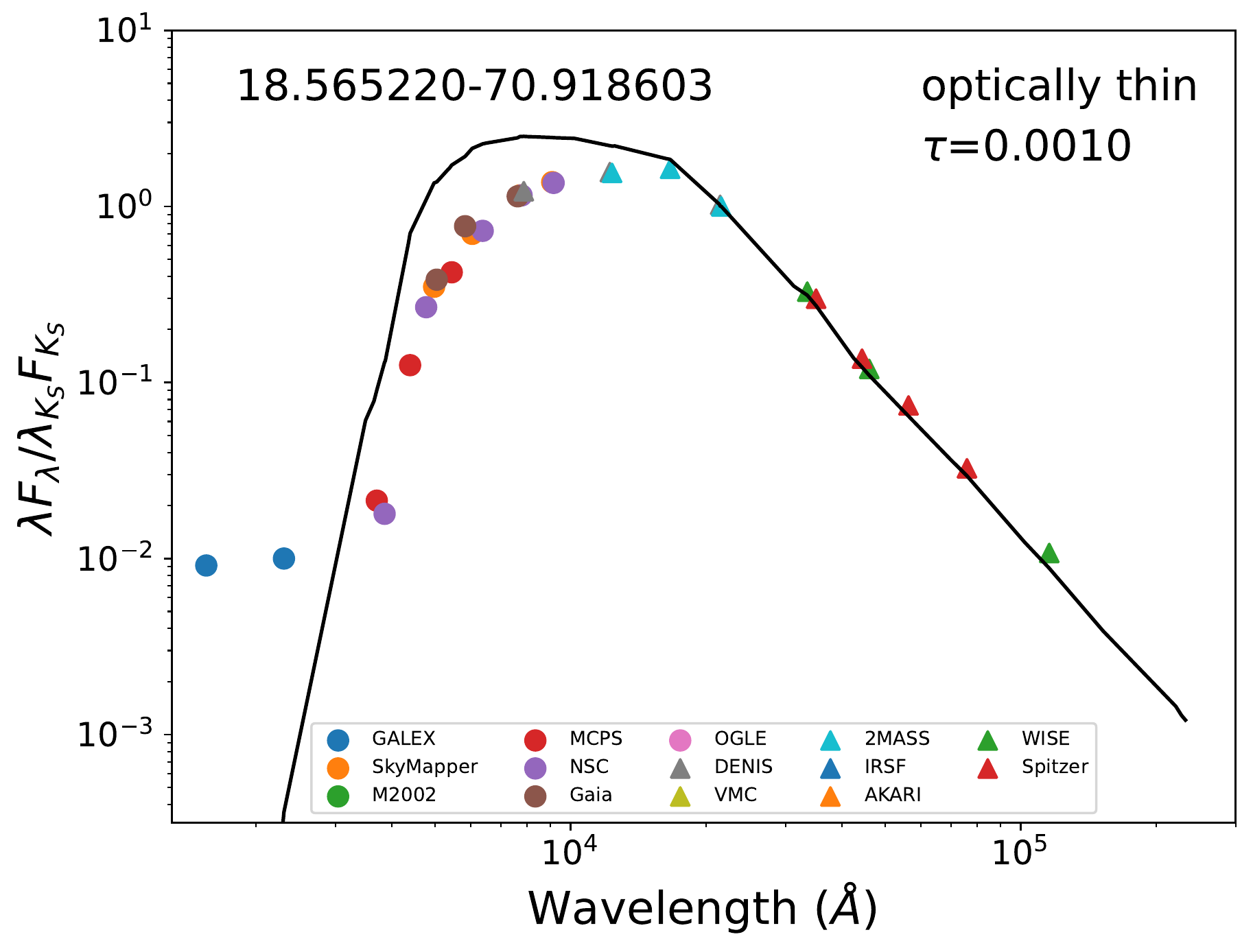}
\caption{Examples of the targets with UV detection, which likely indicate a hot binary companion. 
\label{b_example}}
\end{figure*}

\begin{figure}
\center
\includegraphics[scale=0.32]{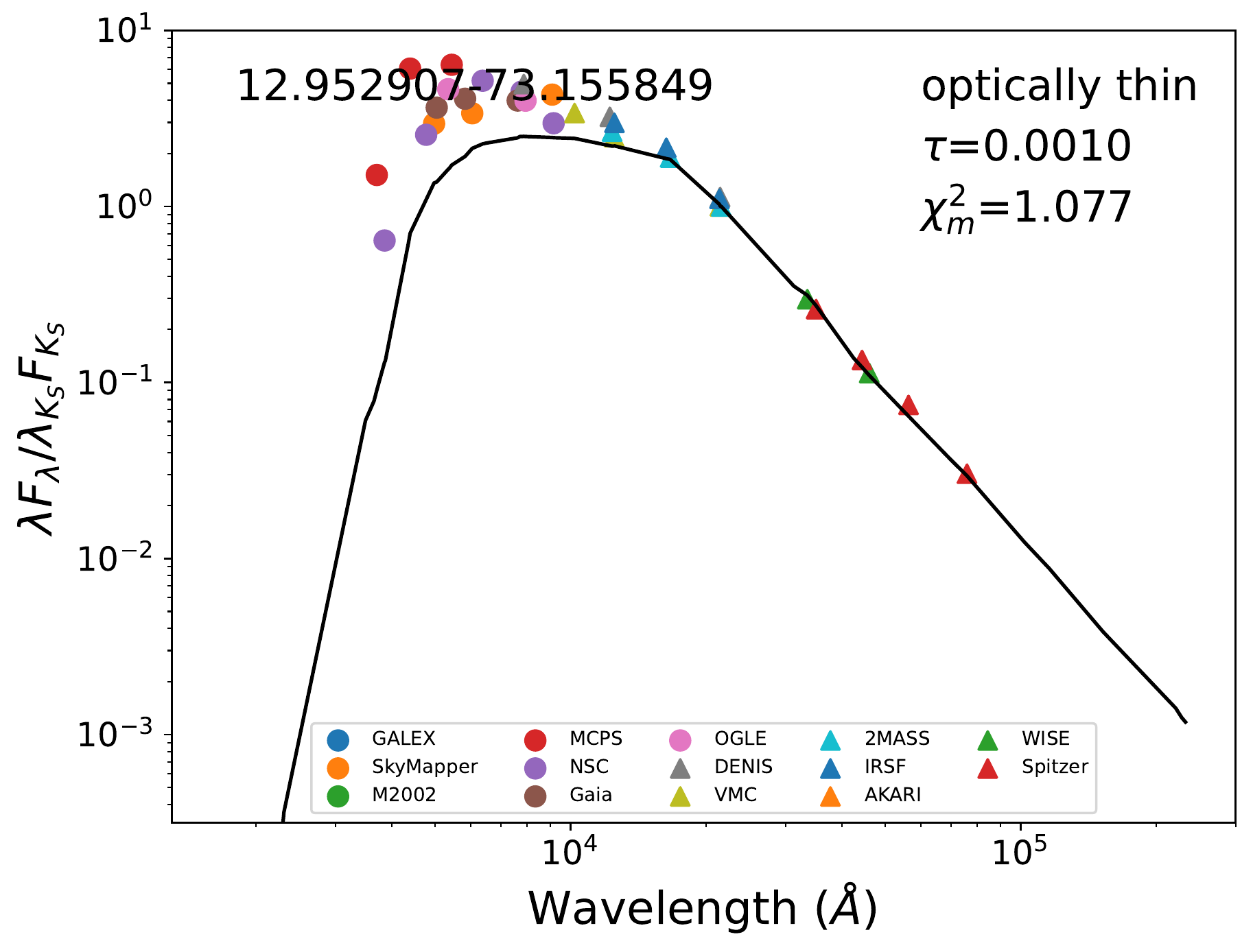}
\includegraphics[scale=0.32]{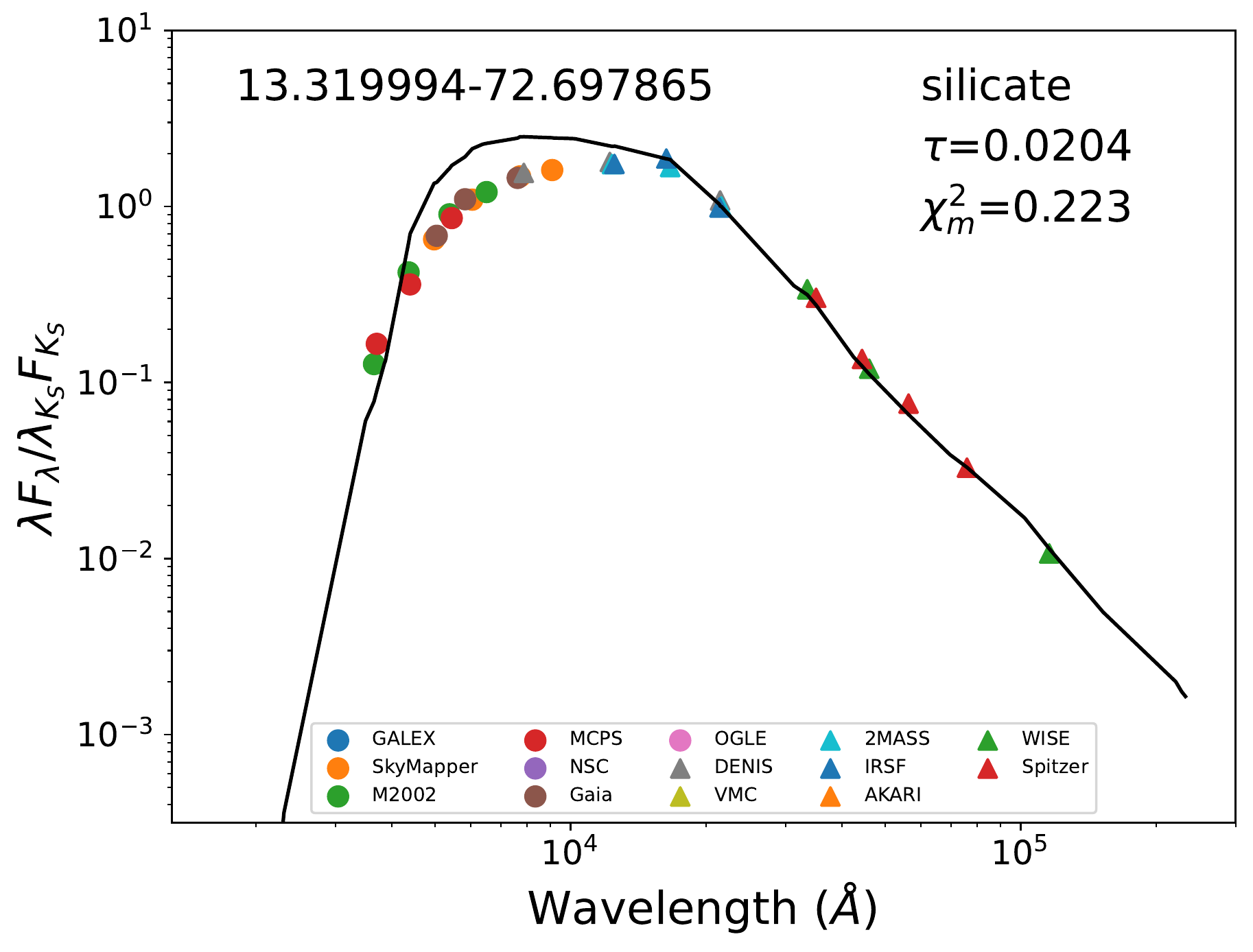}
\caption{Examples of the targets with unusual optical excess/depression, which may indicate a binary companion. 
\label{bc_example}}
\end{figure}

Furthermore, we detected UV flux in 9 targets from the final sample (see examples in Figure~\ref{b_example}), which could be an indicator of OB-type companions of binary RSGs as mentioned in \citet{Neugent2018, Neugent2019, Neugent2020}, \citet{Neugent2021}, and \citet{Patrick2022}. The large difference of (presumable) binary fraction (e.g., $<$1\% for our case and $>$15\% in other studies) is most likely due to the low resolution and sensitivity of GALEX. Moreover, even without the UV detection (again, because of the limitation of GALEX data), some targets ($\sim$7\% in the final sample) also show abnormal optical excess or depression in the blue end similar to the UV detected targets (see examples in Figure~\ref{bc_example}), which may be the indication of the existence of a binary companion.

\section{Summary}

In order to investigate the MLR of RSGs at relatively low-metallicity, we assemble an initial RSGs sample of 2,303 targets taken from \citet{Yang2020} and \citet{Ren2021} in the SMC with 53 different bands of photometric data. The sample is further constrained into a final sample, which contains 2,121 targets, by removing the targets in the overlapping region between AGBs and RSGs on the 2MASS CMD.

The radiatively-driven wind model of the 1D radiative transfer code DUSTY is used to create a grid, covering a wide range of dust parameters, in order to match the observed SEDs from our sample. In all, 17,820 O-rich theoretical models are generated by DUSTY, such that ``normal'' and ``dusty'' grids are half-and-half. We first select the best model for each individual target by calculating the minimum $\chi_{m}^2$ from the normal grids. Then the dusty targets (six targets) were again fitted with the dusty grids. The resulting MLRs from DUSTY then are converted to real MLRs based on the scaling relation.

For our final RSG sample (assuming our sample is quite close to complete), a total MLR of $6.16\times10^{-3}$ $M_\sun$ yr$^{-1}$ is measured, while a typical MLR of $\sim10^{-6}$ $M_\sun$ yr$^{-1}$ is witnessed for the majority of the RSGs. Our result of the total MLR can be converted into a DPR of $\sim6\times10^{-6}$ $M_\sun$ yr$^{-1}$, which is in general agreement with previous studies by a few factors, while discrepancies can be raised from many factors, such as sample sizes, adopted optical constants, model assumptions, etc. Meanwhile, it also indicates that a few dusty targets contributed a significant part of the MLR, e.g., about 3\% of the dusty targets has contributed about 65\% of the MLR. The H-R and luminosity versus MAD diagrams of the final sample indicates the positive relation between luminosity and MLR. Meanwhile, the luminosity versus MLR diagrams show a ``knee-like'' shape with enhanced mass-loss taking place around $\log_{10}(L/L_\sun)\approx4.6$, which may be due to the degeneracy of luminosity, pulsation, low surface gravity, convection, and other factors.

The complexity of mass-loss estimation based on the SED is discussed from almost all aspects covering both methodology and physics. It indicates that MLRs derived based on photometric data can be only considered as a mixture of “snap shot” with large uncertainties. 

We derive our MLR relation by using a third-order polynomial fitting the final sample. We also compare our result with many previous empirical MLR prescriptions, finding that most prescriptions have a similar trend within two orders of magnitude. Specifically, \citet{vanLoon2005} and \citet{Feast1992} are the two most resembling ones with respect to our relation. The classic \citet{Reimers1975} and \citet{deJager1988} (\citealt{Nieuwenhuijzen1990}) are underestimated at both faint and bright ends. \citet{Beasor2020} is in good agreement with our relation at the very bright end but largely underestimated at the medium and faint end of luminosity. In general, based on a much larger sample, our MLR prescription captures the change of slope in MLR that the rest studies can not, and provides a more accurate relation at the cool and luminous region of the H-R diagram at low-metallicity compared to previous works.

Finally, a small fraction of our sample is detected in the UV band, which could be an indication of OB-type companions of binary RSGs. Similarly, some targets without the UV detection also show abnormal optical excess/depression, which may also indicate the existence of a binary companion.

\begin{acknowledgements}

We would like to thank the anonymous referee for many constructive comments and suggestions. This study has received funding from the European Research Council (ERC) under the European Union's Horizon 2020 research and innovation programme (grant agreement number 772086). B.W.J, M.Y. and S.W. gratefully acknowledge support from the National Natural Science Foundation of China (Grant No.12133002 and 12003046). H.T. is Supported by Beijing Natural Science Foundation No. 1214028. We acknowledge the science research grants from the China Manned Space Project with No.CMS-CSST-2021-A08. This work is also supported by National Key R\&D Program of China No. 2019YFA0405501.\\

We thank Philip Massey and Alistair Walker for providing valuable comments on the total system throughput, Zeljko Ivezic for finding the DUSTY V4 bug, Nicholas Cross for identifying the VMC pipeline bug, Marta Sewilo for the discussion of \textit{Spitzer} data quality flag, Jing Tang and Tianmeng Zhang for helpful discussion on the evolution of massive stars and supernova. \\

This work is based in part on observations made with the Spitzer Space Telescope, which is operated by the Jet Propulsion Laboratory, California Institute of Technology under a contract with NASA.\\ 

This publication makes use of data products from the Wide-field Infrared Survey Explorer, which is a joint project of the University of California, Los Angeles, and the Jet Propulsion Laboratory/California Institute of Technology. It is funded by the National Aeronautics and Space Administration.\\

This research has made use of the NASA/IPAC Infrared Science Archive, which is operated by the Jet Propulsion Laboratory, California Institute of Technology, under contract with the National Aeronautics and Space Administration.\\

This work has made use of data from the European Space Agency (ESA) mission {\it Gaia} (\url{https://www.cosmos.esa.int/gaia}), processed by the {\it Gaia} Data Processing and Analysis Consortium (DPAC, \url{https://www.cosmos.esa.int/web/gaia/dpac/consortium}). Funding for the DPAC has been provided by national institutions, in particular the institutions participating in the {\it Gaia} Multilateral Agreement.\\

This research has made use of the Spanish Virtual Observatory (https://svo.cab.inta-csic.es) project funded by MCIN/AEI/10.13039/501100011033/ through grant PID2020-112949GB-I00 \citep{Rodrigo2012, Rodrigo2020}.\\

This research has made use of the VizieR catalogue access tool, CDS, Strasbourg, France (DOI : 10.26093/cds/vizier). The original description of the VizieR service was published in \citet{Ochsenbein2000}.\\

This research has made use of the Tool for OPerations on Catalogues And Tables (TOPCAT; \citealt{Taylor2005}).

\end{acknowledgements}

\clearpage

\end{CJK*}


\begin{thebibliography}{}
\bibitem[Adams et al.(2017)]{Adams2017} Adams, S.~M., Kochanek, C.~S., Gerke, J.~R., et al.\ 2017, \mnras, 468, 4968
\bibitem[Allamandola et al.(1989)]{Allamandola1989} Allamandola, L.~J., Tielens, A.~G.~G.~M., \& Barker, J.~R.\ 1989, \apjs, 71, 733. 
\bibitem[Andrae et al.(2010)]{Andrae2010} Andrae, R., Schulze-Hartung, T., \& Melchior, P.\ 2010, arXiv:1012.3754
\bibitem[Beasor \& Davies(2016)]{Beasor2016} Beasor, E.~R., \& Davies, B.\ 2016, \mnras, 463, 1269
\bibitem[Beasor \& Davies(2018)]{Beasor2018} Beasor, E.~R. \& Davies, B.\ 2018, \mnras, 475, 55
\bibitem[Beasor et al.(2020)]{Beasor2020} Beasor, E.~R., Davies, B., Smith, N., et al.\ 2020, \mnras, 492, 5994
\bibitem[Beck et al.(1992)]{Beck1992} Beck, H.~K.~B., Gail, H.-P., Henkel, R., et al.\ 1992, \aap, 265, 626
\bibitem[Bessell et al.(2011)]{Bessell2011} Bessell, M., Bloxham, G., Schmidt, B., et al.\ 2011, \pasp, 123, 789 
\bibitem[Bianchi et al.(2017)]{Bianchi2017} Bianchi, L., Shiao, B., \& Thilker, D.\ 2017, \apjs, 230, 24
\bibitem[Bouwman et al.(2000)]{Bouwman2000} Bouwman, J., de Koter, A., van den Ancker, M.~E., et al.\ 2000, \aap, 360, 213
\bibitem[Bouwman et al.(2001)]{Bouwman2001} Bouwman, J., Meeus, G., de Koter, A., et al.\ 2001, \aap, 375, 950
\bibitem[Brunner et al.(2018)]{Brunner2018} Brunner, M., Maercker, M., Mecina, M., et al.\ 2018, \aap, 614, A17
\bibitem[Bouchet et al.(1985)]{Bouchet1985} Bouchet, P., Lequeux, J., Maurice, E., et al.\ 1985, \aap, 149, 330
\bibitem[Boyer et al.(2009)]{Boyer2009} Boyer, M.~L., McDonald, I., van Loon, J.~T., et al.\ 2009, \apj, 705, 746
\bibitem[Boyer et al.(2011)]{Boyer2011} Boyer, M.~L., Srinivasan, S., van Loon, J.~T., et al.\ 2011, \aj, 142, 103
\bibitem[Boyer et al.(2012)]{Boyer2012} Boyer, M.~L., Srinivasan, S., Riebel, D., et al.\ 2012, \apj, 748, 40 
\bibitem[Cannon et al.(2021)]{Cannon2021} Cannon, E., Montarg{\`e}s, M., de Koter, A., et al.\ 2021, \mnras, 502, 369
\bibitem[Cardelli et al.(1989)]{Cardelli1989} Cardelli, J.~A., Clayton, G.~C., \& Mathis, J.~S.\ 1989, \apj, 345, 245
\bibitem[Cioni et al.(2011)]{Cioni2011} Cioni, M.-R.~L., Clementini, G., Girardi, L., et al.\ 2011, \aap, 527, A116
\bibitem[Chatys et al.(2019)]{Chatys2019} Chatys, F.~W., Bedding, T.~R., Murphy, S.~J., et al.\ 2019, \mnras, 487, 4832
\bibitem[Dalcanton et al.(2012)]{Dalcanton2012} Dalcanton, J.~J., Williams, B.~F., Melbourne, J.~L., et al.\ 2012, \apjs, 198, 6. 
\bibitem[Davies et al.(2013)]{Davies2013} Davies, B., Kudritzki, R.-P., Plez, B., et al.\ 2013, \apj, 767, 3 
\bibitem[Davies et al.(2017)]{Davies2017} Davies, B., Kudritzki, R.-P., Lardo, C., et al.\ 2017, \apj, 847, 112 
\bibitem[Davies \& Beasor(2018)]{Davies2018} Davies, B., \& Beasor, E.~R.\ 2018, \mnras, 474, 2116 
\bibitem[de Jager et al.(1988)]{deJager1988} de Jager, C., Nieuwenhuijzen, H., \& van der Hucht, K.~A.\ 1988, \aaps, 72, 259
\bibitem[Draine \& Lee(1984)]{Draine1984} Draine, B.~T. \& Lee, H.~M.\ 1984, \apj, 285, 89
\bibitem[Dye et al.(2018)]{Dye2018} Dye, S., Lawrence, A., Read, M.~A., et al.\ 2018, \mnras, 473, 5113
\bibitem[Elitzur \& Ivezi{\'c}(2001)]{Elitzur2001} Elitzur, M. \& Ivezi{\'c}, {\v{Z}}.\ 2001, \mnras, 327, 403
\bibitem[Ekstr{\"o}m et al.(2012)]{Ekstrom2012} Ekstr{\"o}m, S., Georgy, C., Eggenberger, P., et al.\ 2012, \aap, 537, A146 
\bibitem[Ekstr{\"o}m et al.(2013)]{Ekstrom2013} Ekstr{\"o}m, S., Georgy, C., Meynet, G., Groh, J., \& Granada, A.\ 2013, EAS Publications Series, 60, 31 
\bibitem[Epchtein et al.(1997)]{Epchtein1997} Epchtein, N., de Batz, B., Capoani, L., et al.\ 1997, The Messenger, 87, 27
\bibitem[Feast(1992)]{Feast1992} Feast, M.~W.\ 1992, Instabilities in Evolved Super- and Hypergiants, 18
\bibitem[Freedman et al.(2020)]{Freedman2020} Freedman, W.~L., Madore, B.~F., Hoyt, T., et al.\ 2020, \apj, 891, 57
\bibitem[Gaia Collaboration et al.(2016)]{Gaia2016} Gaia Collaboration, Prusti, T., de Bruijne, J.~H.~J., et al.\ 2016, \aap, 595, A1 
\bibitem[Gaia Collaboration et al.(2018)]{Gaia2018} Gaia Collaboration, Brown, A.~G.~A., Vallenari, A., et al.\ 2018, \aap, 616, A1 
\bibitem[Gaia Collaboration et al.(2021)]{Gaia2021} Gaia Collaboration, Brown, A.~G.~A., Vallenari, A., et al.\ 2021, \aap, 649, A1
\bibitem[Gail et al.(1984)]{Gail1984} Gail, H.-P., Keller, R., \& Sedlmayr, E.\ 1984, \aap, 133, 320
\bibitem[Gail \& Sedlmayr(1999)]{Gail1999} Gail, H.-P. \& Sedlmayr, E.\ 1999, \aap, 347, 594
\bibitem[Gail et al.(2020)]{Gail2020} Gail, H.-P., Tamanai, A., Pucci, A., et al.\ 2020, \aap, 644, A139
\bibitem[Gilkis et al.(2021)]{Gilkis2021} Gilkis, A., Shenar, T., Ramachandran, V., et al.\ 2021, \mnras, 503, 1884.
\bibitem[Goldman et al.(2017)]{Goldman2017} Goldman, S.~R., van Loon, J.~T., Zijlstra, A.~A., et al.\ 2017, \mnras, 465, 403
\bibitem[Gordon et al.(2009)]{Gordon2009} Gordon, K.~D., Bot, C., Muller, E., et al.\ 2009, \apjl, 690, L76.
\bibitem[Gordon et al.(2011)]{Gordon2011} Gordon, K.~D., Meixner, M., Meade, M.~R., et al.\ 2011, \aj, 142, 102
\bibitem[Groenewegen et al.(1994)]{Groenewegen1994} Groenewegen, M.~A.~T., de Jong, T., \& Gaballe, T.~R.\ 1994, \aap, 287, 163
\bibitem[Groenewegen et al.(2009)]{Groenewegen2009} Groenewegen, M.~A.~T., Sloan, G.~C., Soszy{\'n}ski, I., \& Petersen, E.~A.\ 2009, \aap, 506, 1277
\bibitem[Groenewegen(2012)]{Groenewegen2012} Groenewegen, M.~A.~T.\ 2012, \aap, 543, A36
\bibitem[Groenewegen \& Sloan(2018)]{Groenewegen2018} Groenewegen, M.~A.~T., \& Sloan, G.~C.\ 2018, \aap, 609, A114 
\bibitem[Gustafsson et al.(2008)]{Gustafsson2008} Gustafsson, B., Edvardsson, B., Eriksson, K., et al.\ 2008, \aap, 486, 951
\bibitem[Hanner(1988)]{Hanner1988} Hanner, M.\ 1988, In NASA, Washington, Infrared Observations of Comets Halley and Wilson and Properties of the Grains p 22-49 (SEE N89-13330 04-89)
\bibitem[Harper et al.(2001)]{Harper2001} Harper, G.~M., Brown, A., \& Lim, J.\ 2001, \apj, 551, 1073
\bibitem[H{\"o}fner \& Andersen(2007)]{Hofner2007} H{\"o}fner, S. \& Andersen, A.~C.\ 2007, \aap, 465, L39.
\bibitem[H{\"o}fner(2008)]{Hofner2008} H{\"o}fner, S.\ 2008, \aap, 491, L1. 
\bibitem[H{\"o}fner, \& Olofsson(2018)]{Hofner2018} H{\"o}fner, S., \& Olofsson, H.\ 2018, \aapr, 26, 1
\bibitem[Humphreys \& Davidson(1979)]{Humphreys1979} Humphreys, R.~M., \& Davidson, K.\ 1979, \apj, 232, 409 
\bibitem[Humphreys(2010)]{Humphreys2010} Humphreys, R.~M.\ 2010, Hot and Cool: Bridging Gaps in Massive Star Evolution, 425, 247 
\bibitem[Humphreys et al.(2020)]{Humphreys2020} Humphreys, R.~M., Helmel, G., Jones, T.~J., et al.\ 2020, \aj, 160, 145
\bibitem[Ita et al.(2010)]{Ita2010} Ita, Y., Onaka, T., Tanab{\'e}, T., et al.\ 2010, \pasj, 62, 273
\bibitem[Ivezic \& Elitzur(1995)]{Ivezic1995} Ivezic, Z. \& Elitzur, M.\ 1995, \apj, 445, 415
\bibitem[Ivezic \& Elitzur(1997)]{Ivezic1997} Ivezic, Z. \& Elitzur, M.\ 1997, \mnras, 287, 799
\bibitem[{Ivezic {et~al.}(1999)Ivezic, Nenkova, \& Elitzur}]{Ivezic1999} Ivezic, Z., Nenkova, M., \& Elitzur, M. 1999, User Manual for {DUSTY}
\bibitem[Javadi et al.(2013)]{Javadi2013} Javadi, A., van Loon, J.~T., Khosroshahi, H., et al.\ 2013, \mnras, 432, 2824
\bibitem[Jones et al.(2017)]{Jones2017} Jones, O.~C., Woods, P.~M., Kemper, F., et al.\ 2017, \mnras, 470, 3250
\bibitem[Josselin et al.(2000)]{Josselin2000} Josselin, E., Blommaert, J.~A.~D.~L., Groenewegen, M.~A.~T., Omont, A., \& Li, F.~L.\ 2000, \aap, 357, 225
\bibitem[Kato et al.(2007)]{Kato2007} Kato, D., Nagashima, C., Nagayama, T., et al.\ 2007, \pasj, 59, 615 
\bibitem[Kee et al.(2021)]{Kee2021} Kee, N.~D., Sundqvist, J.~O., Decin, L., et al.\ 2021, \aap, 646, A180
\bibitem[Keller et al.(2007)]{Keller2007} Keller, S.~C., Schmidt, B.~P., Bessell, M.~S., et al.\ 2007, \pasa, 24, 1  
\bibitem[Kiss et al.(2006)]{Kiss2006} Kiss, L.~L., Szab{\'o}, G.~M., \& Bedding, T.~R.\ 2006, \mnras, 372, 1721
\bibitem[Kim et al.(1994)]{Kim1994} Kim, S.-H., Martin, P.~G., \& Hendry, P.~D.\ 1994, \apj, 422, 164
\bibitem[Kochanek et al.(2008)]{Kochanek2008} Kochanek, C.~S., Beacom, J.~F., Kistler, M.~D., et al.\ 2008, \apj, 684, 1336
\bibitem[Kudritzki \& Reimers(1978)]{Kudritzki1978} Kudritzki, R.~P. \& Reimers, D.\ 1978, \aap, 70, 227
\bibitem[Levesque et al.(2005)]{Levesque2005} Levesque, E.~M., Massey, P., Olsen, K.~A.~G., et al.\ 2005, \apj, 628, 973 
\bibitem[Levesque(2010)]{Levesque2010} Levesque, E.~M.\ 2010, \nar, 54, 1
\bibitem[Levesque(2017)]{Levesque2017} Levesque, E.~M.\ 2017, Astrophysics of Red Supergiants, by Levesque, Emily M.. ISBN: 978-0-7503-1329-2. IOP ebooks. Bristol, UK: IOP Publishing, 2017
\bibitem[Liu et al.(2017)]{Liu2017} Liu, J., Jiang, B.~W., Li, A., et al.\ 2017, \mnras, 466, 1963
\bibitem[MacGregor \& Stencel(1992)]{Macgregor1992} MacGregor, K.~B., \& Stencel, R.~E.\ 1992, \apj, 397, 644
\bibitem[Massey(1998)]{Massey1998} Massey, P.\ 1998, \apj, 501, 153
\bibitem[Massey(2002)]{Massey2002} Massey, P.\ 2002, \apjs, 141, 81
\bibitem[Massey \& Olsen(2003)]{Massey2003} Massey, P., \& Olsen, K.~A.~G.\ 2003, \aj, 126, 2867
\bibitem[Massey et al.(2005)]{Massey2005} Massey, P., Plez, B., Levesque, E.~M., et al.\ 2005, \apj, 634, 1286 
\bibitem[Massey(2013)]{Massey2013} Massey, P.\ 2013, \nar, 57, 14 
\bibitem[Mathis et al.(1977)]{Mathis1977} Mathis, J.~S., Rumpl, W., \& Nordsieck, K.~H.\ 1977, \apj, 217, 425
\bibitem[Matsuura et al.(2013)]{Matsuura2013} Matsuura, M., Woods, P.~M., \& Owen, P.~J.\ 2013, \mnras, 429, 2527
\bibitem[Mauron \& Josselin(2011)]{Mauron2011} Mauron, N., \& Josselin, E.\ 2011, \aap, 526, A156 
\bibitem[McDonald et al.(2011)]{McDonald2011} McDonald, I., Boyer, M.~L., van Loon, J.~T., et al.\ 2011, \apj, 730, 71
\bibitem[Montarg{\`e}s et al.(2021)]{Montarges2021} Montarg{\`e}s, M., Cannon, E., Lagadec, E., et al.\ 2021, \nat, 594, 365
\bibitem[Morrissey et al.(2007)]{Morrissey2007} Morrissey, P., Conrow, T., Barlow, T.~A., et al.\ 2007, \apjs, 173, 682 
\bibitem[Melbourne et al.(2012)]{Melbourne2012} Melbourne, J., Williams, B.~F., Dalcanton, J.~J., et al.\ 2012, \apj, 748, 47.
\bibitem[Meynet et al.(2015)]{Meynet2015} Meynet, G., Chomienne, V., Ekstr{\"o}m, S., et al.\ 2015, \aap, 575, A60.
\bibitem[Min et al.(2009)]{Min2009} Min, M., Dullemond, C.~P., Dominik, C., et al.\ 2009, \aap, 497, 155
\bibitem[Murakami et al.(2007)]{Murakami2007} Murakami, H., Baba, H., Barthel, P., et al.\ 2007, Publications of the Astronomical Society of Japan, 59, S369.
\bibitem[Nanni et al.(2013)]{Nanni2013} Nanni, A., Bressan, A., Marigo, P., et al.\ 2013, \mnras, 434, 2390
\bibitem[Neugent et al.(2018)]{Neugent2018} Neugent, K.~F., Levesque, E.~M., \& Massey, P.\ 2018, \aj, 156, 225
\bibitem[Neugent et al.(2019)]{Neugent2019} Neugent, K.~F., Levesque, E.~M., Massey, P., et al.\ 2019, \apj, 875, 124
\bibitem[Neugent et al.(2020)]{Neugent2020} Neugent, K.~F., Levesque, E.~M., Massey, P., et al.\ 2020, \apj, 900, 118
\bibitem[Neugent(2021)]{Neugent2021} Neugent, K.~F.\ 2021, \apj, 908, 87
\bibitem[Nidever et al.(2018)]{Nidever2018} Nidever, D.~L., Dey, A., Olsen, K., et al.\ 2018, \aj, 156, 131 
\bibitem[Nidever et al.(2021)]{Nidever2021} Nidever, D.~L., Dey, A., Fasbender, K., et al.\ 2021, \aj, 161, 192
\bibitem[Nieuwenhuijzen \& de Jager(1990)]{Nieuwenhuijzen1990} Nieuwenhuijzen, H. \& de Jager, C.\ 1990, \aap, 231, 134
\bibitem[Ochsenbein et al.(2000)]{Ochsenbein2000} Ochsenbein, F., Bauer, P., \& Marcout, J.\ 2000, \aaps, 143, 23. doi:10.1051/aas:2000169
\bibitem[Ohnaka et al.(2008)]{Ohnaka2008} Ohnaka, K., Driebe, T., Hofmann, K.-H., et al.\ 2008, \aap, 484, 371
\bibitem[Onaka et al.(2007)]{Onaka2007} Onaka, T., Matsuhara, H., Wada, T., et al.\ 2007, Publications of the Astronomical Society of Japan, 59, S401
\bibitem[Onken et al.(2019)]{Onken2019} Onken, C.~A., Wolf, C., Bessell, M.~S., et al.\ 2019, \pasa, 36, e033
\bibitem[Ossenkopf et al.(1992)]{Ossenkopf1992} Ossenkopf, V., Henning, T., \& Mathis, J.~S.\ 1992, \aap, 261, 567
\bibitem[Patrick et al.(2022)]{Patrick2022} Patrick, L.~R., Thilker, D., Lennon, D.~J., et al.\ 2022, \mnras, 513, 5847
\bibitem[Paxton et al.(2011)]{Paxton2011} Paxton, B., Bildsten, L., Dotter, A., et al.\ 2011, \apjs, 192, 3 
\bibitem[Paxton et al.(2013)]{Paxton2013} Paxton, B., Cantiello, M., Arras, P., et al.\ 2013, \apjs, 208, 4 
\bibitem[Paxton et al.(2015)]{Paxton2015} Paxton, B., Marchant, P., Schwab, J., et al.\ 2015, \apjs, 220, 15 
\bibitem[Paxton et al.(2018)]{Paxton2018} Paxton, B., Schwab, J., Bauer, E.~B., et al.\ 2018, \apjs, 234, 34 
\bibitem[Pearson(1900)]{Pearson1900} Pearson, K.\ 1900, Philosophical Magazine Series 1, 50, 11
\bibitem[Reid et al.(1990)]{Reid1990} Reid, N., Tinney, C., \& Mould, J.\ 1990, \apj, 348, 98
\bibitem[Reimers(1975)]{Reimers1975} Reimers, D.\ 1975, Memoires of the Societe Royale des Sciences de Liege, 8, 369
\bibitem[Ren et al.(2019)]{Ren2019} Ren, Y., Jiang, B.-W., Yang, M., \& Gao, J.\ 2019, \apjs, 241, 35 
\bibitem[Ren et al.(2021)]{Ren2021} Ren, Y., Jiang, B., Yang, M., et al.\ 2021, \apj, 923, 232
\bibitem[Riebel et al.(2012)]{Riebel2012} Riebel, D., Srinivasan, S., Sargent, B., et al.\ 2012, \apj, 753, 71
\bibitem[Rodrigo et al.(2012)]{Rodrigo2012} Rodrigo, C., Solano, E., \& Bayo, A.\ 2012, IVOA Working Draft 15 October 2012.
\bibitem[Rodrigo \& Solano(2020)]{Rodrigo2020} Rodrigo, C. \& Solano, E.\ 2020, XIV.0 Scientific Meeting (virtual) of the Spanish Astronomical Society, 182
\bibitem[Roman-Duval et al.(2014)]{RomanDuval2014} Roman-Duval, J., Gordon, K.~D., Meixner, M., et al.\ 2014, \apj, 797, 86.
\bibitem[Ruffle et al.(2015)]{Ruffle2015} Ruffle, P.~M.~E., Kemper, F., Jones, O.~C., et al.\ 2015, \mnras, 451, 3504 
\bibitem[Salasnich et al.(1999)]{Salasnich1999} Salasnich, B., Bressan, A., \& Chiosi, C.\ 1999, \aap, 342, 131
\bibitem[Sargent et al.(2010)]{Sargent2010} Sargent, B.~A., Srinivasan, S., Meixner, M., et al.\ 2010, \apj, 716, 878
\bibitem[Sargent et al.(2011)]{Sargent2011} Sargent, B.~A., Srinivasan, S., \& Meixner, M.\ 2011, \apj, 728, 93
\bibitem[Schlafly \& Finkbeiner(2011)]{Schlafly2011} Schlafly, E.~F. \& Finkbeiner, D.~P.\ 2011, \apj, 737, 103
\bibitem[Scicluna et al.(2015)]{Scicluna2015} Scicluna, P., Siebenmorgen, R., Wesson, R., et al.\ 2015, \aap, 584, L10.
\bibitem[Skrutskie et al.(2006)]{Skrutskie2006} Skrutskie, M.~F., Cutri, R.~M., Stiening, R., et al.\ 2006, \aj, 131, 1163 
\bibitem[Smartt(2009)]{Smartt2009} Smartt, S.~J.\ 2009, \araa, 47, 63
\bibitem[Smartt(2015)]{Smartt2015} Smartt, S.~J.\ 2015, \pasa, 32, e016
\bibitem[Smith et al.(2001)]{Smith2001} Smith, N., Humphreys, R.~M., Davidson, K., et al.\ 2001, \aj, 121, 1111.
\bibitem[Smith(2014)]{Smith2014} Smith, N.\ 2014, \araa, 52, 487
\bibitem[Smith et al.(2015)]{Smith2015} Smith, N., Mauerhan, J.~C., Cenko, S.~B., et al.\ 2015, \mnras, 449, 1876
\bibitem[Soraisam et al.(2018)]{Soraisam2018} Soraisam, M.~D., Bildsten, L., Drout, M.~R., et al.\ 2018, \apj, 859, 73
\bibitem[Srinivasan et al.(2011)]{Srinivasan2011} Srinivasan, S., Sargent, B.~A., \& Meixner, M.\ 2011, \aap, 532, A54
\bibitem[Srinivasan et al.(2016)]{Srinivasan2016} Srinivasan, S., Boyer, M.~L., Kemper, F., et al.\ 2016, \mnras, 457, 2814
\bibitem[Sun et al.(2021)]{Sun2021} Sun, M., Jiang, B., Yuan, H., et al.\ 2021, \apjs, 254, 38
\bibitem[Sylvester et al.(1994)]{Sylvester1994} Sylvester, R.~J., Barlow, M.~J., \& Skinner, C.~J.\ 1994, \mnras, 266, 640. 
\bibitem[Sylvester et al.(1998)]{Sylvester1998} Sylvester, R.~J., Skinner, C.~J., \& Barlow, M.~J.\ 1998, \mnras, 301, 1083
\bibitem[Tantalo et al.(2022)]{Tantalo2022} Tantalo, M., Dall'Ora, M., Bono, G., et al.\ 2022, \apj, 933, 197. doi:10.3847/1538-4357/ac7468
\bibitem[Taylor(2005)]{Taylor2005} Taylor, M.~B.\ 2005, Astronomical Data Analysis Software and Systems XIV, 347, 29 
\bibitem[Udalski et al.(1992)]{Udalski1992} Udalski, A., Szymanski, M., Kaluzny, J., Kubiak, M., \& Mateo, M.\ 1992, \actaa, 42, 253 
\bibitem[Ueta \& Meixner(2003)]{Ueta2003} Ueta, T. \& Meixner, M.\ 2003, \apj, 586, 1338
\bibitem[Vanbeveren et al.(1998)]{Vanbeveren1998} Vanbeveren, D., De Loore, C., \& Van Rensbergen, W.\ 1998, \aapr, 9, 63
\bibitem[van Loon(2000)]{vanLoon2000} van Loon, J.~T.\ 2000, \aap, 354, 125
\bibitem[van Loon et al.(2005)]{vanLoon2005} van Loon, J.~T., Cioni, M.-R.~L., Zijlstra, A.~A., \& Loup, C.\ 2005, \aap, 438, 273 
\bibitem[van Loon et al.(2006a)]{vanLoon2006a} van Loon, J.~T., Marshall, J.~R., Cohen, M., et al.\ 2006, \aap, 447, 971. 
\bibitem[van Loon(2006b)]{vanLoon2006b} van Loon, J.~T.\ 2006, Stellar Evolution at Low Metallicity: Mass Loss, Explosions, Cosmology, 353, 211
\bibitem[Verhoelst et al.(2009)]{Verhoelst2009} Verhoelst, T., van der Zypen, N., Hony, S., et al.\ 2009, \aap, 498, 127 
\bibitem[Wang \& Chen(2019)]{Wang2019} Wang, S., \& Chen, X.\ 2019, \apj, 877, 116
\bibitem[Wang et al.(2021)]{Wang2021} Wang, T., Jiang, B., Ren, Y., et al.\ 2021, \apj, 912, 112
\bibitem[Werner et al.(2004)]{Werner2004} Werner, M.~W., Roellig, T.~L., Low, F.~J., et al.\ 2004, \apjs, 154, 1 
\bibitem[Willson(2000)]{Willson2000} Willson, L.~A.\ 2000, \araa, 38, 573.
\bibitem[Wolf et al.(2018)]{Wolf2018} Wolf, C., Onken, C.~A., Luvaul, L.~C., et al.\ 2018, \pasa, 35, e010 
\bibitem[Wright et al.(2010)]{Wright2010} Wright, E.~L., Eisenhardt, P.~R.~M., Mainzer, A.~K., et al.\ 2010, \aj, 140, 1868-1881 
\bibitem[Woosley \& Weaver(1986)]{Woosley1986} Woosley, S.~E. \& Weaver, T.~A.\ 1986, \araa, 24, 205. doi:10.1146/annurev.aa.24.090186.001225
\bibitem[Yang \& Jiang(2011)]{Yang2011} Yang, M., \& Jiang, B.~W.\ 2011, \apj, 727, 53 
\bibitem[Yang \& Jiang(2012)]{Yang2012} Yang, M., \& Jiang, B.~W.\ 2012, \apj, 754, 35
\bibitem[Yang et al.(2018)]{Yang2018} Yang, M., Bonanos, A.~Z., Jiang, B.-W., et al.\ 2018, \aap, 616, A175 
\bibitem[Yang et al.(2019)]{Yang2019} Yang, M., Bonanos, A.~Z., Jiang, B.-W., et al.\ 2019, \aap, 629, A91 
\bibitem[Yang et al.(2020)]{Yang2020} Yang, M., Bonanos, A.~Z., Jiang, B.-W., et al.\ 2020, \aap, 639, A116
\bibitem[Yang et al.(2021a)]{Yang2021a} Yang, M., Bonanos, A.~Z., Jiang, B., et al.\ 2021, \aap, 646, A141
\bibitem[Yang et al.(2021b)]{Yang2021b} Yang, M., Bonanos, A.~Z., Jiang, B., et al.\ 2021, \aap, 647, A167
\bibitem[Yoon \& Cantiello(2010)]{Yoon2010} Yoon, S.-C., \& Cantiello, M.\ 2010, \apjl, 717, L62
\bibitem[Zaritsky et al.(2002)]{Zaritsky2002} Zaritsky, D., Harris, J., Thompson, I.~B., et al.\ 2002, \aj, 123, 855
\bibitem[Zhang et al.(2012)]{Zhang2012} Zhang, T., Wang, X., Wu, C., et al.\ 2012, \aj, 144, 131



\end{thebibliography}
\end{document}